\documentclass[a4paper,11pt]{article}
\pdfoutput=1
\usepackage{jheppub,subcaption}
\usepackage{amsmath,amssymb,tabularx,graphicx,cancel,subcaption,xspace,flafter,ifpdf,todonotes,slashed,cleveref}
\usepackage{placeins}
\usepackage[force]{feynmp-auto}
\def\spa#1.#2{\left\langle#1\,#2\right\rangle}
\def\spb#1.#2{\left[#1\,#2\right]}
\def\spaa#1.#2.#3{\langle\mskip-1mu{#1}
                  | #2 | {#3}\mskip-1mu\rangle}
\def\spbb#1.#2.#3{[\mskip-1mu{#1}
                  | #2 | {#3}\mskip-1mu]}
\def\spab#1.#2.#3{\langle\mskip-1mu{#1}
                  | #2 | {#3}\mskip-1mu\rangle}
\def\spba#1.#2.#3{\langle\mskip-1mu{#1}^+
                  | #2 | {#3}^+\mskip-1mu\rangle}
\def\spav#1.#2.#3{\|\mskip-1mu{#1}
                  | #2 | {#3}\mskip-1mu\|^2}
\def\jc#1.#2.#3{j^{#1}_{#2#3}}
\def\blfootnote{\xdef\@thefnmark{}\@footnotetext}

\newcommand{\as}{\ensuremath{\alpha_s}\xspace}
\newcommand{\HEJ}{{\tt HEJ}\xspace}
\newcommand{\HEJone}{{\tt HEJ1}\xspace}
\newcommand{\HEJtwo}{{\tt HEJ2}\xspace}

\newcommand{\NLO}{{NLO}\xspace}
\newcommand{\NNLO}{{NNLO}\xspace}
\newcommand{\LA}{\ensuremath{\hat s/p_t^2}\xspace}
\newcommand{\MRK}{MRK\xspace}
\newcommand{\QMRK}{QMRK\xspace}
\preprint{\begin{minipage}[t]{\widthof{SAGEX-20-28 }}{
      DESY 20-233\\
      DCPT/20/134\\
      IPPP/20/67\\
      MCnet-20-27\\
      SAGEX-20-28
    }
  \end{minipage}
}
\title{Combined subleading high-energy logarithms and NLO accuracy for W production in association with multiple jets}
\author[a]{Jeppe~R.~Andersen,}
\author[a]{James A.~Black,}
\author[b]{Helen M.~Brooks,}
\author[c]{Emmet P.~Byrne,}
\author[d]{Andreas~Maier}
\author[c]{and Jennifer~M.~Smillie}

\emailAdd{jeppe.andersen@durham.ac.uk}
\emailAdd{james.a.black@durham.ac.uk}
\emailAdd{helen.brooks@monash.edu}
\emailAdd{emmet.byrne@ed.ac.uk}
\emailAdd{andreas.martin.maier@desy.de}
\emailAdd{j.m.smillie@ed.ac.uk}

\affiliation[a]{Institute for Particle Physics Phenomenology, University of Durham, Durham, DH1 3LE, UK}
\affiliation[b]{School of Physics and Astronomy, Monash University, Clayton, VIC 3800, Australia}
\affiliation[c]{Higgs Centre for Theoretical Physics, University of Edinburgh,\\
  Peter Guthrie Tait Road, Edinburgh, EH9 3FD, UK}
\affiliation[d]{Deutsches Elektronen-Synchrotron, DESY, Platanenallee 6,
  15738 Zeuthen, Germany}

\abstract{% \todo{This is the first calculation outside of NLL BFKL of
  Large logarithmic corrections in $\hat s/p_t^2$ lead to substantial variations
  in the perturbative predictions for inclusive $W$-plus-dijet processes at
  the Large Hadron Collider.  This instability can be cured by summing the
  leading-logarithmic contributions in $\hat s/p_t^2$ to all orders in
  $\alpha_s$.  As expected though, leading logarithmic accuracy is
  insufficient to guarantee a suitable description in regions of phase space
  away from the high energy limit.

  We present (i) the first calculation of all partonic channels contributing
  at next-to-leading logarithmic order in $W$-boson production in association
  with at least two jets, and (ii) bin-by-bin matching to next-to-leading
  fixed-order accuracy.  This new perturbative input is implemented in
  \emph{High Energy Jets}, and systematically improves the description of
  available experimental data in regions of phase space which are formally
  subleading with respect to $\hat s/p_t^2$.

}
\begin{document}
\maketitle
\flushbottom

% \newpage
% \tableofcontents
% \newpage
\section{Introduction}
\label{sec:intro}
% LL gives good description in the leading phase space
% FKL constitutes small component in sub-leading phase space
% FO matching
% LL of sub-leading channel

The continuing success of the Large Hadron Collider (LHC) is challenging the
theoretical particle physics community to continuously improve
the precision of theory predictions.
One challenge to our standard approaches is that the
complexity of a perturbative scattering process increases for increasing
centre-of-mass energy and fixed transverse scales.
For example, LHC studies of additional jet activity from dijet systems in regions of
phase space with large dijet invariant mass or rapidity separation~\cite{Aad:2011jz,Aad:2014pua} reveal an
amount of additional jet activity above and beyond what can be described by
fixed-order \NLO approaches. While predictions at \NNLO exist for dijet
production, such advanced predictions have not been applied to these
analyses.
Even at the 1.96~TeV Tevatron, analyses of additional
jet activity in e.g.~$W$-production in association with dijets at
D0~\cite{Abazov:2013gpa} consistently revealed a
tension between data and the standard set of predictions in the same regions of
phase space.

The issues identified with NLO predictions for very exclusive observables may not show up in the
inclusive two-jet distributions in e.g.~$m_{j_1j_2}$ or $\Delta y_{j_1 j_2}$, but
large variations and discrepancies are found, for example, in the description of
additional QCD radiation in $W+$dijets and $Z+$dijets, see e.g.~\cite{Aaboud:2017emo}. Such studies are
highly relevant in their own right, but are also necessary for the development
of the systematic description of perturbative corrections to the QCD channel of
Higgs boson+dijets production. This is particularly true within the phase space cuts used for the study of
the Vector Boson Fusion channel. The similarity in the radiative
corrections to the various processes is caused by the universality of the QCD radiation pattern
arising from a colour octet exchange between jets.

The colour-octet exchange
emphasises the contribution from real-emission, higher-order perturbative
corrections and is also accompanied by a tower of logarithms from virtual
corrections. Both sources of perturbative corrections are included in the
BFKL equation~\cite{Fadin:1975cb,Kuraev:1976ge,Kuraev:1977fs,Balitsky:1978ic},
which captures the dominant logarithms ($\ln\hat s/p_t^2$, where $\hat s$ is the
partonic centre-of-mass energy and $p_t$ is the transverse momentum scale) that govern
the high-energy limit of the on-shell scattering cross sections.
Such logarithms are not systematically included in the standard perturbative
methods for obtaining predictions for LHC observables.

The dominant logarithmic corrections to the $n$-jet production rate of the form
$\alpha_s^m \ln^m\hat s/p_t^2$ (and, as introduced in this paper, specific
subleading contributions of order $\alpha_s^{m+1} \ln^m\hat s/p_t^2$) are,
however, systematically included in the calculations of the on-shell partonic
scattering amplitudes within the framework of \emph{High Energy
  Jets} (\HEJ)~\cite{Andersen:2009nu,Andersen:2009he,Andersen:2011hs,Andersen:2012gk}. The
framework is based on a systematic power-expansion of the $n$-body on-shell scattering
matrix element, which controls the logarithmic corrections from real
emissions, coupled with the Lipatov Ansatz for the structure of the virtual corrections.
The virtual corrections
not only cancel the infrared poles from the real corrections, but also
contribute to the finite part of the matrix element. This approach captures
the logarithmic corrections from both real and virtual corrections, and
indeed both are necessary for obtaining full logarithmic accuracy.\footnote{This
is in contrast to the resummation of soft-collinear logarithms in the
standard formulation of a traditional leading-colour parton shower, where the
assumed Sudakov form of the virtual corrections keeps the shower
\emph{unitary}, allowing for a probabilistic interpretation of emission.}

In \HEJ, the sum over $n$ and the integration over
each $n$-body phase space point is performed explicitly using Monte Carlo sampling,
and as such the predictions are made at the partonic level with direct access
to the 4-momenta of each of the $n$ particles. The framework merges
fixed-order (currently leading order), high-multiplicity matrix elements with
an all-order description of the dominant logarithms. The formalism has been
implemented for several processes, and compares favourably to data in studies
of the jet activity from dijet systems
production~\cite{Aad:2011jz,Chatrchyan:2012gwa,Aad:2014pua},
also in association with electro-weak
bosons~\cite{Abazov:2013gpa,Aad:2014qxa,Aaboud:2017fye}. These studies
indicate that for large invariant mass, and large rapidity differences, the
high-energy logarithms are important, and their systematic inclusion improves
the accuracy of the theoretical prediction.

In this paper, for the first time, we apply the \HEJ formalism to well-defined
partonic channels in inclusive $W+$dijet production which are
formally \emph{subleading} in the high energy limit. This resummation
of the LL corrections to partonic channels contributing at subleading level
for $\as^3$ and $\as^4$ does not achieve full NLL accuracy for the rates of
$W+$dijets, since it does not include the sub-leading corrections to the channels
contributing also at LL level. %Such a calculation is beyond the scope of the
%current paper.
However, the new contribution is well-defined,
and as we will demonstrate, the inclusion of
such subleading channels leads to a significantly improved description of
transverse observables, and also to a lesser dependence of the predictions
upon the matching to fixed-order.

In \cref{sec:HEScatter}, we commence by reviewing the scaling of the amplitude
and the definition of leading and subleading channels in the high energy limit.
We then define and explain the generic building blocks that are required for the
amplitude-based approach used in the \HEJ formalism.  In~\cref{sec:subleading} we
present the calculation of the new building blocks which are necessary for the
inclusion of subleading channels. In~\cref{sec:macthmerge} we present a new
method of matching which increases the fixed-order accuracy of our predictions
for measured distributions to next-to-leading order.  Finally we present
comparisons of our results to experimental data
in~\cref{sec:results} and our conclusions in~\cref{sec:conclusions}.

%%% Local Variables:
%%% mode: latex
%%% TeX-master: "main"
%%% End:

\section{High Energy Scattering}
\label{sec:HEScatter}

\subsection{Scaling of amplitudes and cross sections in the High Energy Limit}
\label{sec:HEscaling}
% Discussion a la Higgs paper.
% \begin{itemize}
%  \item Talk about t channel propagators and spin. Include ``blob'' diagrams. Figure 1?
%  \item Talk about connection with final states/ channels/ proccesses.
%  \item Introduce FKL / non-FKL terminology.
%  \item Distinguish between \textit{processes} which are suppressed,  versus real (soft) and virtual corrections that are needed for a full NLL calculation. Explain why including subleading \textit{channels}
%  is a well-defined thing to do, and moreover is worthwhile.
% \end{itemize}

% \hrule

The all-order resummation in \HEJ improves the perturbative description of
scatterings by
systematically including the dominant contributions from the
logarithms in \LA which arise at each order in the coupling.  The
centre-of-mass energy of LHC collisions is such that these logarithms can become
large and spoil the convergence of the perturbative series, particularly in
analyses where a large rapidity separation or large
invariant mass between jets is required.  Within \HEJ the calculation of the
logarithmic contributions to the scattering amplitude is organised in terms
of virtual and real corrections just like for standard fixed-order
perturbation theory. The logarithmic expansion of the cross section relies on two concepts:
\begin{enumerate}
\item The kinematic regions important for the logarithmic accuracy,
\item The systematic approximation to the scattering amplitude, ensuring the
  logarithmic accuracy of the cross section in these regions.
\end{enumerate}

\subsubsection{Kinematic regions}
\label{sec:KinematicRegions}
\begin{figure}[tb]
  \centering
  \includegraphics[width=0.4\textwidth]{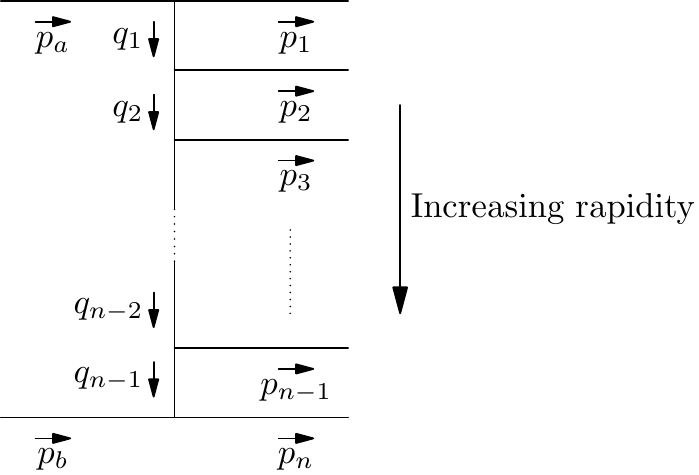}
  \caption{An illustration of the definition of the $t$-channel momenta in
    non-overlapping channels or planar diagrams.}
  \label{fig:ts}
\end{figure}

Leading Logarithmic accuracy (LLA) in \LA is controlled by the
Multi-Regge Kinematic (\MRK) limit of the $2\to n$ scattering process, where
the invariant mass between each set of particles is large, while all
transverse scales are similar:
\begin{align}
  \forall i,j\in\{1,\ldots,n\}, i\not=j: \quad s_{ij} \to \infty, \qquad |p_{i\perp}|\sim|p_{j\perp}|.
  \label{eq:MRKlimit1}
\end{align}
The set of $t$-momenta in the so-called \emph{non-overlapping channels}, or planar
diagrams (see \cref{fig:ts}), are defined as
\begin{align}
  q_i=p_a-\sum_{l=1}^i p_l.
  \label{eq:qis}
\end{align}
In the \MRK limit
\begin{align}
  t_i=q_i^2\to -q_{i\perp}^2.
  \label{eq:tis}
\end{align}
The reason that the logarithmic accuracy is controlled by the regions of \MRK
and Quasi Multi-Regge Kinematics (\QMRK) will become clear after the discussion of the scaling of the
amplitude in these limits.
In fact, the MRK limit requires also the transverse part of these $t$-channel
momenta to be finite in the MRK limit, such that the conditions in
eq.~\eqref{eq:MRKlimit1} are supplemented with the requirement
\begin{align}
  \forall i,j:\ |q_{\perp i}|\sim|p_{\perp j}|
  \label{eq:MRKtis}
\end{align}
such that \emph{all} transverse scales relevant for the evaluation of the
amplitudes are similar.

If we let the indices increase with the ordering of the particles in
rapidity, the \MRK limit can also be expressed as
\begin{align}
  \begin{split}
    &\forall i\in\{1,\ldots,n-1\}:  \ y_i \ll y_{i+1} \ \\
    &\forall i\in\{1,\ldots,n-1\},j\in\{1,\ldots,n\}: |q_{\perp
      i}|\sim|p_{\perp j}|\sim p_{\perp},
  \end{split}
\label{eq:mrklimit}
\end{align}
for some fixed $p_\perp$, and where the so-called \emph{strong ordering} of
rapidities means the limit of infinite rapidity separation between each
neighbouring pair of particles.  Next-to-Leading Logarithmic accuracy (NLLA) in \LA receives
contributions also from all the $n-1$ \QMRK limits, where
the requirement of a large separation is relaxed for exactly one
pair $(j, j+1)$ of particles with arbitrary $j \in \{1,\ldots,n-1\}$:
\begin{align}
\forall i\in\{1,\ldots,n-1\}, i\not=j: \quad y_i \ll y_{i+1},\qquad
  |q_{\perp i}|\sim p_{\perp}.
\label{eq:qmrklimit}
\end{align}
The factorising of scattering amplitudes in the QMRK will be further
discussed in sections~\ref{sec:ScalingAgain} and
\ref{sec:HEJamplitudes}. With respect to the kinematic limits, in the QMRK
the pair of momenta $(j,j+1)$ is treated as one momentum in the MRK
considerations, and the limit requires just the size of the sum of transverse momenta
$(j,j+1)$ to be small, and therefore the induced change in $|q_i|$ to be small.

\subsubsection{Scaling of Scattering Amplitudes in the limits of \MRK and \QMRK}
\label{sec:ScalingAgain}
We will start this section with a discussion of the historical roots of the
application of Regge theory in the study of the strong force. The scaling of
amplitudes for the scattering of specific hadrons is described effectively
using Regge theory for multi-particle production~\cite{Brower:1974yv},
describing the scaling with energy due to exchange of various specific
mesons, e.g.~$\pi^0$ or $\rho$. A partial wave expansion of the scattering
amplitude exposes a connection between the spin of the exchanged particles
and the scaling of the amplitude with energy. For the scattering $2\to n$
with momenta $ab\to1\cdots n$, the scaling with $s$ in the MRK limit of the
scattering amplitude is
\begin{align}
  \mathcal{M} \sim s_{12}^{\alpha_1(t_1)} \ldots
  s_{n-1,n}^{\alpha_{n-1}(t_{n-1})} \gamma\,,
\label{eq:reggescaling}
\end{align}
where the $s_{ij}$ are the invariant masses of rapidity-ordered pairs of
particles.  The $t_i$ are defined as $t_i=q_i^2$ with $q_i$ defined in \cref{eq:qis}. The factor $\gamma$ depends on
transverse scales only, which are suppressed compared to  $s$ in the MRK and
QMRK limits. The exponent
of the invariant $s_{i,i+1}$, $\alpha_i(t_i)$, is the effective spin of any
possible particle exchanged in the $i^\mathrm{th}$ $t$-channel. The dominant
contributions at large $s_{i, i+1}$ will therefore be generated from processes
where the particles exchanged in the $t_i$-channel have the largest possible
spin.

It is perhaps surprising that the partial wave analysis and the scaling of
scattering amplitudes applies also to processes in the gauge theory of
QCD. This insight was obtained in the seminal
work~\cite{Fadin:1975cb,Kuraev:1976ge,Kuraev:1977fs} of Fadin, Kuraev and
Lipatov. Na\"ively, the quarks and gluons of QCD allow for a scaling of the
amplitude itself with invariants of power $\frac 1 2$ and $1$ respectively -- and therefore
of the cross section with twice these powers\footnote{These powers receive
  radiative corrections.}. The analysis should of course not be tied to
gauge-dependent subsets of Feynman diagrams, since the freedom of the gauge
choice will allow for contributions to be moved between diagrams with
different assignments of particles in the propagators. A simple and
illustrative example~\cite{Andersen:2009he} is provided by the process of
$qg\to qg$. One of the three contributing Feynman diagrams contains a gluon
exchange in the $t$-channel (meaning the propagator momentum is $t=p_a-p_1$,
where $p_a$ and $p_1$ is the momentum of the incoming and outgoing quark
respectively). A suitable choice for the gauge vector in an axial gauge renders
the contribution from this diagram 0, such that the process requires the
calculation of just two Feynman diagrams, with propagators of
$s=(p_a+p_b)^2$ and $u=(p_a-p_2)^2$. However, the scaling with $s$ of the
gauge-invariant amplitude is of course not affected by such gauge
choices. For a given process one finds that if a planar diagram exists
between the rapidity-ordered states, and this diagram has a gluon exchange in
the $t$-channel, then the amplitude for the process will (at tree-level) scale as $s_{ij}^1$,
and therefore the contribution to the cross section as $s_{ij}^2$. If no such
planar diagram exists (and therefore the particle of largest spin exchanged for the
planar diagrams is a quark), then the scaling of the amplitude in the MRK
limit is $s_{ij}^{1/2}$ and the cross section therefore scales as $s_{ij}$. The reason for resorting to
planar diagrams in this discussion is to ensure the
assignment of momenta in the propagators in terms of the rapidity-ordered
external momenta is unchanged when a different particle assignment is discussed.

\subsubsection{Identifying Leading and Next-to-Leading Logarithmic
  Contributions to the Cross Section}
\label{sec:LLandNLLToCrossSection}
We will now discuss how the logarithmic corrections to the cross section
arise. In this section we focus just on the logarithms using the scaling
arguments for the scattering amplitudes, as they are captured by the
BFKL formalism~\cite{Fadin:1975cb,Kuraev:1976ge,Kuraev:1977fs,Fadin:1996nw}. This presentation will additionally serve to highlight the
improvements made in \HEJ, while maintaining the logarithmic accuracy.

The cross section for the contribution from the partonic $2\to n$ scattering
to e.g.~dijet production is calculated by the explicit phase space integral
of the scattering amplitudes, and the flux factors involving the parton
distribution functions.
\begin{align}
  \begin{split}
    \label{eq:dijet}
    \sigma^{n}_{2j}=&\sum_{f_{i1}, f_{i2}}\ %\sum_{n=2}^\infty\
    \prod_{i=1}^n\left(\int_{p_{j\perp}=0}^{p_{j\perp}=\infty}
      \frac{\mathrm{d}^2\mathbf{p}_{j\perp}}{(2\pi)^3}\
      \int \frac{\mathrm{d} y_j}{2}
    \right)\
    \frac{\overline{|\mathcal{M}^{f_{i1} f_{i2}\to f_1 f_2\cdots f_{n}}(\{ p_j\})|}^2}{\hat s^2} \\
    &\times\ \ x_a f_{A,f_{i1}}(x_a, Q_a)\ x_b f_{B,f_{i2}}(x_b, Q_b)\ (2\pi)^4\ \delta^2\!\!\left(\sum_{k=1}^n
      \mathbf{p}_{k\perp}\right )\ \mathcal{O}_{2j}(\{p_j\}),
  \end{split}
\end{align}
where the sum is over the flavours of incoming partons, the phase space
integrals are the standard Lorentz-invariant measure, the square of the
scattering amplitude is summed and averaged over colours and spins, the flux
factor is given by the momentum fractions and parton distribution functions as
$x_a f_{A,f_{i1}}(x_a, Q_a)\ x_b f_{B,f_{i2}}(x_b, Q_b)/\hat s^2$, and the
remaining delta-functional ensures conservation of transverse momentum. The
momenta of the incoming partons are reconstructed by momentum
conservation. The last factor, $\mathcal{O}_{2j}(\{p_j\}),$ imposes the
requirement of two jets in the event.

For large $\Delta y$, the simple $2\to2$ processes are dominated by those
permitting a gluon assignment in a planar diagram, and the square of the
scattering amplitudes scale as
\begin{align}
  |\mathcal{M}_2|^2\to s_{ab}^2/t^2\to s_{ab}^2 \Gamma_2,
\end{align}
where $\Gamma_2$ depends on just transverse scales, since $t\to -p_{1\perp}^2=-p_{2\perp}^2$.

Let us now consider the perturbative real emission corrections to the
scattering for a given kinematic configuration. The following discussion
applies to the perturbative corrections to any configuration, but to be
specific consider a scattering which at Born level is $2\to2$. Let $\Delta y$
denote the distance in rapidity between the two jets. We are particularly
interested in the case of large $\Delta y\sim\log(\hat s/p_\perp^2)$. The
$2\to3$ real radiation phase space with all momenta well separated in
rapidity is systematically increasing for increasing $\Delta y$. This is the
MRK limit. According to the discussion in the previous
section, for the cases where a gluon exchange can be assigned to the planar
diagram connecting the rapidity-ordered assignments of flavours, the square
of the $2\to3$ scattering amplitude $|M_3|^2$ will scale as
$|M_3|^2\propto s_{12}^2 s_{23}^2\Gamma_3$, where $\Gamma_3$ depends on
transverse momenta only (which do not increase in the MRK limit). In the MRK
limit,
\begin{align}
  \label{eq:sabs12s23}
 s_{ab}\propto s_{13}\propto ( s_{12}\ s_{23})/|p_{2\perp}|^2,
\end{align}
which is most easily shown using light-cone coordinates. Therefore, the matrix element
\begin{align}
  |\mathcal{M}_3|^2\propto s_{12}^2\ s_{23}^2\ \Gamma_3\propto s_{ab}^2\
  \Gamma
\label{eq:m3sq}
\end{align}
where $\Gamma$ depends on transverse scales only. The contribution to the
cross section from eq.~\eqref{eq:dijet} would then be found as an integration
over $|\mathcal{M}_3|^2/\hat s^2$ with $\hat s^2=s_{ab}^2$ over the rapidity
of $p_2$. As will be demonstrated now, the power expansion (in $s_{ab}$) of
the square of the scattering amplitude leads to a logarithmic (in $s_{ab}$) expansion of
the cross section. In the limit studied of large $\Delta y$ between the two primary
jets, the contribution to the parton momenta fractions $x_a, x_b$ are
dominated by the contribution from the forward and backward partons
respectively, and ignoring the change induced by the additional emission, the
correction to the cross section is
\begin{align}
  \label{eq:LLcorrection}
  \as\ \Delta y\ \int_{p_{2\perp}=0}^{p_{2\perp}=\infty}
      \mathrm{d}^2\mathbf{p}_{2\perp}\  \Gamma.
\end{align}
As it stands, the integral over transverse scales is divergent; however, this
divergence will be regulated by virtual corrections, leaving a finite
remainder. The result is that the configurations which display the scaling
in eq.~\eqref{eq:m3sq}, will lead to a perturbative correction scaling
parametrically as $\as\Delta y\propto\as\log (s/p_\perp^2)$. This behaviour
is found for each order in \as for the configurations which allow for
spin-1 $t$-channel exchanges between each particle in the rapidity-ordered
planar diagrams. Since $\log(s/p_\perp^2)\to\Delta y$ in the MRK limit, such
configurations therefore lead to leading logarithmic $\log(s/p_\perp^2)$
corrections. As an example, a $2\to 3$ flavour assignment leading to such a
scaling is $qQ\to qgQ$ or $gg\to ggg$ (where the flavours correspond to the
ordering of momenta $p_a p_b\to p_1 p_2 p_3$ on figure~\ref{fig:ts}).

While the
assignment of a quark exchanged in the $t$-channel  compared to a gluon assignment
leads to a suppression in the MRK limit, there is no suppression of one
configuration over the other in the
respective QMRK limit, where one
invariant, e.g. $s_{12}$, can be finite.  Therefore, in the QMRK, subprocesses whose $t$-channel
assignments differ only by a quark or gluon
in that one $t$-channel will contribute at the same level.  The fact that QMRK-contributions are solely NLL then rests with the fact
that the QMRK phase space itself is subleading compared to MRK.

As hinted above, virtual corrections will also contribute to the leading
logarithms. Not just will these virtual corrections cancel the
singularities introduced by the real corrections studied above, they will
also contribute a finite remainder affecting the normalisation of the cross
section and relative contribution from each multiplicity. The leading
logarithmic contribution again arises from well-defined partonic assignments
in the planar diagrams which allow $t$-channel gluon exchanges. These of
course are the configurations attracting leading logarithmic corrections from
the real emissions -- and thus are in need of virtual corrections to cancel
the singularity introduced. The first few orders of the perturbative
corrections can be calculated
explicitly~\cite{DelDuca:1995hf,DelDuca:2001gu,Bogdan:2002sr}, and are
captured by the so-called ``Lipatov
ansatz''~\cite{Fadin:1975cb,Kuraev:1977fs,Balitsky:1978ic}, which captures the
leading logarithmic contribution from the virtual corrections by replacing
each of the factors $s/t$ in the amplitude for a leading-logarithmic contribution with
\begin{align}
  \label{eq:LipatovAnsatz}
  \frac s t \to \frac 1 2 \left[ \left(\frac{-s}{-t}\right)^{j_G(t)}-\left(\frac{s}{-t}\right)^{j_G(t)}\right],
\end{align}
where $j_G(t)={1+\omega(t)}$ is called the Regge trajectory of the gluon.

Sub-leading corrections arise from two sources:
\begin{enumerate}
\item the leading contribution from sub-leading configurations (and the leading-logarithmic corrections to these),
\item next-to-leading logarithmic corrections to leading-logarithmic configurations
\end{enumerate}
In the current paper, we will calculate the full sub-leading contributions from the
first point to all processes contributing to $pp\to W+\ge2$jets. It is therefore relevant to
understand how these next-to-leading logarithmic contributions arise, which
is the focus of the remaining part of the subsection. We note in passing that
part of these contributions to the sub-leading corrections were presented
in the framework of \HEJ in Ref.~\cite{Andersen:2017kfc} in the context of
Higgs-boson-plus-jets.

\begin{figure}[btp]
  \centering
  \includegraphics[width=0.5\textwidth]{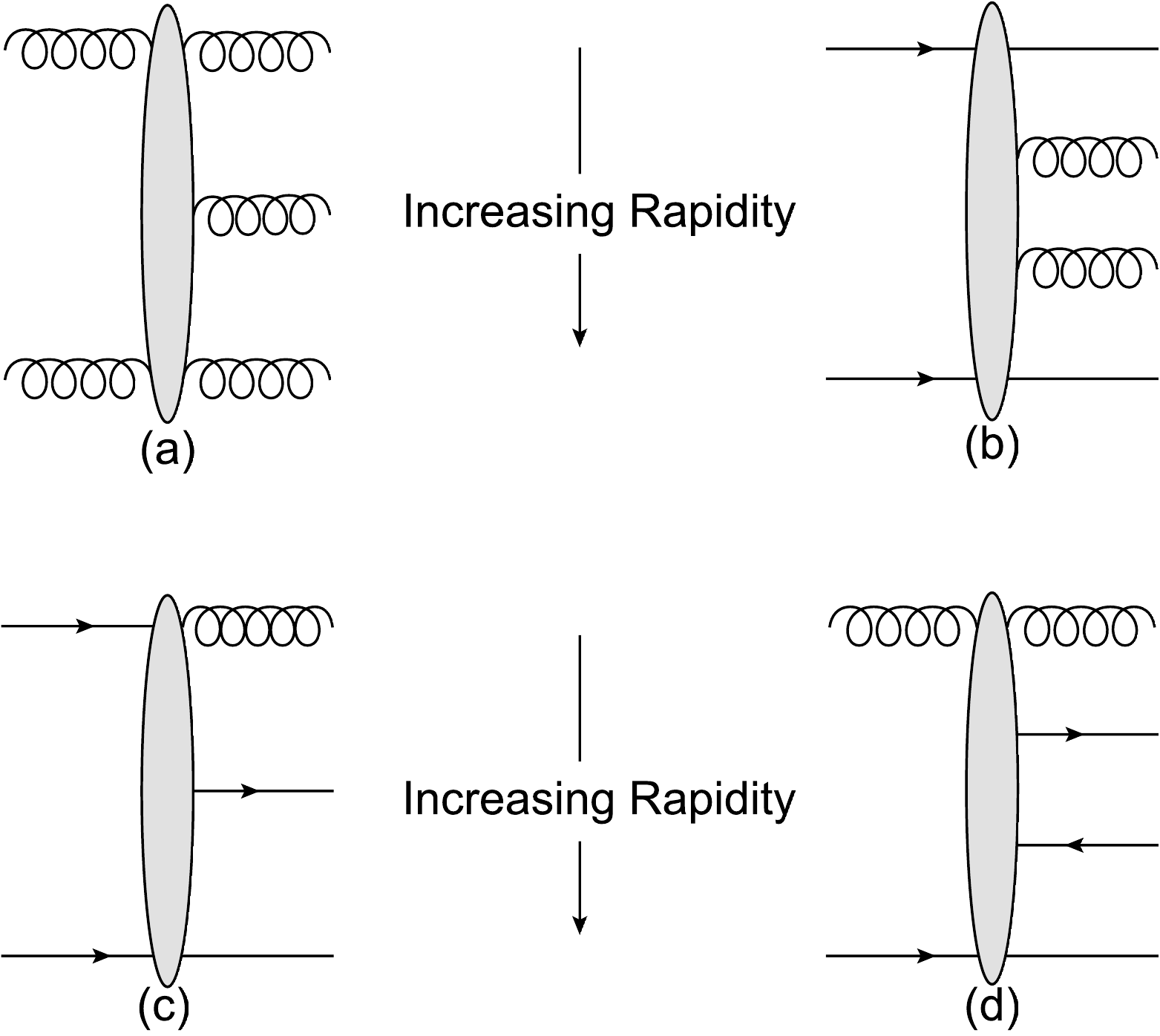}
  \caption{The particle flavour and momentum configurations
    in (a) and (b) allow the maximum number of $t$-channel gluon exchanges
    and so contribute at LL, while those in (c) and (d) require one $t$-channel
    quark-exchange and hence contribute at NLL. }
  \label{fig:fklvsnonfkl}
\end{figure}

The flavour assignments to the planar momentum-flow which contribute at NLL
are identified by revisiting the arguments in
section~\ref{sec:ScalingAgain}. If the exchange in the $t_i$-channel is
mediated by a quark instead of a gluon, the square of the amplitude scales
with one less power $s_{ii+1}=2 p_i\cdot p_{i+1}$. In the example of
$2\to3$ scattering above, the momentum assignment leading to one less power
of $s_{12}$ could be e.g.~$qQ\to gqQ$ or $gg\to qQ g$, which both have a
quark propagator assigned to the momentum $q_1$ in figure~\ref{fig:ts} (LL
and NLL assignments for processes are illustrated in
figure~\ref{fig:fklvsnonfkl}). The scaling of such configurations is
\begin{align}
  \label{eq:NLLscaling2to3}
  |\mathcal{M}_3^{NLL}|^2\propto s_{12}\ s_{23}^2\ \Gamma^{NLL}_3\propto s_{23} s_{ab}\
  \Gamma^{NLL},
\end{align}
which is one power of $s_{ab}$ down in the MRK from the scaling of the LL
contribution studied in eq.~\eqref{eq:m3sq}. Since
$\log(s_{ab}/p_t^2)\to\Delta y$, this leads to a perturbative correction of
order
\begin{align}
  \label{eq:NLL2to3correction}
  \as\ \int_{p_{2\perp}=0}^{p_{2\perp}=\infty}
      \mathrm{d}^2\mathbf{p}_{2\perp}\  \Gamma^{NLL},
\end{align}
which is sub-leading to the contribution in eq.~\eqref{eq:LLcorrection} from
the LL configurations. The all-order corrections of order $\alpha_s\Delta y$
to these $\alpha_s$ corrections can be systematically calculated by including
the LL corrections to the gluon $t$-channel exchanges in terms of further
real gluon emissions, and the virtual corrections according to the Lipatov Ansatz.

\begin{figure}[tb]
  \centering
  \begin{subfigure}{0.49\textwidth}
   \includegraphics[width=\textwidth]{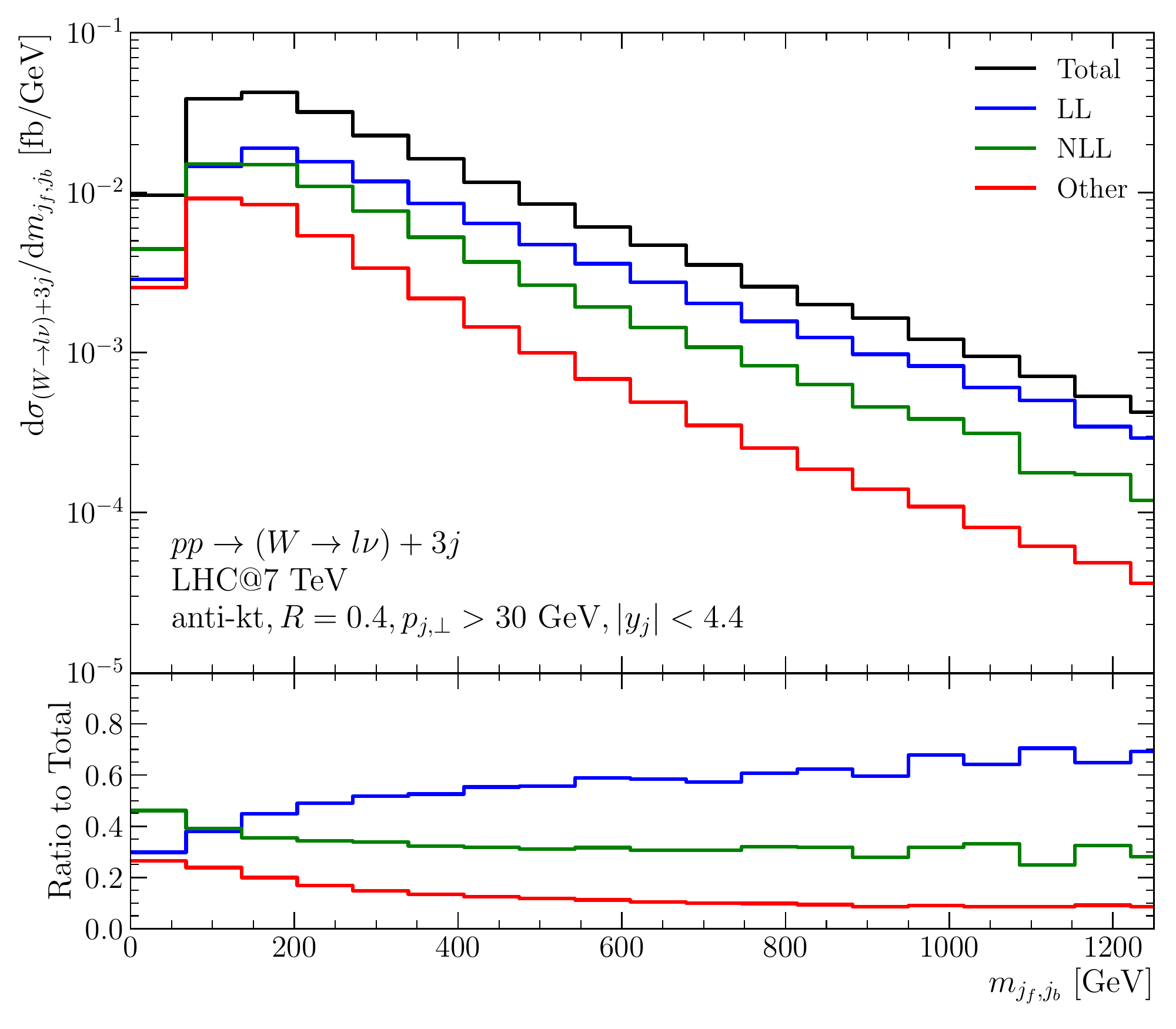}
  \caption{}
  \end{subfigure}
\begin{subfigure}{0.49\textwidth}
  \centering
   \includegraphics[width=\textwidth]{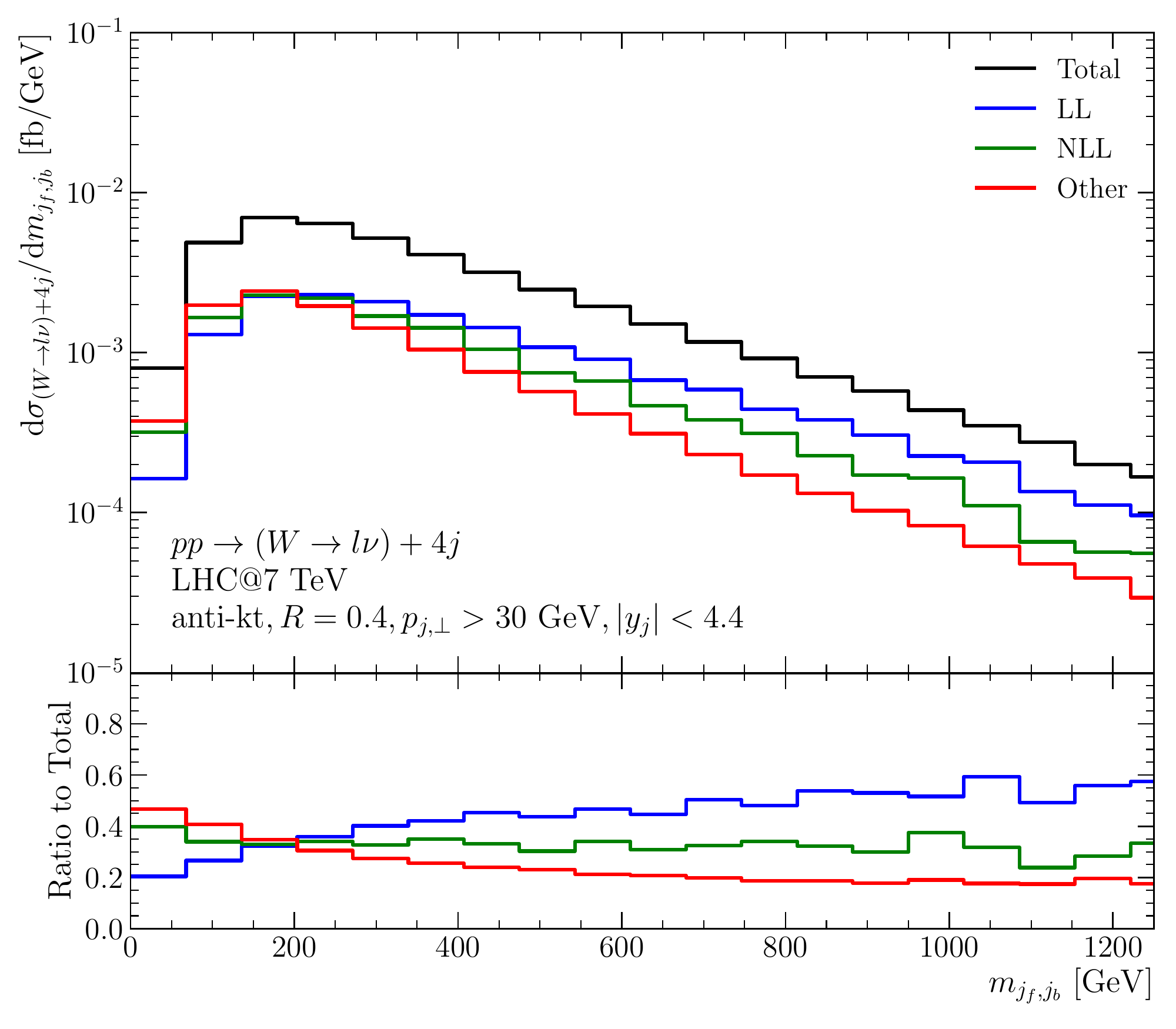}
   \caption{}
  \end{subfigure}
   \caption{The leading order (a) $W+3$-jet and (b) $W+4$-jet cross sections as a
    function of the invariant mass between the most forward and backward jets,
    $m_{\rm fb}$.  In addition to the total (black), we also show the split into
    the contributions from the LL
    configurations (blue), NLL configurations (green) and other
    configurations (red).  As $m_{\rm fb}$ increases, the LL
    configurations increase in dominance but the NLL configurations
    remain significant.}
  \label{fig:mfbcomponents}
\end{figure}

\begin{figure}[tb]
  \centering
  \begin{subfigure}{0.49\textwidth}
   \includegraphics[width=\textwidth]{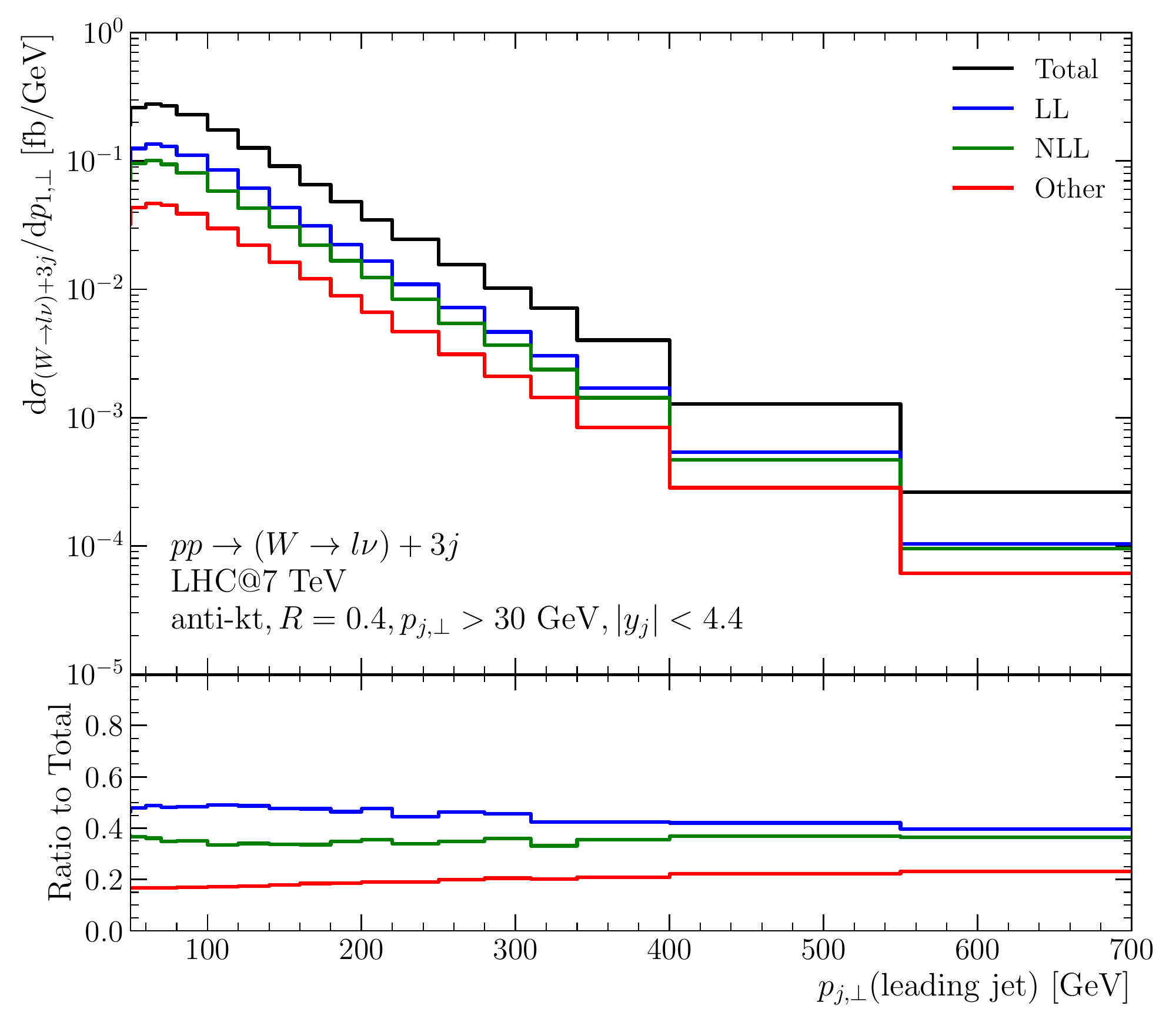}
  \caption{}
  \end{subfigure}
\begin{subfigure}{0.49\textwidth}
  \centering
   \includegraphics[width=\textwidth]{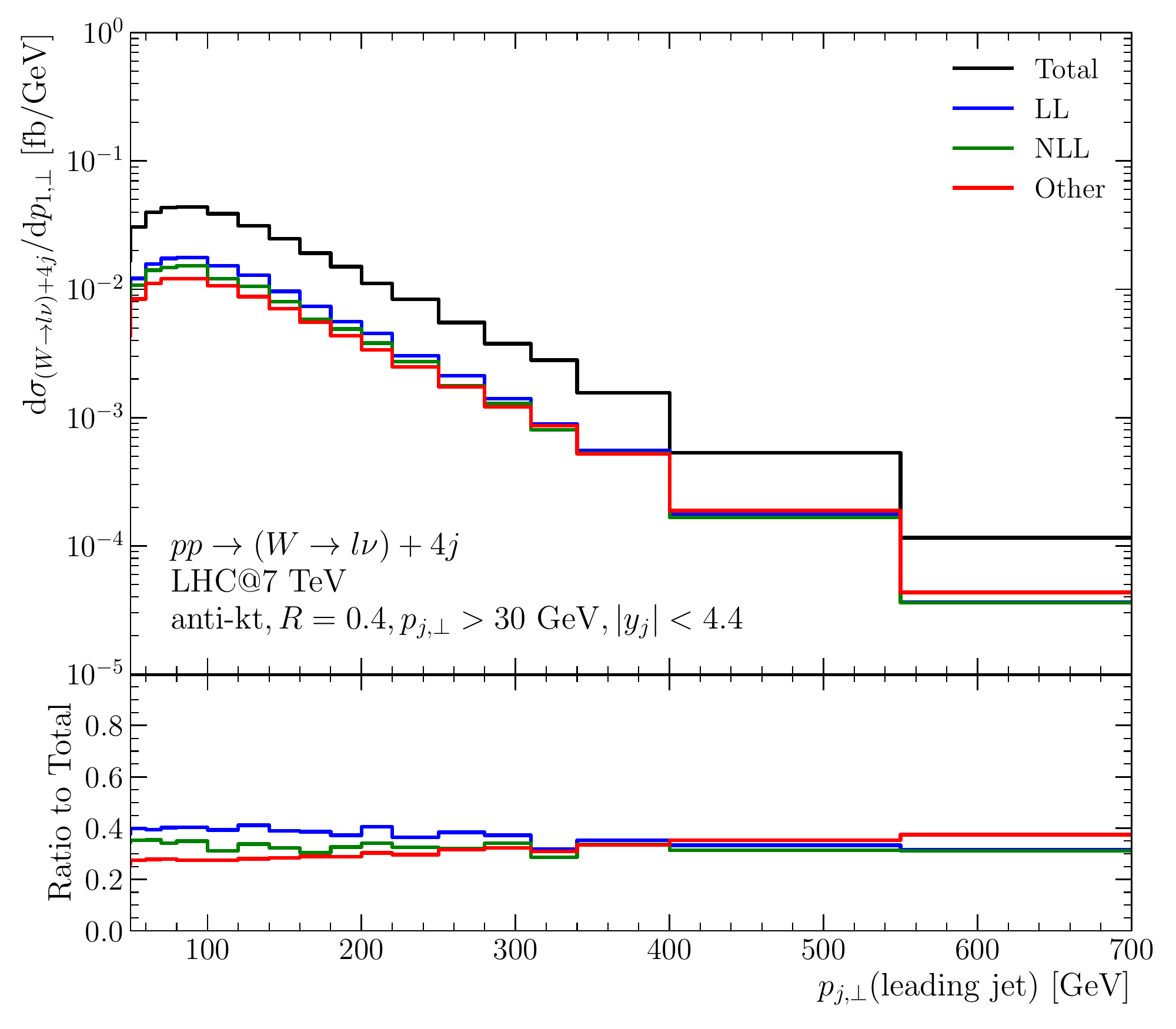}
   \caption{}
  \end{subfigure}
   \caption{The leading order (a) $W+3$-jet and (b) $W+4$-jet cross sections as a
    function of the transverse momentum of the hardest jet, $p_{\perp,1}$.  In
    addition to the total (black), we also show the split into the contributions
    from the LL
    configurations (blue), NLL configurations (green) and other
    configurations (red).  As $p_{\perp,1}$ increases, the LL
    configurations decrease in importance underlining the importance of
    also applying resummation to the NLL contributions.}
  \label{fig:pt1components}
\end{figure}

The importance of these NLL configurations may be demonstrated by considering
contributions from LL and NLL configurations to the Born-level processes for
$2\to3$ and $2\to4$.  In \cref{fig:mfbcomponents} we illustrate the
differential distributions for (a) $W+3$-jet and (b) $W+4$-jet production as
a function of the invariant mass between the most forward and backward jets,
split into the LL component (blue), the NLL component (green) and other
configurations (red).  The results are all obtained using the full SM
scattering amplitudes, and classifying the contributions according to the
flavour and momentum configurations of the states, as summarised in \cref{tab:LLNLLOther}. These plots were made for
7~TeV proton-proton collisions and the jets were required to have
$p_\perp>30$~GeV and $|y_j|<4.4$, but the behaviour is not sensitive to the
details of such choices.
\begin{table}[btp]
  \centering
  \begin{tabular}{|c|l|l|}
    \hline
    & LL processes & NLL processes \\ \hline
    $pp\to W + 3j$ & $qg \to W q'gg$, &
$qg \to W gq'g$, $qQ\to W gq'Q$, \\ &$qQ \to W q'gQ$
& $gg\to W q \bar q' g$, $gQ\to W q \bar q'Q$, $gQ \to W q \bar q Q'$
                                        \\ \hline
    $pp\to W + 4j$ & $qg \to W q'ggg$, & $qg \to W
                                                                   gq'gg$, $qQ\to
                                                                   W gq'gQ$, $qQ\to W q' g Q g$,\\
&$qQ \to W q'ggQ$&  $gg\to W q \bar q' gg$, $gQ\to W q \bar q'gQ$, $gQ\to W
                   q\bar q g Q'$,\\
&& $Qg\to WQ' q\bar q g$,  $Qg\to W Q q\bar q' g$, $gg\to W gq\bar q'g$, \\
&& $Q\tilde Q \to W Q' q\bar q \tilde Q$, $Q\tilde Q \to W Q q\bar q' \tilde Q$\\ \hline
  \end{tabular}
  \caption{A summary of the particle configurations which enter the LL and NLL
    lines in \cref{fig:mfbcomponents} and \cref{fig:pt1components}.  All particles except for the $W$ are listed in rapidity order.  $q$, $Q$
    etc.~may refer to quarks or anti-quarks.  Each process
    refers to itself and its symmetric counterpart, i.e.~$qg \to W q'gg$ is
    shorthand for $qg \to W q'gg + gq\to W ggq'$.  All other
    subprocesses not listed above are included in the line
    labelled "Other".}
  \label{tab:LLNLLOther}
\end{table}

As expected from arguments above, the LL configurations dominate as
$m_{\rm fb}$ increases, while the relative contributions from the other configurations decrease.  However, even
at $m_{\rm fb} = 1$~TeV, the sub-leading contributions still contribute roughly 30\% in the $W+3$-jet case and
almost 40\% in the 4-jet case.  A large $m_{\rm fb}$ is only part of the
requirement of the MRK limit - indeed, large $m_{\rm fb}$ requires just one
but not all invariant masses large.

The importance of accurately describing the NLL configurations is even
more stark for transverse momentum distributions as $d\sigma/dp_{\perp,1}$,
with $p_{\perp,1}$ the transverse momentum of the hardest jet, as in \cref{fig:pt1components}.
There is no correlation expected between the MRK limit (where the LL
configurations will dominate) and the transverse
momenta, and therefore there is no systematic suppression of sub-leading channels.  In fact
one sees that as $p_{\perp,1}$ increases, the contribution from sub-leading
channels increases to
60\% for $W+3$-jets and 70\% for $W+4$-jet production.

Previously, the sub-leading channels had been included in the formalism of
HEJ just through fixed-order matching; meaning that the NLL configurations of
3j and 4j channels did not receive the sophisticated all-order treatment of
the LL channels. The all-order treatment of the NLL channels is made possible
by the calculations presented in the current paper. As such, the all-order resummation will
be applied to a large part of the total cross section, increasing from 40\% to
80\% in inclusive $W+3$-jet production, and from 30\% to 70\% in inclusive $W+4$-jet production.

This illustrates the importance of an effective description of NLL
components, including all-order high-energy corrections, in order to improve
the accuracy of the resummed predictions in sub-asymptotic regions of phase
space.  We will see that this leads to a much reduced dependence on
fixed-order matching, and an improvement in our description of LHC data.

In order to set up the framework for the new calculations in
\cref{sec:subleading}, we outline the structure of a \HEJ amplitude in the
next subsection.

\subsection{High Energy Factorisation of the On-Shell Scattering Amplitudes}
\label{sec:HEJamplitudes}

In the high energy limit of a $2\rightarrow n$ QCD partonic scattering process,
namely where \textit{all} partons are strongly ordered in rapidity, one finds
that the matrix elements factorise into a product of expressions, each
exhibiting dependence on a much reduced subset of external
momenta~\cite{DelDuca:1999ha}.  This factorisation holds even after the addition
of a colour-singlet such as a Higgs, $W$ or a $Z/\gamma^*$ boson to the
scattering.

\begin{figure}[btp]
  \centering
  \includegraphics[width=0.8\textwidth]{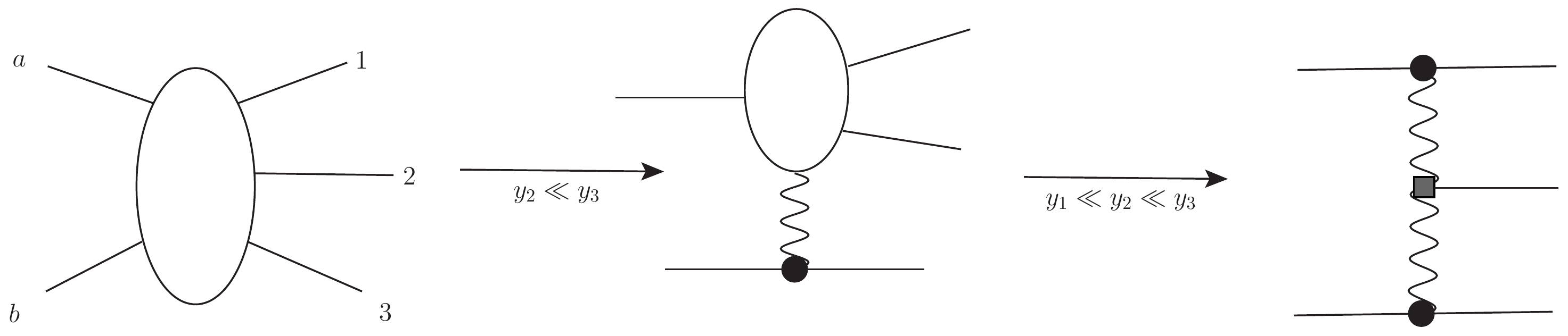}

  $i\mathcal{M}=\ f(p_a,p_b,p_1,p_2,p_3)$, \hspace{0.4cm}
  $f_1(p_a,p_1,p_2)f_2(p_b,p_3)$, \hspace{0.6cm}
  $g_1(p_a,p_1)g_2(q_1,p_2)g_3(p_b,p_3)$
  \caption{In the limit that particles are well-separated in rapidity, QCD
    matrix elements factorise into independent pieces dependent only on a subset
    of momenta, as illustrated.  This is
    true in the full high-energy limit where \emph{all} particles are
    well-separated in rapidity (right-hand side), but also when the limit only
    applies to a subset of particles (middle).}
  \label{fig:mefactorise}
\end{figure}

In fact, a stronger statement than the one above can be made: the amplitude will factorise even
if only a subset of the final state partons are strongly ordered in rapidity.
The dependence of the amplitude upon partons obeying strong ordering will remain isolated in factors that remain
simple, whereas the dependence upon partons where the rapidity-ordering condition is relaxed will appear in
factors that are more complex and exhibit co-dependence on a greater number of external partons.
Notably, if the strong ordering condition is relaxed between a pair of neighbouring partons, there will appear
one less $t$-channel colour-octet propagator. This behaviour is illustrated in
\cref{fig:mefactorise}, where the separation of the analytical expressions of the
amplitude is also given in terms of unspecified functions $f$, $f_i$ and $g_j$.
This motivates the concept of \emph{local momenta} for each component, which is
the relevant momentum subset within which there is no strong-rapidity ordering assumed.

The factorisation of the amplitude is extremely powerful because the kinematic
dependence of external legs is isolated in only a small number of factors, which
prevents a significant increase in complexity as the number of outgoing
particles increases. This has the advantage that is easy to demonstrate that a
minimal combination of components exhibit the scaling behaviour expected from
\cref{eq:reggescaling}.  Furthermore, the factorisation implies that the pieces
are not only independent of the other momenta in the process, but also of the
details of the rest of the process.  In the example in \cref{fig:mefactorise} it means that
$f_2(p_b,p_3)=g_3(p_b,p_3)$ and in general means that components for high-multiplicity
processes can be derived from low-multiplicity ones.

In Ref.~\cite{DelDuca:1999ha}, the factorised components of the amplitude (the impact
factors) were derived in terms of scalar momentum components after these are
approximated by their dominant components in the high-energy limit.  In the \HEJ
formalism, we have found that using contractions of vector currents and
tensors allows us to keep the required properties of factorisation, while making
sufficiently few approximations so that \HEJ amplitudes still satisfy~\cite{Andersen:2009nu}:
\begin{itemize}
\item gauge invariance in all phase space (not just in the high-energy limit),
\item crossing symmetry between incoming and outgoing particles, and
\item conservation of energy and momentum.
\end{itemize}
The fact that the factorised structure remains allows us to apply the
resummation using the Lipatov Ansatz and generalisations of the Lipatov vertices used
within BFKL descriptions~\cite{Andersen:2009nu,Andersen:2012gk}.
%
% In \cref{sec:subleading}, where we relax the conditions
% of strong ordering, the resulting components derived will exhibit scaling
% behaviour that is suppressed by powers of invariant mass relative to the leading
% channels, and hence are formally subleading.  In preparation for
% \cref{sec:subleading}, in the rest of this section we review the form of each
% of the factorised expressions for the simplest case of an FKL configuration.
%
% Following the discussion in section~\ref{sec:HEscaling}, an FKL configuration
% can always be written $f_a f_b\to X+f_a \cdot ng \cdot f_b$ in that rapidity order.  Current available
% options for $X$ are a Higgs, $W$ or $Z$ boson or it can be absent for pure jet
% production. An extension to the available processes is planned in the near future.
%
Specifically, in the \HEJ formalism, the matrix element for a leading-log configuration
$f_a f_b\to X+f_a \cdot ng \cdot f_b$ has the schematic form
\begin{align}
  \label{eq:final}
    \begin{split}
    \overline{\left|\mathcal{M}^{\rm HEJ,reg}_{f_af_b\to Xf_a \cdot ng\cdot f_b}
      \right|}^2 =&\ \mathcal{B}_{f_a,f_b,X}(p_a,p_b,p_1,p_{n+2},\{p\}_{X}) \\ &\ \cdot \prod_{i=1}^n
    \mathcal{V}(p_a,p_b,p_1,p_{n+2},q_i,q_{i+1}) \\ &\ \cdot \prod_{j=1}^{n+1}\mathcal{W}(q_j,y_j,y_{j+1}).
  \end{split}
\end{align}
The first factor is the \emph{process-dependent} Born-level function, $\mathcal{B}$,
which does not depend on the momenta of any additional gluons which are produced.  The
notation $\{p\}_{X}$ represents the local momenta associated with the
production of $X$, if present.
This is then supplemented by
vertex functions $\mathcal{V}$ for each of the $n$ additional gluons
which depend on the momenta of the incoming and outermost outgoing particles, and the
derived $t$-channel momentum $q_i$
defined as
in fig.~\ref{fig:ts}.  The final factors, $\mathcal{W}$, represent the finite
contribution from the combination of the virtual corrections and unresolved real
emissions.  There is one for each $q_j$ which depends only on that momentum and
the rapidity difference between the emissions on either side.  $\mathcal{V}$ and
$\mathcal{W}$ are independent of the process-type (i.e.~they are the same for
all choices of $f_a$, $f_b$ and $X$).

In this paper, we will concentrate on the first of these factors, $\mathcal{B}$,
where we derive new results.  Our treatment of real and virtual
corrections and the cancellation of the divergences between them (which leads to
$\mathcal{V}$ and $\mathcal{W}$) is identical to previous studies and is
implemented using the methods of \HEJ2~\cite{Andersen:2018tnm}.  Their analytic
construction was first described in detail in Ref.~\cite{Andersen:2011hs}.

The function $\mathcal{B}$ is constructed as a contraction, $S$, of two ``generalised
currents'' divided by corresponding $t$-channel momenta, written as the product $T_X$, and multiplied by
suitable couplings and colour factors as follows:
\begin{align}
  \label{eq:Bdef}
  \mathcal{B}_{f_a,f_b,X}(p_a,p_b,p_1,p_{n+2},\{p\}_{X}) = (g_s^2)^2 \
  \frac{K_{f_a}K_{f_b} }{4(N_C^2-1)}\ ||S_{f_af_b\to Xf_a\ldots f_b} ||^2\ \frac{1}{T_X}.
\end{align}
As a simple example, in pure
QCD ($X=0$), the spinor structure is just a helicity averaged contraction of two spinor currents,
\begin{equation}
S^{{h_a,h_b,h_1,h_{n+2}}}_{f_af_b\to f_a\ldots f_b} = \bar{u}^{h_1}(p_1)
\gamma^\mu u^{h_a}(p_a)\
g_{\mu\nu}\ \bar{u}^{h_{n+2}}(p_{n+2}) \gamma^\nu u^{h_b}(p_b),
\label{eq:Sqcd}
\end{equation}
where $h_i$ is the helicity of parton $i$.  We use double-bar notation to write this quantity summed over all helicity combinations as
\begin{align}
  \label{eq:Sdouble}
  \left\| S_{f_af_b\to f_a...f_b} \right\|^2 = \sum_{\{h_i\}} |
  S^{{h_a,h_b,h_1,h_{n+2}}}_{f_af_b\to f_a\ldots f_b}(p_1,p_{n+2},p_a,p_b) |^2.
\end{align}
The required $t$-channel momenta here are
$T_0=t_1t_{n+1}$.  The factors $K_{f_i}$ are colour factors which depend on the
incoming particles. If $i$ is a quark, we simply
have $K_q=C_F$. If $i$ is a gluon, we have a more involved factor depending on
colour factors and the fraction of incoming light-cone momenta being carried by the most
forward/backward gluon~\cite{Andersen:2009he}.  In the strict high energy limit,
we recover the result from BFKL, $K_g\rightarrow C_A$.

The function $\mathcal{B}$ is used to describe the production of hard perturbative particles and
give a skeleton to which the other components are added.  It would diverge in
the limit that the momentum of one of the external partons (here $p_1$ or $p_{n+2}$)
goes to zero as the formalism does not contain the corresponding virtual
corrections for these (which would form part of the full next-to-leading
logarithmic corrections).  In order to enforce this important distinction
between particles in $\mathcal{B}$ and particles produced via vertices in $\mathcal{V}$ in our event generator
implementation, we require the outer partons to be part of the
most forward/backward hard jet respectively, and to carry a significant fraction
of the total jet momentum.\footnote{The exact fraction can be set by the user
  through the parameter \texttt{max ext soft pt fraction},
  see~\cite{Andersen:2019yzo}.  We recommend choosing a value between 90-95\%
  for the contribution from the hard perturbative particle.}  The high-energy
resummation is then applied to the phase space region \emph{between} these two
particles in rapidity.

% Treating the number of additional gluons as a variable between $2$ and $\infty$,
% the expression may be integrated to give inclusive dijet
% cross sections which are accurate to leading logarithm in
% $\hat{s}/p_\perp$. Fixed-order samples up to the highest available multiplicity
% are matched and merged according to the efficient procedure described
% in~\cite{Andersen:2018tnm}.
In the rest of this section we discuss the
construction of the function $\mathcal{B}$ for other processes, highlighting the
features which will be important when we come to the new results in this paper.

\subsection{Process-dependent Contractions of Effective Currents}
\label{sec:currents}

The function $\mathcal{B}_{f_a,f_b,X}(p_a,p_b,p_1,p_{n+2},\{p\}_{X}) $
in~\cref{eq:final} represents the skeleton or Born process which is then
supplemented with all-order high-energy resummation in \HEJ.
It takes the form shown in~\cref{eq:Bdef} where the key component is the
function, $S_{f_af_b\to Xf_1  \dots f_2}$.  Using currents as described in
section~\ref{sec:HEJamplitudes}, we define the following general structure:
\begin{equation}
  \label{eq:Sabstract}
  S_{f_af_b\to Xf_1  \dots f_2} =
  j^{h_a,\dots,h_1}_{\mu}(p_a,p_1,\{p\}_{X})\  X^{\mu \nu}(\{p\}_{X})\  j^{h_b,\dots,h_n}_{\nu}(p_b,p_{n+2},\{p\}_{X}).
\end{equation}
This spinor structure is sufficient to describe all existing \HEJ processes and
the new processes described in this paper.
For example, we recover the pure QCD expressions given in \cref{eq:Sqcd} by taking:
\begin{align}
  X^{\mu \nu} = g^{\mu \nu}, \qquad   j_\mu^{h}(\tilde p,p) = \bar u^h(p) \gamma_\mu u^h(\tilde p).
  \label{eq:Sdefqq}
\end{align}

\begin{figure}[btp]
   \includegraphics[width=\textwidth]{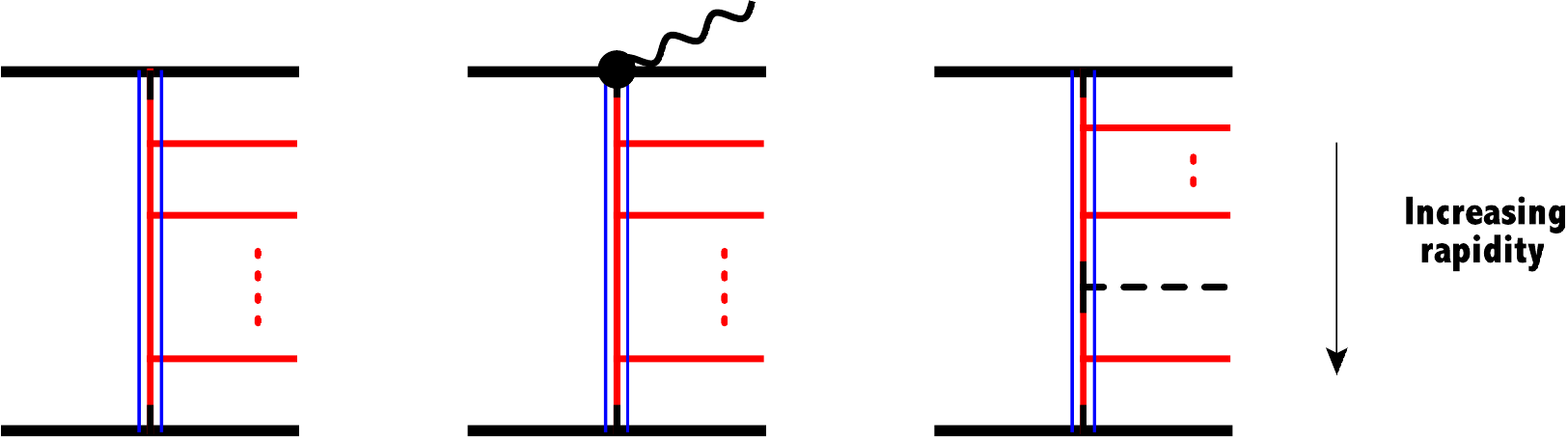}

  \hspace{1.1cm} (a) \hspace{3.75cm} (b) \hspace{3.7cm} (c)
   \caption{These diagrams illustrate the schematic structure of an amplitude in
    \HEJ (\cref{eq:final}) for (a) pure jets, (b) $W$ plus jets and (c) Higgs boson plus jets.
    The black thick lines represent the skeleton or Born process in each case,
    described by the function $\mathcal{B}$.  The external red lines represent the
    production of resolved real gluons, each with a factor $\mathcal{V}$.  The blue lines indicate
    the range of rapidity over which the virtual corrections are applied, which
    are encoded in the $\mathcal{W}$ factors.  Dotted red lines indicate that the
    number of additional gluons is not fixed and the dashed black line in (c) indicates
    the Higgs boson.}
  \label{fig:existingcomponents}
\end{figure}
Figure~\ref{fig:existingcomponents}(a) illustrates that the function
$\mathcal{B}$ is used to describe the outer ends of the process (the parts
represented by thick lines), while being independent of the exact structure in
between.  The resummation is applied between these ends in rapidity,
illustrated by the thin blue lines.  This is therefore the region where we may have
an arbitrary number of resolved, real gluons, shown in red.

\begin{figure}[b]
  \centering
  \begin{minipage}[b]{0.2\textwidth}
    \includegraphics[width=\textwidth]{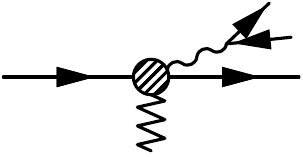}
  \end{minipage}
  \begin{minipage}[b]{0.1\textwidth}
    \centering{=}
    \vspace{0.7cm}
  \end{minipage}
  \begin{minipage}[b]{0.2\textwidth}
    \includegraphics[width=\textwidth]{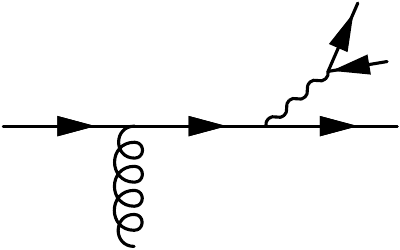}
  \end{minipage}
  \begin{minipage}[b]{0.1\textwidth}
    \centering{+}
    \vspace{0.7cm}
  \end{minipage}
  \begin{minipage}[b]{0.2\textwidth}
    \includegraphics[width=\textwidth]{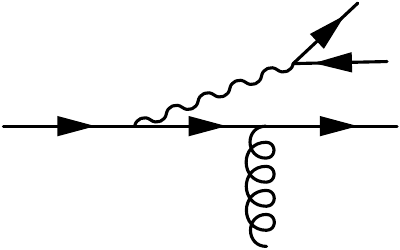}
  \end{minipage}
  \caption{The $j_W$ current, \cref{eq:Weffcur1}, is constructed from the two diagrams which
    contribute to the production of a $W$-boson from a given quark line.}
  \label{fig:jW}
\end{figure}

The addition of an electroweak boson does not change the analysis of the scaling
of multijet amplitudes given in~\cref{eq:reggescaling}.  Therefore, the
leading-logarithmic contributions to $W$-plus-dijet production are FKL
configurations of coloured particles and, in particular, the non-extremal
partons are all gluons.  The $W$ boson is therefore emitted from one of the
extremal legs as in~\cref{fig:existingcomponents}(b).  Without loss of generality, if it couples to the $p_a-p_1$ end
of the chain, we find for $W\to \ell \bar{\ell}$
\begin{align}
S_{f_af_b\to Wf_a^\prime  \dots f_b} =
  j_{W,\mu}(p_a,p_1,p_\ell,p_{\bar{\ell}})\  g^{\mu \nu}\  j^{h_b,\dots,h_n}_{\nu}(p_b,p_{n+2})
\end{align}
where $j_{W,\mu}(p_a,p_1,p_\ell,p_{\bar{\ell}})$ is the exact sum of the
two contributions shown in ~\cref{fig:jW}:
\begin{align}
  \label{eq:Weffcur1}
  j_W^\mu(p_a,p_1,p_\ell,p_{\bar{\ell}}) =&\ \frac{g_W^2}{2}\
     \frac1{p_W^2-M_W^2+i\ \Gamma_W M_W}\ \bar{u}^-(p_\ell) \gamma_\alpha
                                               v^-(p_{\bar\ell})\nonumber \\
& \cdot \left( \frac{ \bar{u}^-(p_1) \gamma^\alpha ( \slashed{p}_1 +
  \slashed{p}_W) \gamma^\mu u^-(p_a)}{(p_1+p_W)^2} +
\frac{ \bar{u}^-(p_1)\gamma^\mu (\slashed{p}_a - \slashed{p}_W)\gamma^\alpha u^-(p_a)}{(p_a-p_W)^2} \right).
\end{align}
The necessary components to form $\mathcal{B}$ in~\cref{eq:Bdef} are then:
\begin{align}
  \label{eq:combinecouplings}
  \begin{split}
    ||S_{f_af_b\to Wf_a^\prime...f_b}||^2 &=
    \sum_{h_b=h_{n+2}} \left|j_{W}^\mu (p_a,p_{\ell},p_{\bar{\ell}}, p_1)
      j_\mu^{h_b}(p_b,p_{n+2}) \right|^2, \\ T_W&= q_1^2 q_{n+1}^2=
    (p_a-p_1-p_\ell-p_{\bar\ell})^2 (p_b-p_{n+2})^2.
  \end{split}
\end{align}
No approximation has been made in the Born process and hence this gives the
exact expression for the $2\to 2$ amplitude.  For certain combinations of
initial quark flavours, it may be possible for a $W$ boson to be emitted from
either end of the chain.  Both are sampled within \HEJ.  The interference
between the two is suppressed by both kinematic and flavour effects and hence is
neglected.  % \footnote{If instead, we consider the production of a $Z$ boson or virtual
% photon, neither suppression is present and we sum both contributions at
% amplitude-level to immediately include this effect, see~\cite{Andersen:2016vkp}.}

% The structure of amplitudes with a Higgs boson is somewhat different because it
% couples through a top loop to the gluons in the effective $t$-channel. The outer
% currents remain the simple spinor
% strings of the pure QCD case, but the tensor which contracts them, $X^{\mu\nu}$
% takes a non-trivial form.  In the infinite top mass limit, the effective vertex of the Higgs'
% coupling to $t$-channel gluons gives~\cite{Andersen:2009he}:
% \begin{equation}
%  X^{\mu \nu} = V_H^{\mu \nu}(q_j,q_{j+1}) = \left(\frac{\alpha_s}{3\pi v}
%  \right)\left(g^{\mu \nu} q_j\cdot q_{j+1}  - q^\mu_j q^\nu_{j+1}\right).
% \end{equation}
% The factorised structure of amplitudes within \HEJ allows inclusion of finite
% quark mass effects and full loop propagator effects to arbitrary high
% multiplicities~\cite{Andersen:2018kjg}.  These may be encoded through a further
% modification to the tensor $X^{\mu\nu}$.  As shown in
% figure~\ref{fig:existingcomponents}(c), the region of resummation is still
% defined by the rapidities of the external coloured particles (which is also the
% case when the Higgs boson is emitted outside of these in rapidity).

The structure above is easily extended.  The treatment of
$Z/\gamma^*$-plus-dijet production is very similar to $W$-plus-dijet production
with additional helicities and interference effects
included~\cite{Andersen:2016vkp}.  The description of Higgs-boson-plus-dijet
production is described in detail in \cite{Andersen:2017kfc,Andersen:2018kjg}
and is the first application where the tensor $X^{\mu\nu}$ is non-trivial, as
illustrated in~\cref{fig:existingcomponents}(c), where the black Higgs vertex in
the middle of the red region of resummation forms part of $\mathcal{B}$.

In this section, we have described the \HEJ construction of
all leading-logarithmic contributions in $\hat{s}/p_{\perp}^2$ to inclusive
$W$-plus-dijet
production.  In the next section we present the calculation of the new
components required to describe the well-defined subset of the
next-to-leading-logarithmic contributions which we have identified as the most
important to improve the description away from the strict high-energy limit.

%%% Local Variables:
%%% mode: latex
%%% TeX-master: "main"
%%% End:

\section{Amplitudes for Subleading Processes}
\label{sec:subleading}
In the previous section we have outlined the construction of the necessary
amplitudes to describe the leading-logarithmic contributions in
$\hat{s}/p_t^2$ to inclusive $(X+)$-dijet processes within the \HEJ
framework.  In this section we present our new calculations which provide the
necessary components to calculate the leading-logarithmic contribution for all
 $W+3$-jet and $W+4$-jet subprocesses which contribute at
next-to-leading log level to the inclusive $W+$dijet cross section.  Until
now, these processes have been included in \HEJ only through matching to fixed
order without any further all-order corrections.  These new results therefore allow
all-order high-energy resummation to be applied to a much larger
fraction of the total cross section and significantly reduce the dependence of
the \HEJ predictions on fixed-order matching.  We illustrate the numerical
impact of these new components in~\cref{sec:impact-including-nll}, after presenting the new calculations
in~\cref{sec:calculation-new-nll,sec:new-nll-components-4}.

\subsection{New NLL Components: Inclusive 3-jet Processes}
\label{sec:calculation-new-nll}

We are seeking to describe the leading-logarithmic terms of subprocesses whose
particle flavour and momentum configurations give next-to-leading logarithmic
contributions to the inclusive cross section.  This means that at the level of
the matrix-element-squared, their contribution should be suppressed by one power
of $s_{ij}$ compared to the leading-logarithmic channels.  The scaling behaviour
in \cref{eq:reggescaling} then implies that we need to describe processes with
one less effective $t$-channel exchange of a gluon than the maximum number.  An
example of such a process is to take a LL configuration and swap the rapidity
order of an outgoing quark and the gluon next to it, to give a single quark in
the $t$-channel, as shown for 3 partons in \cref{fig:fklvsnonfkl}(c).  In the
rest of this section, we derive the necessary components to describe all
3-jet subprocesses which contribute at this order.

We take the following $W+3$-jet process as the first new case:
\begin{align}
  \label{eq:Wunodef}
  q(p_a)Q(p_b) \longrightarrow (W\to)\ell(p_\ell) \bar{\ell}(p_{\bar\ell})
  g(p_1) q^\prime(p_2) Q(p_3),
\end{align}
where as usual we order $y_1<y_2<y_3$, and we take $q, q^\prime , Q$ to be
different quark flavours such that the vertex $q \to q^\prime W$ exists for a choice
of sign for the $W$ charge.  If we apply strong rapidity
ordering among all particles, \cref{eq:reggescaling} gives the following scaling
of the amplitude in the MRK limit:
\begin{align}
  \label{eq:scaleWuno}
  \mathcal{M} \sim (s_{12})^{1/2}(s_{23})^1 \gamma.
\end{align}
This is suppressed by a half power of $s_{12}$ compared to the LL configuration
where $y_{q^\prime}<y_g$, leading  to an NLL contribution to the inclusive
dijet cross section as required.  We stress that for this particle assignment and configuration it
is the LL contribution.  This scaling argument holds whenever we have an LL configuration up
to a single gluon produced backward of a quark or a single gluon produced forward
of a quark; we refer to such a process as production of an `unordered' gluon.

We remove the requirement of strong ordering between $p_1$ and $p_2$, but
keep it for the other colour-charged particles: $y_1,y_2 \ll y_3$, as in the
middle diagram of \cref{fig:mefactorise}.  This is an example of Quasi-Multi
Regge Kinematics (QMRK), where strong rapidity ordering is only enforced in a
subset of particles.  In this case, the scaling with $s_{12}$ is not prescribed
(nor is that invariant necessarily large), and the scaling depends only on
the invariants which are still controlled by strong ordering, in this case
$s_{23}$.  We expect the factorised structure to follow that of
\cref{fig:mefactorise}.  A new feature of the $j_{W{\rm uno}}$ current is that it
must allow for different colour structures as the outgoing gluon and $t$-channel
gluon may occur in either order.  We therefore now write the colour matrix for the
non-gluon end explicitly and expect the spinor structure within
the corresponding skeleton function $\mathcal{B}_{qQW}^{\rm uno W}$ should now be
\begin{equation}
  \label{eq:SabsWuno}
  S_{qQ\to Wgq^\prime Q}^{\rm unoW} =
  j_{W{\rm uno}\,\mu}^d(p_a,p_1,p_2,p_\ell,p_{\bar\ell})\  g^{\mu
    \nu}\ T^d_{3b}\  j^{h_b,h_3}_{\nu}(p_b,p_{3}),
\end{equation}
where $j_\nu^{h_b,h_3}$ is the same current as the QCD process,
\cref{eq:Sdefqq} and $T^m_{ij}$ represents fundamental colour matrices
between quark state $i$ and $j$ with adjoint index $m$.  The new current,
$j^d_{W{\rm uno}}$, is non-zero only for the left-handed helicities $h_a=h_1=-$ and hence we have
suppressed its helicity indices.  It is derived from the
sum of all leading-order Feynman
diagrams for the process given in \cref{eq:Wunodef}. A factorised form can
be obtained by dropping the terms kinematically suppressed in the QMRK limit; full details are given in
\cref{sec:WunoDeriv}.  We find
\begin{align}\label{eq:wunocurrent}
\begin{split}
  j^{d\,\mu}_{W \rm uno}(p_a,p_1,p_2,p_\ell,p_{\bar{\ell}}) =& \ -i  \varepsilon_{\nu}(p_1)\
  \bar{u}^-(p_\ell) \gamma_\rho v^-(p_{\bar \ell}) \\ & \quad \times\
  \left(T^1_{2i} T^d_{ia} (\tilde U_1^{\nu\mu\rho}-\tilde L^{\nu\mu\rho}) +
    T^d_{2i} T^1_{ia} (\tilde U_2^{\nu\mu\rho}+\tilde L^{\nu\mu\rho}) \right),
  \end{split}
\end{align}
where expressions for $\tilde U_{1,2}^{\nu\mu\rho}$ and $\tilde L^{\nu\mu\rho}$
may be found in \cref{eq:U1tensor,eq:U2tensor,eq:Ltensor}.  It is
straightforward to check that the Ward identity for the
external gluon is satisfied as the tensors obey
\begin{equation}
p_{1\nu} \tilde L^{\nu\mu\rho} = p_{1\nu} \tilde U_1^{\nu\mu\rho} = -p_{1\nu} \tilde U_2^{\nu\mu\rho}.
\end{equation}
The resulting amplitude is therefore gauge invariant in \emph{all} of phase
space, and not only in the QMRK limit.  In
the derivation, approximations are only made to the values of the momenta at the
opposite end ($p_b$ and $p_3$).  This increases the region of validity of the
expression, but more than that, means that the approximations made are so
minimal that crossing symmetry between initial and final states is preserved at
the level of the derived currents, which we return to below.

\begin{figure}[btp]
  \includegraphics[width=\textwidth]{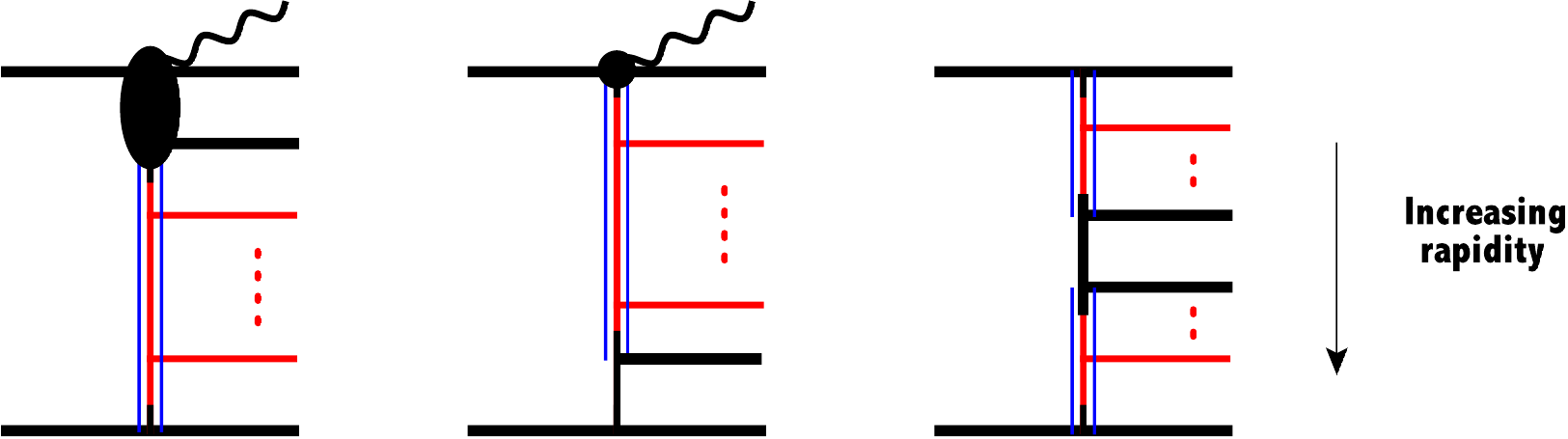}

  \hspace{1.1cm} (a) \hspace{3.75cm} (b) \hspace{3.7cm} (c)
  \caption{These diagrams illustrate the resummation structure of the new
    processes we describe after the calculation of the new components in this
    section: (a) $qQ\to (W\to)\ell \bar{\ell} gq^\prime \ldots Q$, (b)
    $qQ\to (W \to) \ell \bar{\ell} q^\prime \ldots Qg$ and (c)
    $qQ\to q \ldots \tilde{q}\bar{\tilde{q}} \ldots Q$. The thick black lines
    show the components described by the skeleton function $\mathcal{B}$ and the
    external red lines indicate possible real gluons.  The blue lines show the
    region where strong ordering and therefore resummation is applied.  The
    decay products of the $W$ have been omitted for clarity.}
  \label{fig:newstructure}
\end{figure}

We define the following contractions:
\begin{align}
  \label{eq:XYdefsforuno}
  \begin{split}
    \tilde X &=  \varepsilon^{h_1}_{\nu}(p_1)\ \left[\bar{u}^-(p_\ell) \gamma_\rho v^-(p_{\bar
        \ell}) \right] \big[\bar{u}^{h_3}(p_3) \gamma_\mu ^{h_b}(p_b) \big]
    \left( \tilde U_1^{\nu\mu\rho}-\tilde L^{\nu\mu\rho} \right), \\
    \tilde Y &= \varepsilon^{h_1}_{\nu}(p_1)\ \left[\bar{u}^-(p_\ell) \gamma_\rho v^-(p_{\bar
        \ell}) \right] \big[\bar{u}^{h_3}(p_3) \gamma_\mu ^{h_b}(p_b) \big]
    \left( \tilde U_2^{\nu\mu\rho}+\tilde L^{\nu\mu\rho} \right).
  \end{split}
\end{align}
Then, from \cref{eq:wunocurrent}, the final \HEJ expression for the tree-level process $qQ\to e \nu_e gq^\prime Q$ is
\begin{align}
  \label{eq:coloursquareWuno}
    \left| \overline{\mathcal{M}_{qQ\to Wgq'Q}^{\HEJ\ \rm{tree}}} \right|^2
      = g_s^6\frac{C_F^2 }{4(N_C^2-1)} \frac{\left\|S_{q Q\to Wgq' Q}^{\rm uno}\right\|^2}{(p_a-p_1-p_2-p_\ell-p_{\bar\ell})^2(p_b-p_3)^2},
\end{align}
where
\begin{align}
  \label{eq:SunoW}
  \begin{split}
    \left\|S_{q Q\to Wgq' Q}^{\rm unoW}\right\|^2 &= \sum_{h_b,h_3,h_1}\
    \bigg[C_F|\tilde X + \tilde Y|^2 - C_A\, {\rm
      Re}(\tilde X \tilde Y^*)
  \bigg].
  \end{split}
\end{align}
We now wish to
derive our skeleton function $\mathcal{B}_{qQW}^{\rm uno W}$ for this process, to be used
to describe the process above with an arbitrary number of additional gluons (analogous to e.g.~\cref{eq:Bdef}).  In our formalism,
it is valid to apply resummation over the range of rapidity where strong
rapidity ordering is applied.  In this example, \cref{eq:Wunodef}, that is the range $[y_2,y_{3}]$.
This will be the region where real gluons (governed by vertex functions
$\mathcal{V}$) and virtual corrections (governed by exponential factors
$\mathcal{W}$) are applied, and so at all orders leads to the structure shown in
\cref{fig:newstructure}(a).

For the process with $n$ coloured outgoing particles (i.e.~$(n-3)$ additional real gluons between the quarks):
\begin{align}
  \label{eq:unoWn}
    q(p_a) f_b(p_b) \to (W\to)\ell(p_\ell) \bar{\ell}(p_{\bar\ell})\  g(p_1)
  q'(p_2) \ldots g(p_i) \ldots f_b(p_n),
\end{align}
we then define the following skeleton function
\begin{align}
  \label{eq:BWuno}
  \begin{split}
     \mathcal{B}_{qQW}^{\rm uno W} =&
    g_s^6 \frac{C_F K_{f_b}}{4(N_C^2-1)}\ \frac1{q_2^2 q_{n-1}^2} \left\|S_{q f_b\to Wgq^\prime\ldots f_b}^{\rm
         uno W}\right\|^2 ,\\
     q_2=& p_a-p_\ell - p_{\bar{\ell}}-p_1-p_2, \qquad q_i = q_{i-1}-p_i,\  i >
     2 \\
\left\|S_{q f_b\to Wgq^\prime\ldots f_b}^{\rm uno W}\right\|^2 &=
\sum_{h_b,h_{n},h_1}\
    \bigg[C_F|\tilde X + \tilde Y|^2 - C_A\, {\rm
      Re}(\tilde X \tilde Y^*)
  \bigg],
  \end{split}
\end{align}
where the mapping $p_3\to p_{n}$ is made within $\tilde X$ and $\tilde Y$.
% where now
% \begin{align}
%   \label{eq:Xtildenemission}
%     \tilde X &= \varepsilon^{h_1}_{\nu}(p_1)\ \left[\bar{u}^-(p_\ell) \gamma_\rho v^-(p_{\bar
%         \ell}) \right] \big[\bar{u}^{h_{n+3}}(p_{n+3}) \gamma_\mu ^{h_b}(p_b) \big]
%     \left( \tilde U_1^{\nu\mu\rho}-\tilde L^{\nu\mu\rho} \right), \\
%     \tilde Y &= \varepsilon^{h_1}_{\nu}(p_1)\ \left[\bar{u}^-(p_\ell) \gamma_\rho v^-(p_{\bar
%         \ell}) \right] \big[\bar{u}^{h_{n+3}}(p_{n+3}) \gamma_\mu ^{h_b}(p_b) \big]
%     \left( \tilde U_2^{\nu\mu\rho}+\tilde L^{\nu\mu\rho} \right).
% \end{align}
The tree-level factor of
$1/(p_b-p_3)^2$ has been generalised to $1/q_{n-1}^2$; as in the LL processes in
\cref{sec:currents}, we do this symmetrically to match the factors of $1/q_i^2$
in our prescription for real gluon vertices.  When $n=3$ and $f_b$ is a quark, \cref{eq:BWuno}
reproduces the tree-level result of \cref{eq:coloursquareWuno}.
As before, $\mathcal{B}_{qQW}^{\rm uno W}$ represents the underlying skeleton
process and therefore each outgoing coloured particle will be required to be
hard in the perturbative sense (implemented by requiring each parton to
contribute a significant fraction of momentum to \emph{its own jet}, not
containing another skeleton parton).

As our second process, we consider
\begin{align}
  \label{eq:WunoOppositeEnds}
  q(p_a) Q(p_b) \longrightarrow (W\to)\ell (p_\ell) \bar{\ell}(p_{\bar{\ell}})
  q^\prime(p_1) Q(p_2) g(p_3).
\end{align}
Like the process above, this has a $W$ produced from a quark line and an
unordered gluon, but these are now produced from \emph{different} quark lines.  If there was strong
ordering between $p_2$ and $p_3$, the effective $t$-channel particle between
them would be a quark.
The effective $t$-channel between $p_1$ and $p_2$ is still a gluon, so we expect
$\mathcal{M} \sim (s_{12})^1(s_{23})^{1/2}$.  This is therefore the same logarithmic order
as the first example in this section.  We will now relax the requirement
between $p_2$ and $p_3$ such that the
relevant QMRK limit is now $y_1 \ll y_2, y_3$.  We therefore expect the spinor structure to
take the form
\begin{align}
  \label{eq:SWoppuno}
  S_{qQ\to Wq^\prime Qg}^{\rm uno} = j_{W\,\mu}(p_a,p_1,p_\ell,p_{\bar{\ell}}) T^d_{1a}\
  g^{\mu\nu} \ j_{{\rm uno}\, \nu}^{d\, h_b,h_2,h_3}(p_b,p_2,p_3).
\end{align}
The factorisation of amplitudes in the high-energy limit dictates that $j_{{\rm
    uno}\, \nu}^d$ is independent of the $W$ boson and, indeed, of the rest of
the process.  It is therefore the same unordered current that we find in the
pure jet process and in Higgs boson plus dijets.  This was given
in Ref.~\cite{Andersen:2017kfc}:
\begin{align}
  \label{eq:unoderiv}
  j^{d,h_b,h_2,h_3\, \mu}_{\rm uno}(p_b, p_2, p_3) = -i \varepsilon_\nu(p_3)
  \left(T^3_{2i}T^d_{ia}\left(U_1^{\mu \nu} - L^{\mu \nu}\right) +
  T^d_{2i}T^3_{ia}\left(U_2^{\mu \nu} + L^{\mu \nu}\right) \right),
\end{align}
with $U_{1,2}^{\mu\nu}$ and $L^{\mu\nu}$ given in
\cref{eq:unocomps}.  The colour structure is equivalent to the previous case, as
we should expect.  Again it is easily checked that this current also satisfies
gauge invariance in all of phase space.  One can derive the known scalar
expression for the BFKL NLO impact factor~\cite{DelDuca:1999ha} from
\cref{eq:unoderiv} by taking the further approximations used in that paper.

The resummation pattern of this subprocess is shown in \cref{fig:newstructure}(b). The
construction of the skeleton function $\mathcal{B}_{qQW}^{\rm uno}$ allowing for $(n-3)$ extra gluons between the quarks follows the same pattern
as the previous case, \cref{eq:BWuno}. After defining,
\begin{align}
  \label{eq:XYdefuno}
      X &= \varepsilon^{h_{n}}_{\nu}(p_{n})\ j_{W\,\mu}(p_a,p_\ell,p_{\bar\ell},p_1)
    \left( U_1^{\mu\nu}-L^{\mu\nu} \right), \\
    Y &= \varepsilon^{h_{n}}_{\nu}(p_{n})\ j_{W\,\mu}(p_a,p_\ell,p_{\bar\ell},p_1)
    \left( U_2^{\mu\nu}+ L^{\mu\nu} \right).
\end{align}
 we find
\begin{align}
  \label{eq:BWoppuno}
  \begin{split}
     \mathcal{B}_{qQW}^{\rm uno} =&
    g_s^6 \frac{ C_F^2}{4(N_C^2-1)} \left\|S_{q Q\to Wq^\prime\ldots Qg}^{\rm
         uno}\right\|^2 \ \frac1{q_1^2 q_{n-2}^2},\\
     q_1=& p_a-p_1-p_\ell -p_{\bar{\ell}}, \qquad q_i = q_{i-1}-p_i,\  i \ge
     2 \\
     \left\|S_{q Q\to Wq^\prime\ldots Qg}^{\rm uno}\right\|^2 &= \sum_{h_b,h_{n-1},h_{n}}\ \bigg[
     C_F |X +Y|^2 - C_A\,{\rm Re}(XY^*) \bigg],
  \end{split}
\end{align}
with the substitution $p_2\to p_{n-1}$ and $p_3\to p_{n}$ in $X$ and
$Y$.  Resummation is again applied over the range of strong ordering, which in
this case is
$[y_1,y_{n-1}]$, as illustrated in \cref{fig:newstructure}(b).

We note in passing that we may deduce the equivalent pure jet process without a $W$ boson,
$q(p_a) Q(p_b)\to q(p_1) Q(p_2) g(p_3)$ by replacing $j_W$ in \cref{eq:SWoppuno}
with a quark current:
\begin{align}
  \label{eq:SunowithoutW}
  S^{\rm uno}_{qQ\to qQg} = j_\mu(p_a,p_1) T^d_{1a} g^{\mu\nu} j_{{\rm
  uno}\nu}^{d\, h_b, h_2,h_3}(p_b,p_2,p_3).
\end{align}
The construction of the suitable $\mathcal{B}^{\rm uno}_{qQ}$ function is analogous
to \cref{eq:BWoppuno}.

For our third and fourth processes at this order, we consider $W+3$-jet channels
with a gluon in the incoming state.  At leading log, an incoming gluon implies
that the corresponding extremal outgoing particle must also be a gluon to give
the necessary gluon in the effective $t$-channel.  For the NLL contributions
considered in this section, that restriction no longer applies. Our first
example of this type is where an incoming gluon splits into an outgoing
$\bar{q}q^\prime$ pair and a $W$ boson, e.g.
\begin{align}
  \label{eq:gqqbarWsplit}
  g(p_a)Q(p_b) \to (W\to) \ell(p_\ell) \bar{\ell}(p_{\bar{\ell}})  \bar{q}(p_1) q^\prime(p_2) Q(p_3).
\end{align}
In strong rapidity ordering, there would be an effective $t$-channel quark
between $p_1$ and $p_2$ giving the suppression of $(s_{12})^{1/2}$ seen in the
previous processes in this section.  In the QMRK, $y_1, y_2 \ll y_3$, we
apply resummation over the region of strong ordering $[y_2,y_3]$ as shown in
\cref{fig:newstructure}(a).  At leading-order, there exists crossing symmetry
between this process and the process in \cref{eq:Wunodef}, under
$p_a \leftrightarrow p_1$.  In the derivation of the
$j^{d\, \mu}_{W {\rm uno}}$ current for that process, approximations were only
made to $p_b$ and $p_3$, and therefore this crossing symmetry persists at the
level of the currents in HEJ amplitudes.\footnote{This symmetry is not present in
  the scalar impact factors of ref.~\cite{DelDuca:1999ha} due to the additional
  approximations in that calculation.}  We therefore find
\begin{align}
  \label{eq:jWgqqp}
  j^{d\, \mu}_{Wq\bar{q}^\prime}(p_a,p_1,p_2,p_\ell,p_{\bar\ell}) = j^{d\,
  \mu}_{W {\rm uno}} (p_1,p_a,p_2,p_\ell,p_{\bar\ell}).
\end{align}
% as shown explicitly in \cref{sec:w-ext-curr}.
The full skeleton $\mathcal{B}^{\rm extW}_{gW}$
is then constructed as in \cref{eq:BWuno}.

Finally we consider the case with an incoming gluon splitting into a $q\bar{q}$
pair, where now the $W$ is produced from the \emph{other} quark line:
\begin{align}
  \label{eq:gqqbarWopp}
  q(p_a)g(p_b) \to (W\to) \ell(p_\ell) \bar{\ell}(p_{\bar{\ell}})  q^\prime(p_1) Q(p_2) \bar{Q}(p_3).
\end{align}
This is related by crossing symmetry to the process given in
\cref{eq:WunoOppositeEnds} under $p_b \leftrightarrow p_3$ such that one can derive
\begin{align}
  \label{eq:crossoppends}
  j_{q\bar{q}}^{d\, h_b, h_2, h_3, \mu}(p_b,p_2,p_3) = j_{\rm uno}^{d\,
  h_3,h_2,h_b\, \mu} (p_3,p_2,p_b).
\end{align}
This is then
contracted with a $W$ current at the opposite end, as indicated by the structure
in \cref{fig:newstructure}(b), to give the following spinor structure
\begin{align}
  \label{eq:Bfinal3}
    S_{qg\to         Wq^\prime  \ldots Q\bar{Q}}^{\rm ext} = j_{W\,\mu}(p_a,p_1,p_\ell,p_{\bar{\ell}}) T^d_{1a}\
  g^{\mu\nu} \ j_{q\bar{q}\, \nu}^{d,h_b,h_2,h_3}(p_b,p_2,p_3).
\end{align}
The skeleton function $\mathcal{B}^{\rm ext}_{gW}$ is then constructed as in
\cref{eq:BWoppuno}.  From here, as for the production of an unordered gluon, one can immediately deduce the corresponding process in three-jet
production without a $W$ boson, $q(p_a)g(p_b)\to q(p_1)Q(p_2)\bar{Q}(p_3)$, by
replacing $j_W(p_a,p_1,p_\ell,p_{\bar{\ell}})$ with $j(p_a,p_1)$.

We now wish to illustrate that the new components we have derived do indeed give the correct
scaling for the matrix-element in the full MRK limit.  We will also
compare these approximate HEJ skeleton amplitudes (i.e.~without resummation) to
leading-order results taken from
\texttt{MadGraph5\_aMC@NLO}~\cite{Alwall:2014hca}, to illustrate the quality of
the approximation.

\begin{figure}[btp]
 \centering
 \begin{subfigure}{0.49\textwidth}
   \includegraphics[width=\textwidth]{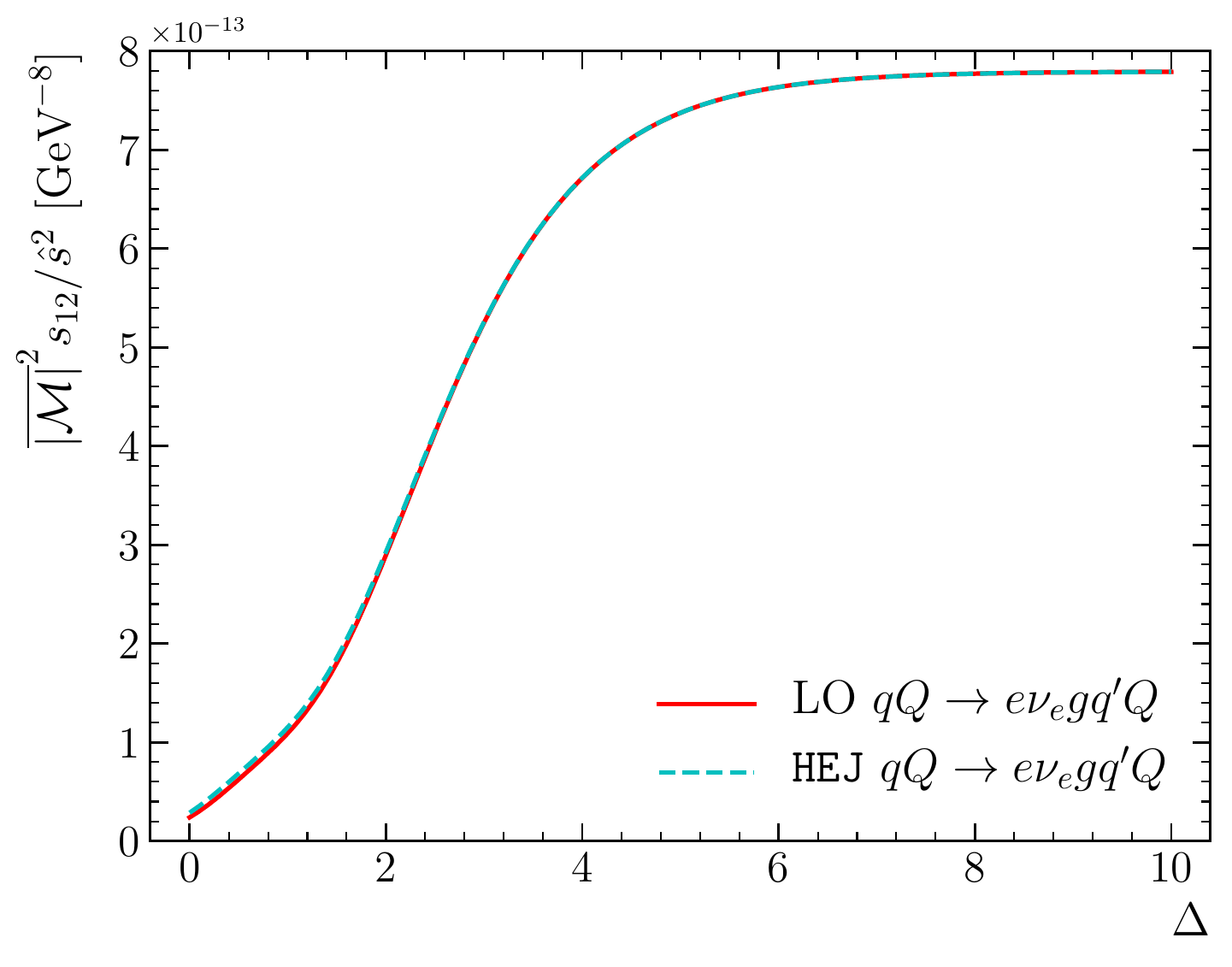}
   \caption{}
   \label{fig:wuno_explore}
 \end{subfigure}
  \begin{subfigure}{0.49\textwidth}
\includegraphics[width=\textwidth]{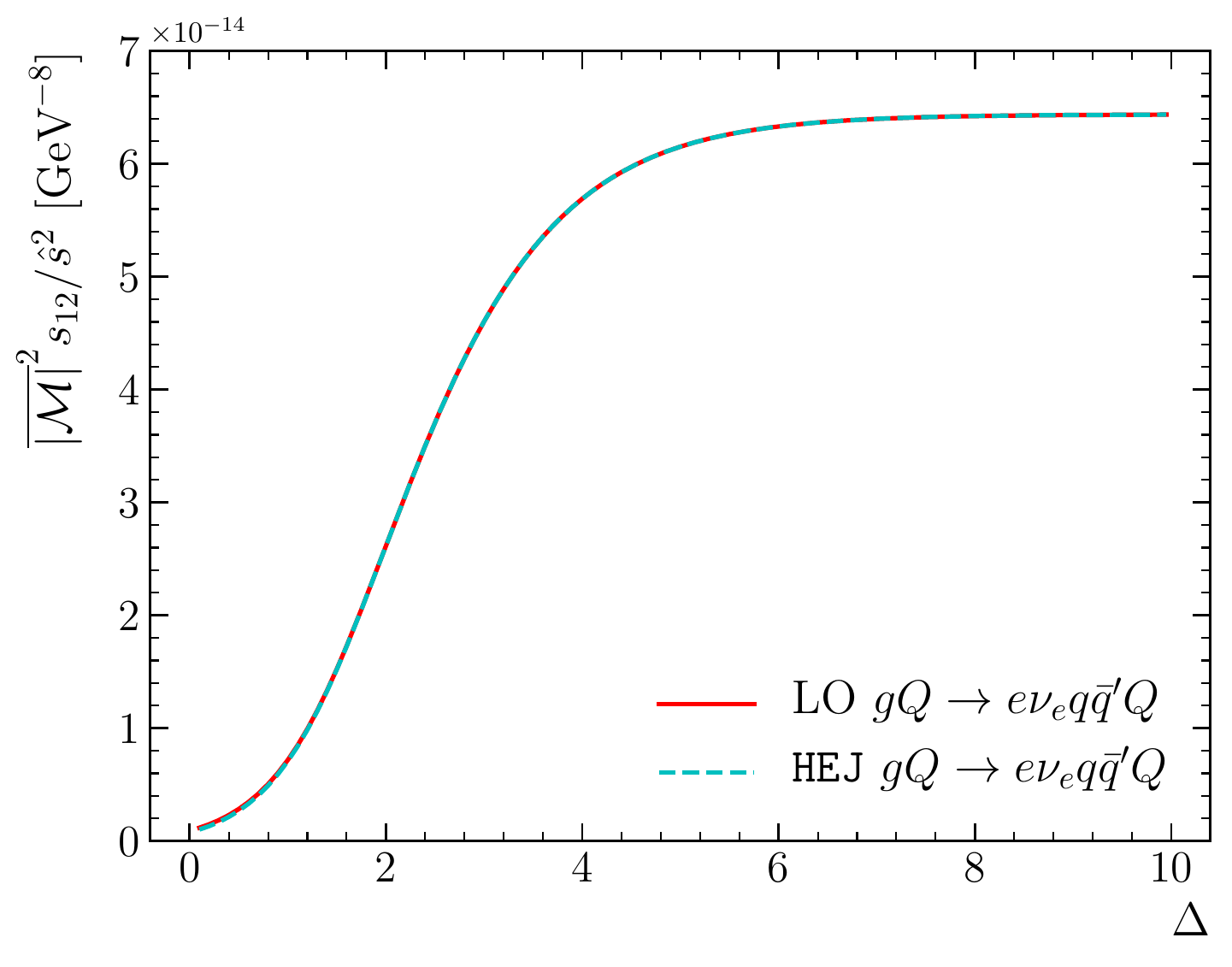}
\caption{}
\label{fig:wexqqbar_explore}
 \end{subfigure}
 \caption{Plots illustrating the scaling of the \texttt{HEJ} approximation at
   tree-level to
   the matrix elements (cyan, dashed) for (\subref{fig:wuno_explore})
   $qQ \rightarrow e \nu_e g q' Q$, and (\subref{fig:wexqqbar_explore})
   $gQ\rightarrow e\nu_eq \bar q' Q$.  The parameter $\Delta$ represents the
   rapidity separation of the coloured particles as described in the text.  Also shown is
   the leading order result (red, solid).  Both
   cases are seen to obey the expected scaling, \cref{eq:QMRKMEsqscaling}, and
   further to give a good description of the full LO matrix element across the
   whole range.}
  \label{fig:3jexplore_uno}
\end{figure}

For our first subleading channels with an unordered gluon, in the MRK
limit we should have
(see \cref{eq:scaleWuno}):
\begin{equation}
 \left| \overline{\mathcal{M}_{qQ\to Wgq'Q}^{\HEJ\ \rm{tree}}} \right|^2 \sim s_{12}s^2_{23}\gamma
\end{equation}
where again $\gamma$ is a finite function of transverse momentum.  In the MRK
limit, this scaling is equivalent to:
\begin{equation}
 \left| \overline{\mathcal{M}_{qQ\to Wgq'Q}^{\HEJ\ \rm{tree}}} \right|^2  \times
 \frac{s_{12}}{\hat{s}^2}\rightarrow
\mathrm{finite\ const}.
\label{eq:QMRKMEsqscaling}
\end{equation}
We illustrate this behaviour in \cref{fig:3jexplore_uno}(a) for  $qQ \rightarrow  e \nu_e g q'
Q$ (\cref{eq:Wunodef}).  The parameter $\Delta$ parameterises the rapidity separation between the
coloured particles, as we have chosen $y_1=-\Delta$, $y_2=0$ and $y_3=\Delta$,
see \cref{sec:phase-space-slices} for the exact parameters.
We can see that the quantity
$\overline{|\mathcal{M}|}^2 \hat{s}_{12}/\hat{s}^2$ does indeed tend to a
non-zero, finite constant at large values of $\Delta$. Moreover, the \HEJ
approximation describes the full leading-order result well across the entire
phase space, deviating only slightly at the lowest values of $\Delta$.  This is a consequence of the minimality of the approximation made
in the derivation.

In \cref{fig:3jexplore_uno}(b), we show the equivalent plots for the subprocess
$gQ\rightarrow e\nu_eq \bar q' Q$ (\cref{eq:gqqbarWsplit}).  The jets have
rapidities as in \cref{fig:3jexplore_uno}(a) such that large values of $\Delta$
represent the MRK limit.  It is
again clear that the expected scaling is seen as both lines
approach a non-zero finite constant.  Moreover, the approximations in the
skeleton \HEJ process give an extremely good description of the leading-order
matrix element for all values of $\Delta$.  This would not have been the case
if the strict limit was applied in all phase space.

%\FloatBarrier

\subsection{New NLL Components: Inclusive 4-Jet Processes}
\label{sec:new-nll-components-4}
In the previous section, we calculated all the components which are necessary
to calculate the leading logarithmic components of all 3-jet subprocesses which
contribute at next-to-leading log to the inclusive $W+3$-jet
cross section.  This automatically includes many of the processes which are
necessary to do the same for $W+4$jets.  For example, the process
\begin{align}
 \label{eq:Wuno4jdef}
  q(p_a)Q(p_b) \longrightarrow (W\to)\ell(p_\ell) \bar{\ell}(p_{\bar\ell})
  g(p_1) q^\prime(p_2) g(p_3) Q(p_4),
\end{align}
is given in \HEJ by:
\begin{align}
  \label{eq:4jexamplealreadyhave}
      \overline{\left|\mathcal{M}^{\HEJ\ \rm{tree}}_{qQ\to Wgq^\prime gQ}
      \right|}^2 =\ \mathcal{B}_{qQW}(p_a,p_b,p_1,p_2,p_4,p_\ell,p_{\bar\ell})  \cdot
    \mathcal{V}(p_a,p_b,p_1,p_{4},q_1,q_{2}),
\end{align}
where $\mathcal{B}_{qQW}(p_a,p_b,p_1,p_2,p_4,p_\ell,p_{\bar\ell})$ has already
been given in \cref{eq:BWuno} for the $W+3$-jet process.  This easily generalises
further to 5, 6 or more jets.  Similarly, one can approximate the other
unordered gluon processes and incoming gluon to quark-anti-quark processes
using the results already derived in \cref{sec:calculation-new-nll}, by taking
the relevant function $\mathcal{B}$ and multiplying by the required number of
vertices $\mathcal{V}$.

\begin{figure}[tbp]
\centering
\includegraphics[height=5.3cm]{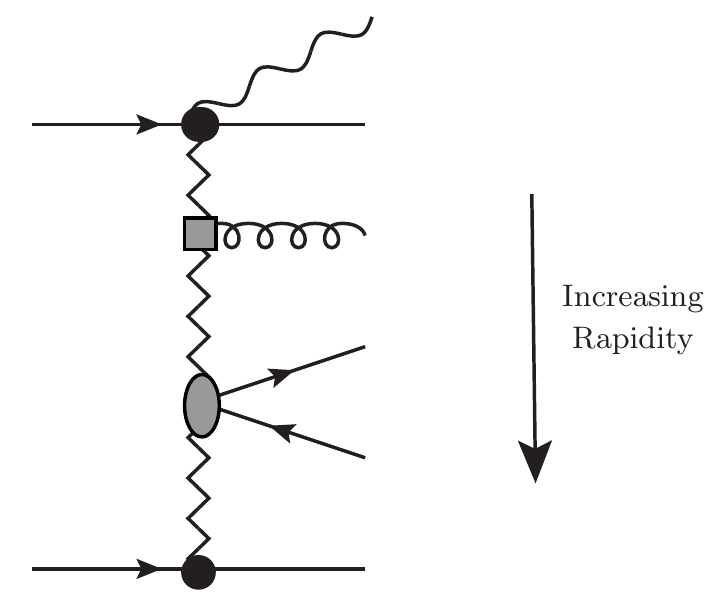}
\hspace{1.5cm}
\includegraphics[height=4.5cm]{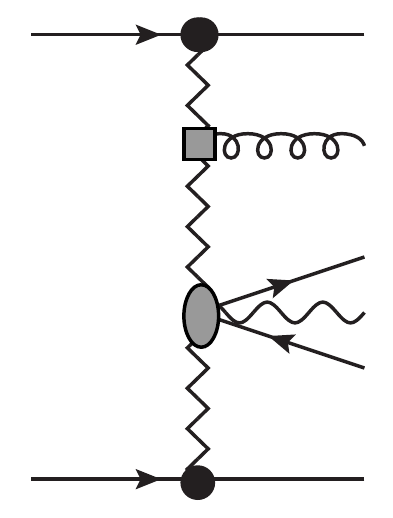}

(a) \hspace{4.4cm} (b) \hspace{2.3cm} \phantom{a}
\caption{Two processes which contribute at next-to-leading log and
  are relevant for four jets and above: (a) $q\tilde q\to Wq'gQ\bar{Q}\tilde q$
  in that rapidity order which contains a central $Q\bar{Q}$ pair and $W$
  produced from a different quark line, and (b) $q\tilde{q}\to qgQ\bar{Q}'W\tilde q$ in that
  rapidity order where the $W$ is produced from the central $Q\bar{Q}'$-pair. The rapidity-ordering applies to the
  coloured particles and not to the $W$ boson or its decay products.}
\label{4jetqqbar}
\end{figure}

There is just one further class of subprocess which formally contributes at NLL.
This is where a $q\bar{q}$ pair is produced in the middle of the rapidity chain,
potentially accompanied by a $W$ boson, see \cref{4jetqqbar}.  This differs from
the processes in \cref{sec:calculation-new-nll} because the extremal ends of the
chain are as in the LL case, but we must now derive a new piece to
describe particles which are intermediate in rapidity.  The structure of the
resummation is as shown in \cref{fig:newstructure}(c).  As the figure suggests,
we do not modify either of the currents in \cref{eq:Bdef}, but instead alter the
contraction between them, $X^{\mu\nu}$.  We will find that this also needs to carry two colour
indices, $d$ and $e$.
We will derive the necessary tensors, $X^{de\, \mu\nu}_{{\rm cen}}(p_2,p_3,q_1,q_3)$ and $X^{de\, \mu
    \nu}_{{\rm cen}W}(p_2,p_3,p_\ell,p_{\bar\ell},q_1,q_3)$, by considering the lowest order processes where
they occur.  For the case where the $W$ is produced from an outer quark line, we
may exploit amplitude factorisation to derive the central $Q\bar{Q}$ tensor from
a process without a $W$ boson:
\begin{align}
  \label{eq:cenproc}
  q(p_a)\ \tilde{q}(p_b)\ &\longrightarrow\ q(p_1)\ Q(p_2)\ \bar{Q}(p_3)\
                            \ \tilde{q}(p_4).
\end{align}
For the case where the $W$ is produced from the central $Q\bar{Q}$ pair, we use
\begin{align}
    q(p_a)\ \tilde{q}(p_b)\ &\longrightarrow\ q(p_1)\ Q(p_2)\ \bar{Q^\prime}(p_3)\
    (W\to)\ell(p_\ell) \bar{\ell}(p_{\bar\ell})\ \tilde{q}(p_4).
\label{eq:processWcen}
\end{align}
As the new central $Q\bar{Q}$ piece contains the quark propagator, we will treat
this as part of the skeleton process, as illustrated in
\cref{fig:newstructure}(c).  This means that we do not impose strong ordering
between the $Q\bar{Q}$-pair and take the following rapidity limit for the coloured particles:
\begin{align}
\label{eq:cenqqbarraporder}
  y_1 \ll y_2,y_3 \ll y_4.
\end{align}
In fact we also do not impose ordering between $y_2$ and $y_3$ so the same
results apply if the anti-quark in the $Q\bar{Q}$-pair is backward of the
quark.  In this limit, we expect the matrix elements corresponding to \cref{eq:cenproc}
and \cref{eq:processWcen} to take the forms:
%\begin{subequations}
\begin{align}
  \label{eq:Mcentral}
    i\mathcal{M}_{\rm cen} &= g_s^4 T^d_{1a} T^e_{4b}\ \frac{j_{\mu}(p_a,p_1)\ X^{de\, \mu
    \nu}_{{\rm cen}}(p_2,p_3,q_1,q_3)\
    j_{\nu}(p_b,p_4)}{t_{a1}t_{b4}}
    \\ % {\rm and} \qquad
  i\mathcal{M}_{\rm cenW} &= g_s^4 T^d_{1a} T^e_{4b}\ \frac{j_{\mu}(p_a,p_1)\ X^{de\, \mu
    \nu}_{{\rm cen}W}(p_2,p_3,p_\ell,p_{\bar\ell},q_1,q_3)\
    j_{\nu}(p_b,p_4)}{t_{a1}t_{b4}}.
                           \label{eq:McentralW}
  \end{align}
%\end{subequations}
We sum together \emph{all} leading-order diagrams in each process and after
applying the QMRK limit (\cref{eq:cenqqbarraporder}), we find (see \cref{sec:cenqqbar,sec:cenWqqbar}):
\begin{subequations}\label{eq:Xcen}
  \begin{align}
    \begin{split}
      X^{de\, \mu \nu}_{\rm cen} &= i T^d_{2q}T^e_{q3} \left[
      X_s^{\mu\nu} + X^{\mu \nu}_{6} \right]
     - i T^e_{2q}T^d_{q3} \left[  X_s^{\mu\nu} + X^{\mu \nu}_{7} \right],
    \end{split}\label{eq:Xcen1}
\\
    \begin{split}
      X^{de\, \mu\nu}_{{\rm cen}W} &=  i T^d_{2q}T^e_{q3} \left[\tilde
        X_s^{\mu\nu} + \tilde X_6^{\mu\nu} \right]
      -i T^e_{2q}T^d_{q3}\left[ \tilde X_s^{\mu\nu} + \tilde X^{\mu\nu}_{7}\right] ,
    \end{split}\label{eq:Xcen2}
  \end{align}
\end{subequations}
where $X_s^{\mu\nu}$, $X^{\mu \nu}_{6}$, $X^{\mu \nu}_{7}$, $\tilde
X^{\mu\nu}_s$, $\tilde X^{\mu\nu}_6$ and $\tilde X^{\mu\nu}_7$ are
defined in
\cref{eq:Xsdefinition,eq:cencomponents,eq:uncrossContr,eq:crossContr,eq:EffectiveVertexWqqbar}.
From \cref{eq:Mcentral}, the final summed and averaged matrix-element-squared for the central
process without a $W$ boson is then
given by
\begin{align}
  \label{eq:cenme2}
  \begin{split}
    \overline{\left|M_{\ q \tilde{q}\to qQ\bar{Q}\tilde{q}}^{\HEJ\ \rm{tree}}\right|}^2 &= \frac{1}{4C_A^2}
    \frac{g_s^8}{t_{a1}^2 t_{b4}^2}\ \sum_{ \left\{ h_i\right\}} \left[C_F( |V|^2 + |W|^2) -
      \frac{1}{C_A} {\rm Re}(V W^*) \right],
  \end{split}
\end{align}
where the sum runs over all allowed helicity combinations and the contracted current structures $V$ and $W$ are defined as
\begin{align}
  \label{eq:P2and3}
  \begin{split}
    V &= j_{\mu}(p_a,p_1) \left( X_s^{\mu\nu} + X^{\mu \nu}_{6} \right)
    j_{\nu}(p_b,p_4), \quad   W = j_{\mu}(p_a,p_1) \left( X_s^{\mu\nu} + X^{\mu
        \nu}_{7} \right) j_{\nu}(p_b,p_4).
  \end{split}
\end{align}
This easily describes the process we need with a $W$ boson produced from an
outer quark line after replacing either $j_\mu(p_a,p_1)$ or $j_\nu(p_b,p_4)$ in
\cref{eq:P2and3} with the $W$ current given in \cref{eq:Weffcur1}. We may similarly adapt \cref{eq:cenme2} to describe the process where a $W$ boson is produced from a central $Q \bar{Q}'$ pair by the simple replacement of $V$ and $W$ with
\begin{align}
	\label{eq:P2and3W}
	\begin{split}
		\widetilde{V} &= j_{\mu}(p_a,p_1) \left( \tilde{X}_s^{\mu\nu} + \tilde{X}^{\mu \nu}_{6} \right)
		j_{\nu}(p_b,p_4), \quad   \widetilde{W} = j_{\mu}(p_a,p_1) \left( \tilde{X}_s^{\mu\nu} + \tilde{X}^{\mu
		\nu}_{7} \right) j_{\nu}(p_b,p_4).
	\end{split}
\end{align}

\begin{figure}[btp]
	\centering
	\includegraphics[width=0.49\textwidth]{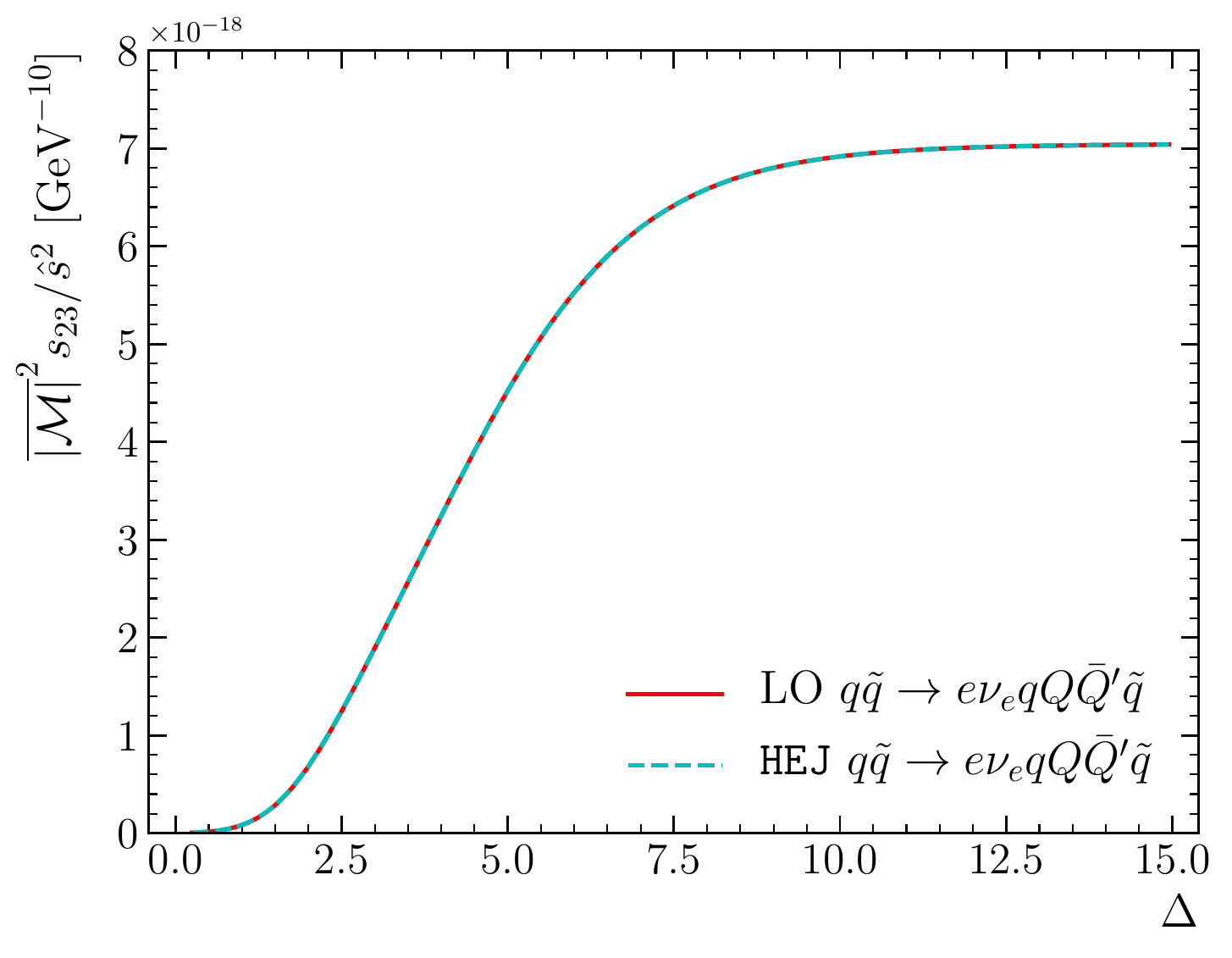}
	\caption{
		Plot illustrating the scaling of the \texttt{HEJ} approximation to
		the matrix element (cyan, dashed) for
		$q \tilde{q} \rightarrow e \nu_e q Q \bar{Q}' \tilde{q}$.  The parameter $\Delta$ represents the
		rapidity separation of the coloured particles as described in the text.  Also shown is
		the leading order result (red, solid).  We see that the expected scaling relation \cref{eq:4jscaling} is obeyed, and that the \HEJ approximation gives a good description of the full LO matrix element across the
		whole range of $\Delta$.}
	\label{fig:4jexplore_qqbar}
\end{figure}

We now wish to demonstrate that this construction does indeed respect the correct
scaling in the full MRK limit.  Specifically,  we should find (cf.~\cref{eq:QMRKMEsqscaling})
\begin{align}
  \label{eq:4jscaling}
  \frac{\overline{\left|M_{q\tilde{q}\to qQ\bar{Q}\tilde{q}}^{\HEJ\ \rm{tree}}\right|}^2
  s_{23}}{\hat s^2} \longrightarrow {\rm finite\ constant}\ne 0.
\end{align}
This scaling behaviour will be unaltered by the production of a $W$ boson from an outer quark or from the central $Q \bar{Q}$ pair. We illustrate this scaling for the process $q \tilde{q} \rightarrow e \nu_e q Q \bar{Q}' \tilde{q}$ in \cref{fig:4jexplore_qqbar}.
We take a slice through phase space where the jets have the following rapidities
\begin{align}
  \label{eq:4jslice}
  y_1=\Delta,\ y_2=\Delta/3,\ y_3=-\Delta/3,\, y_4=-\Delta,
\end{align}
such that the MRK limit is reached in the limit of large $\Delta$, see \cref{sec:phase-space-slices}.  It is clear that indeed the limit of \cref{eq:4jscaling} is
reached both in the \HEJ approximation and in the full leading-order result.
Furthermore, the \HEJ approximation of the skeleton amplitude gives a very close
approximation to the full leading-order result.

As before, we construct the required skeleton functions by generalising the corresponding matrix-element-squared to allow for the production of additional gluons. We again only apply our resummation over the
regions of strong rapidity ordering, which in this case is the rapidity interval
between the most forward and backward coloured particles minus the rapidity
range between the quark and anti-quark.  For the following process,
\begin{align}
  \label{eq:qqbaratn}
  q(p_a) f_b(p_b) \to (W\to)\ell(p_\ell) \bar{\ell}(p_{\bar\ell})\  q'(p_1) g(p_2) \ldots g(p_{i-1}) Q(p_i) \bar{Q}(p_{i+1})
  g(p_{i+2}) \ldots g(p_{n-1}) f_b(p_n),
\end{align}
the resummation range is therefore the sum of $[y_1,y_i]$ and $[y_{i+1},y_n]$,
as illustrated in \cref{fig:newstructure}(c).
The skeleton function for this process is
\begin{align}
  \label{eq:Bcen}
  \begin{split}
    \mathcal{B}_{\rm cen} &= g_s^8 \frac{C_FK_{f_b}}{4(N_c^2-1)}
    \frac{1}{q_1^2 q_{n-1}^2} \left\| S^{\rm
        cen}_{q f_b\to W q' \ldots Q \bar{Q} \ldots f_b} \right\|^2, \\
    \left\|S_{q f_b\to W q' \ldots Q \bar{Q} \ldots f_b}^{\rm cen}\right\|^2 &=
    \frac{1}{2} \frac{1}{q_{i-1}^2q_{i+1}^2}\sum_{\{h_i\}}\ \left[C_F(|V|^2 + |W|^2) -
      \frac{1}{C_A} {\rm Re}(V W^*) \right].
  \end{split}
\end{align}
This expression allows for the incoming particle $b$ to be a quark, anti-quark or
gluon.

We follow the identical steps to construct the skeleton function for processes where
the $W$ boson is produced from the central $Q\bar{Q}'$ pair. In this case the skeleton function is given by
\begin{align}
  \label{eq:BcenW}
  \begin{split}
    \mathcal{B}_{\rm cenW} &= g_s^8 \frac{K_{f_a}K_{f_b}}{4(N_c^2-1)}
    \frac{1}{q_1^2 q_{n-1}^2} \left\| S^{\rm
        cenW}_{f_a f_b\to f_a \ldots WQ \bar{Q}' \ldots f_b} \right\|^2, \\
    \left\|S_{f_a f_b\to f_a \ldots WQ \bar{Q}' \ldots f_b}^{\rm cenW}\right\|^2 &=
    \frac{1}{2} \frac{1}{q_{i-1}^2q_{i+2}^2}\sum_{\{h_i\}}\ \left[C_F(|\widetilde{V}|^2 + |\widetilde{W}|^2) -
      \frac{1}{C_A} {\rm Re}(\widetilde{V} \widetilde{W}^*) \right].
  \end{split}
\end{align}
Real and virtual all-order corrections are then added to \cref{eq:Bcen,eq:BcenW}
as in \cref{eq:final} to give the full all-order \HEJ amplitudes for
these processes at any multiplicity.

\subsection{Numerical Impact of New NLL Components}
\label{sec:impact-including-nll}

In \cref{fig:mfbcomponents,fig:pt1components} in
\cref{sec:HEscaling}, we presented a breakdown of the leading order
cross-section for $pp \to (W\to\ell\nu)+3j$ and $pp \to (W\to\ell\nu)+4j$. The individual contributions
were separated into leading-log (LL) configurations, next-to-leading-log (NLL)
configurations and other (i.e.~further suppressed) configurations and we saw that
NLL contributions accounted for as much as 40\% of the cross section in key
areas of phase space.  In the previous subsections, we have derived all of the
necessary components to construct \HEJ amplitudes for these NLL configurations,
which then allows for an all-order resummation of the dominant high-energy
effects to be applied to this part of the cross section too.

\begin{figure}
  \centering
  \includegraphics[width=0.49\linewidth]{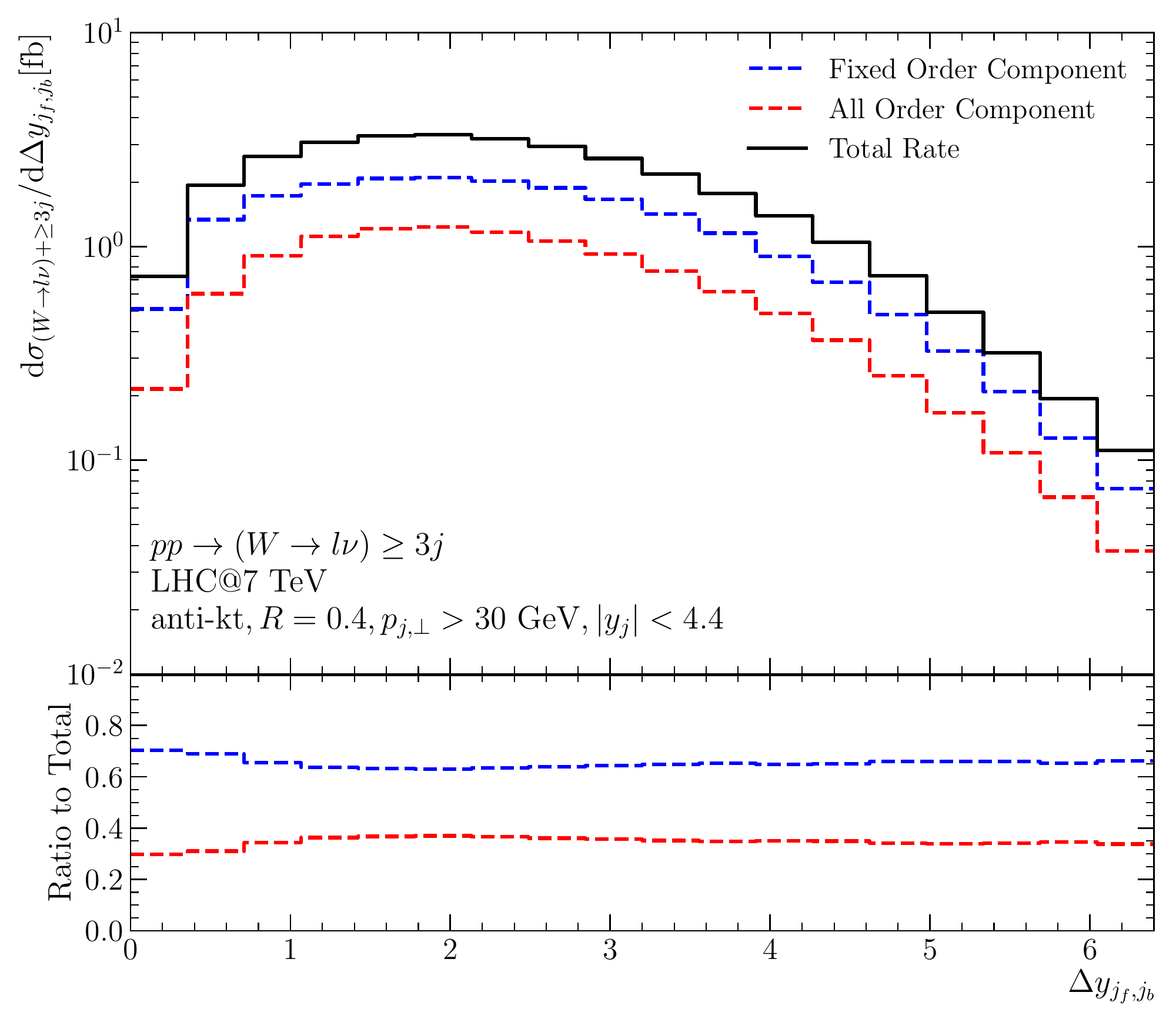}\hfill
  \includegraphics[width=0.49\linewidth]{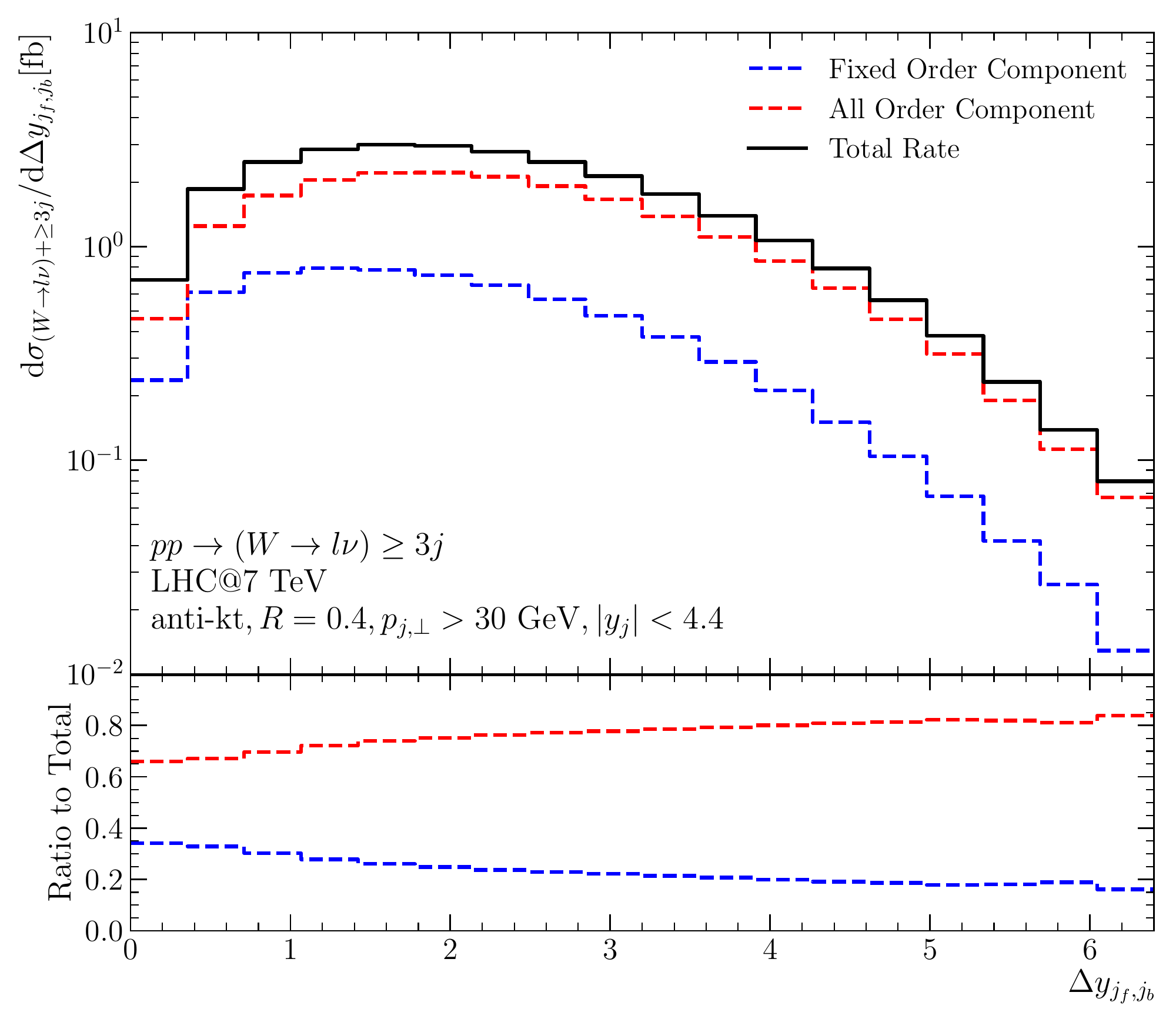}\\
  (a) \hspace{7cm} (b) \\
  \includegraphics[width=0.49\linewidth]{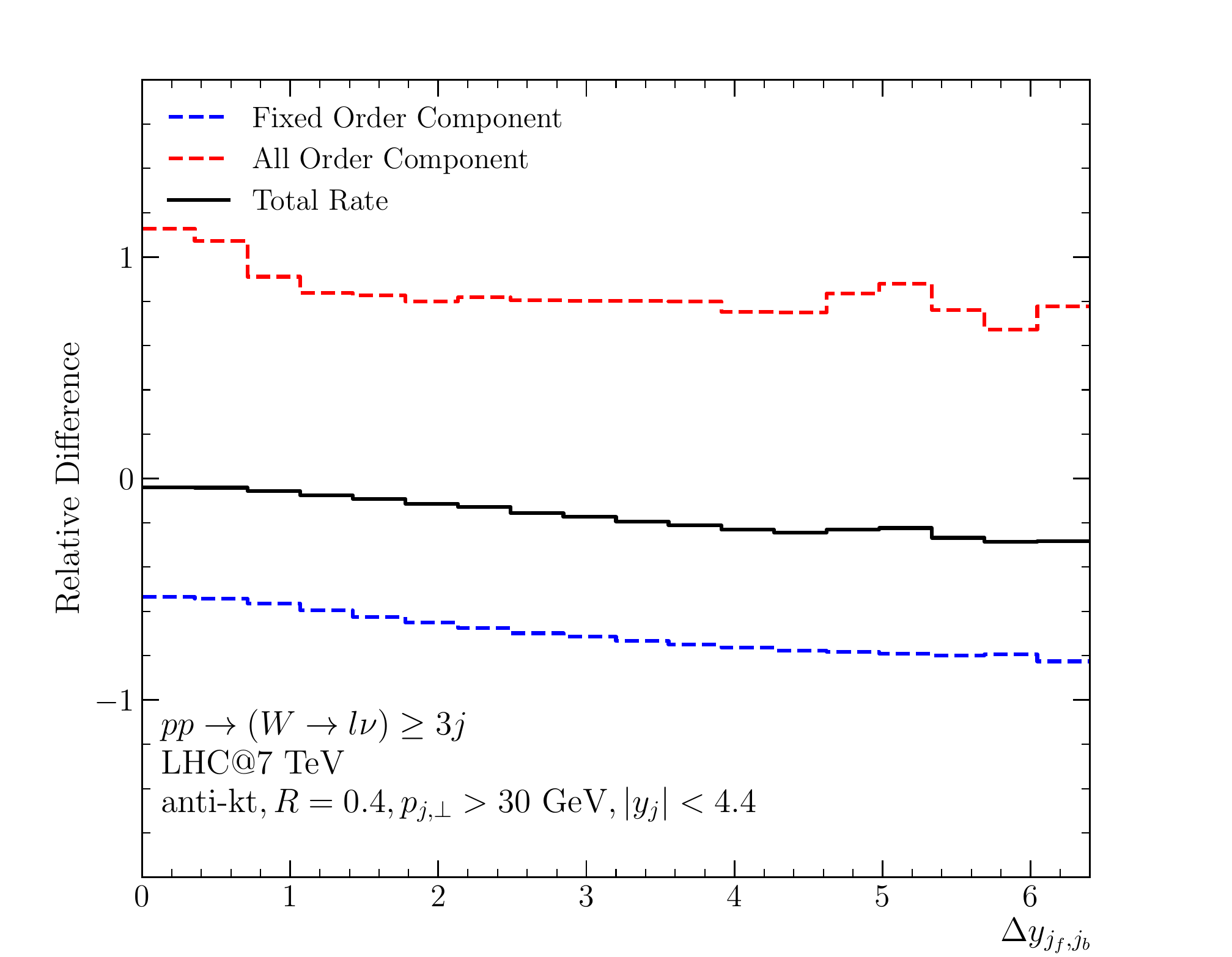}\\
  (c)
  \caption{The differential distribution (black, solid) in the rapidity
    difference between the most-forward and most-backward jets,
    $\Delta y_{j_f,j_b}$, in $pp \to (W\to \ell \nu) +\ge3j$, for (a) case 1, where resummation is applied only to
    LL states, and (b) case 2, where resummation is applied to all LL and
    NLL states.  Also shown in each case is the breakdown into the
    component where all-order resummation is applied (red, dashed) and the
    component which remains described at fixed-order only (blue, dashed). The relative change in each line is shown in (c), using the procedure described in the text.}
  \label{fig:2dyfb}
\end{figure}

\begin{figure}
  \centering
  \includegraphics[width=0.49\linewidth]{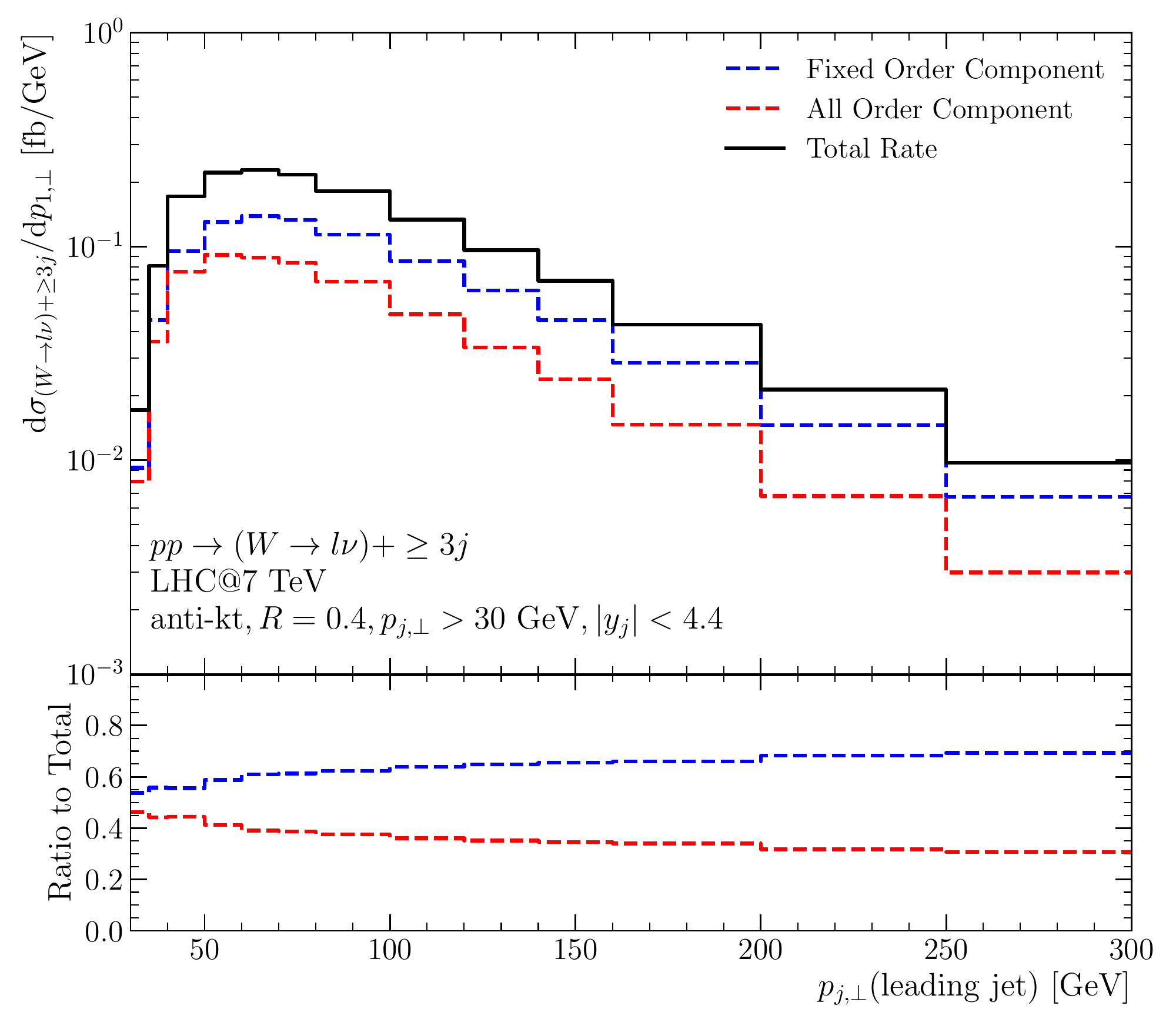}\hfill
  \includegraphics[width=0.49\linewidth]{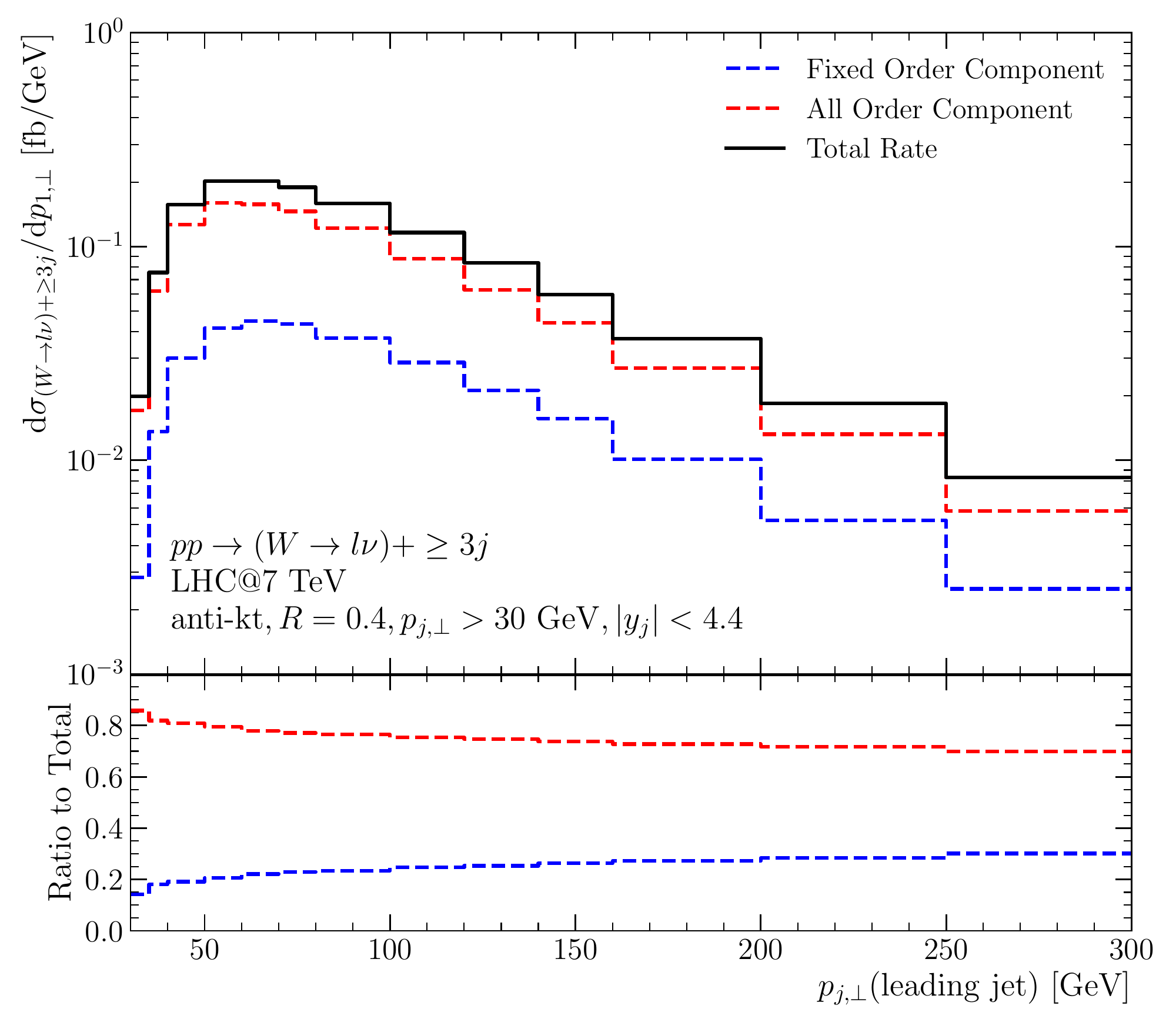}\\
  (a) \hspace{7cm} (b) \\
  \includegraphics[width=0.49\linewidth]{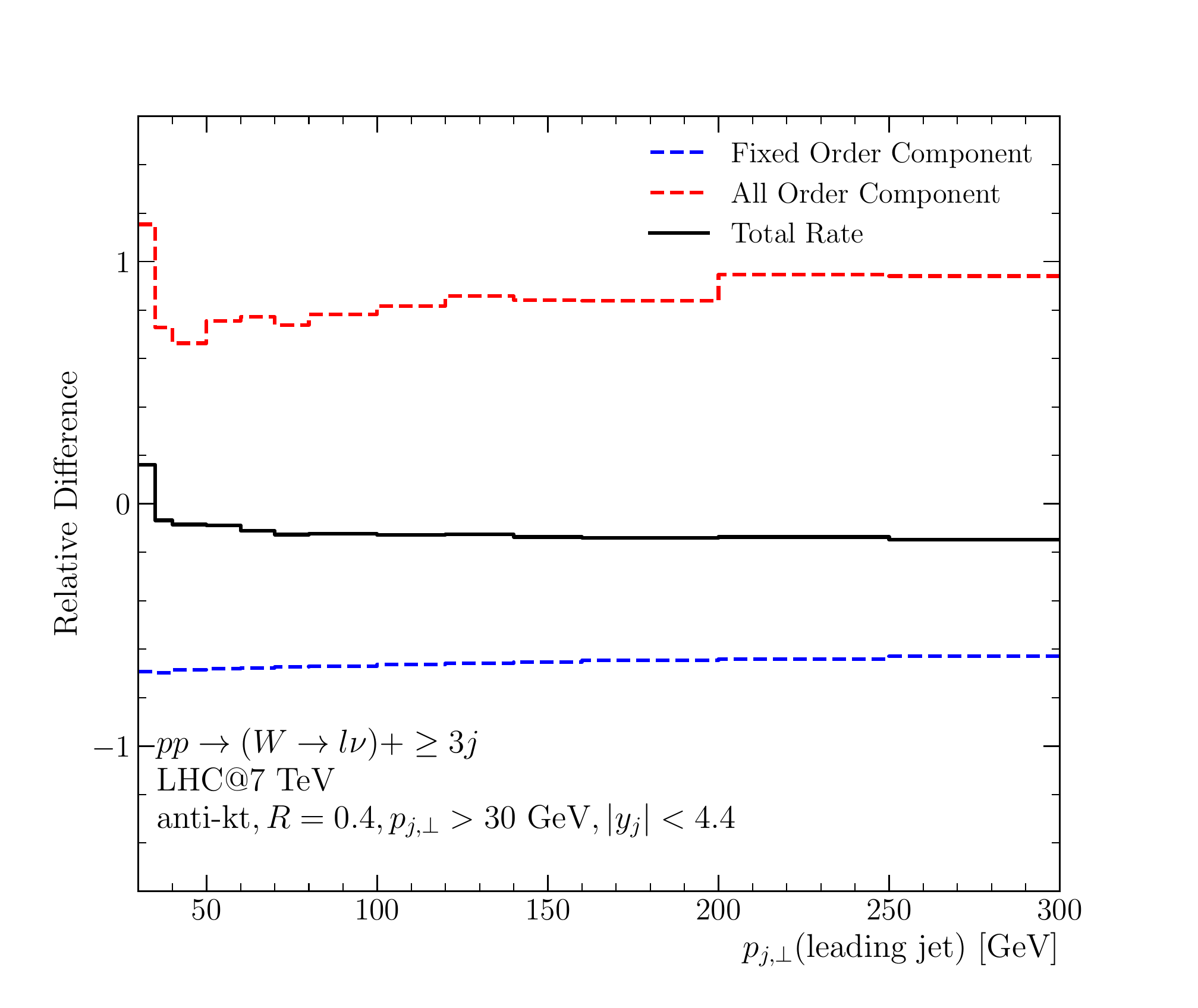}\\
  (c)
  \caption{The differential distribution (black, solid) in the transverse momentum of the
    leading jet in $pp \to (W\to\ell\nu)+\ge 3j$, without and with resummation applied to
    NLL states.  The panels and lines are as in \cref{fig:2dyfb}.}
  \label{fig:2pt2}
\end{figure}

\begin{figure}
  \centering
  \includegraphics[width=0.49\linewidth]{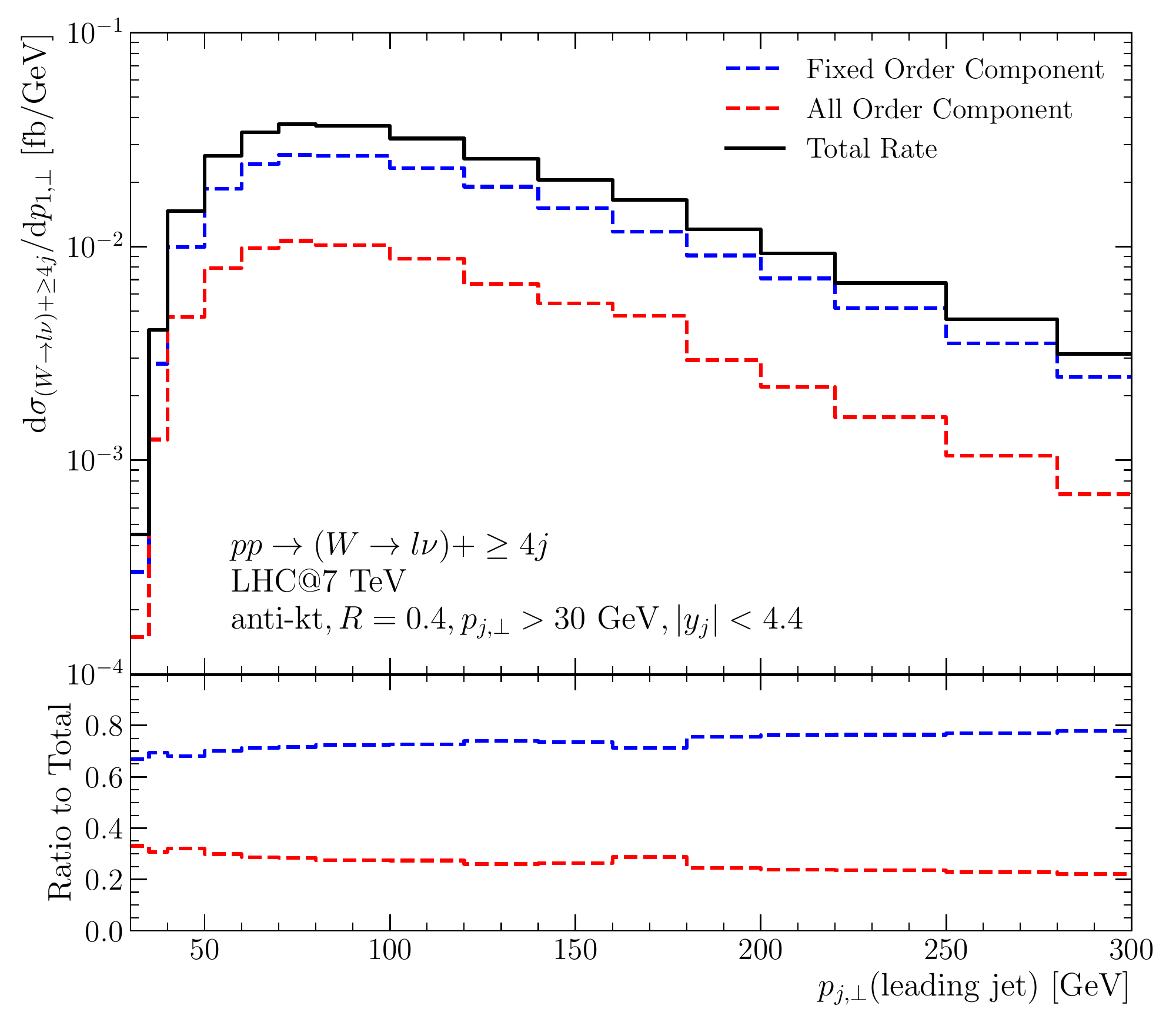}\hfill
  \includegraphics[width=0.49\linewidth]{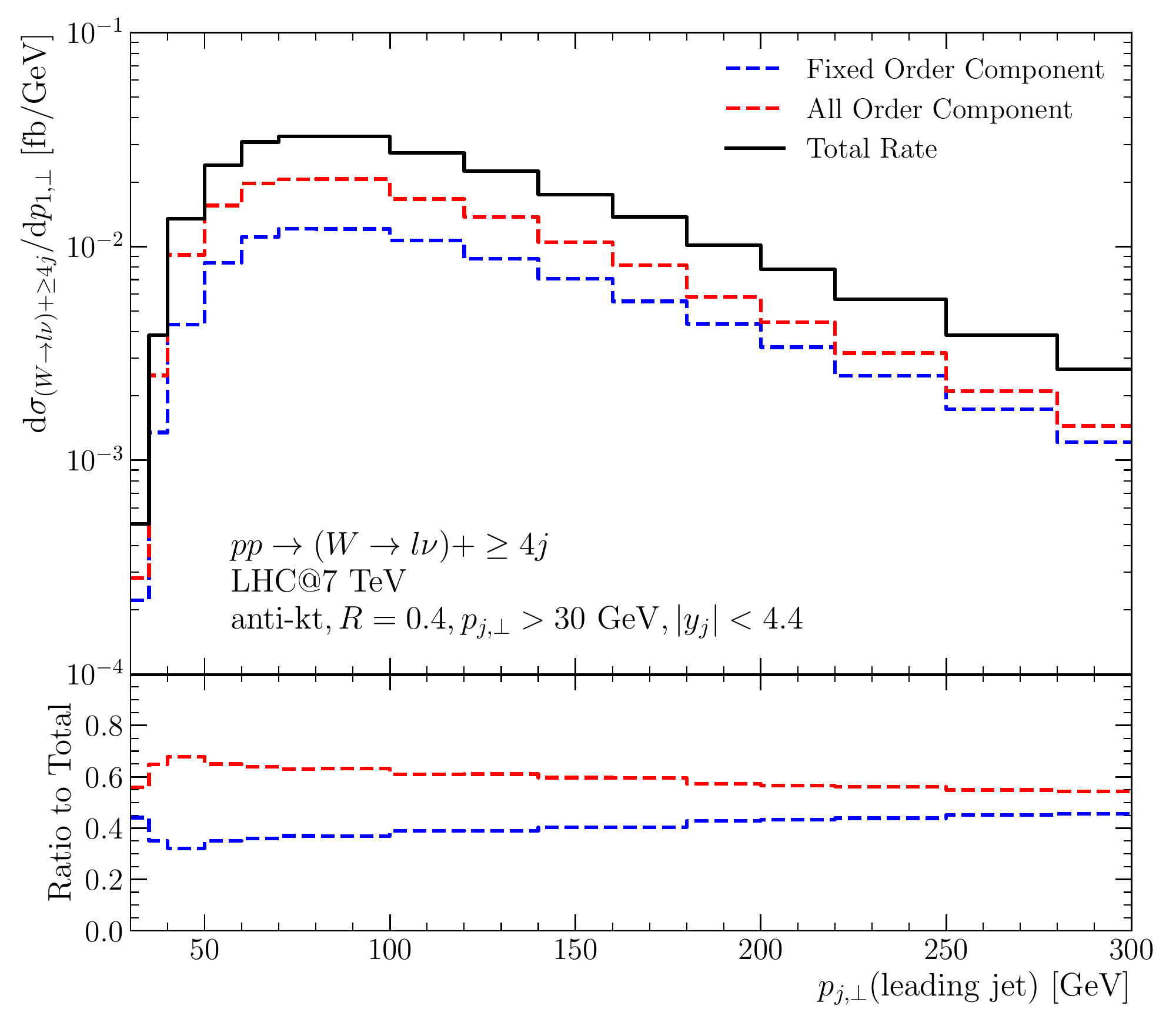}\\
  (a) \hspace{7cm} (b) \\
  \includegraphics[width=0.49\linewidth]{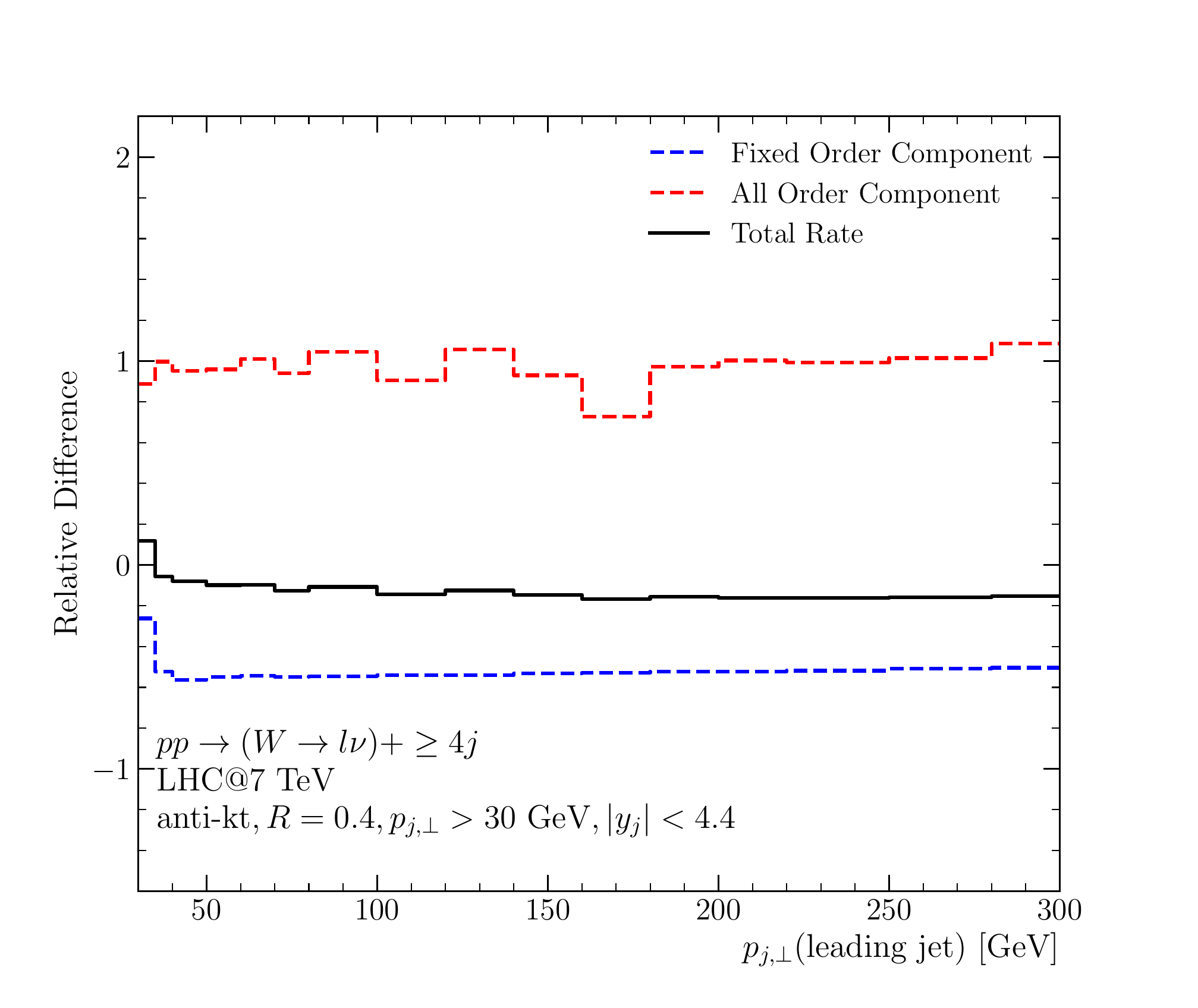}\\
  (c)
  \caption{The differential distribution (black, solid) in the transverse momentum of the
    leading jet in $pp \to (W\to \ell\nu)+\ge 4j$, without and with resummation applied to
    NLL states.  The panels and lines are as in \cref{fig:2dyfb}.}
  \label{fig:2pt3}
\end{figure}

We will now illustrate the numerical impact of these new components in
figs.~\ref{fig:2dyfb}--\ref{fig:2pt3}.  We plot the total differential
cross-section and show its split into all-order and fixed-order components as follows:
\begin{itemize}
\item Case 1: The LO result plus all LL corrections
  ($\alpha_s^{2+k} \log^k(\hat s/p_t^2)$) are included. This is plotted in panel
  (a) of figs.~\ref{fig:2dyfb}--\ref{fig:2pt3}.  Resummation is therefore only
  applied to the LL processes listed in the middle column of \cref{tab:LLNLLOther} and their
  equivalent processes with $\ge 5$-jets, and this contribution is shown by the
  red dashed line marked ``All Order Component''.  All other subprocesses are
  described at fixed-order only and enter the blue dashed line.
\item Case 2: The LO result plus the LL corrections of case 1 are included plus
  the new LL
  corrections to the states starting at $\alpha_s^3$ and
  $\alpha_s^4$ which lead to contributions which scale as $\alpha_s^{3+k}
  \log^k(\hat s/p_t^2)$, calculated in this section. This is plotted in panel
  (b) of figs.~\ref{fig:2dyfb}--\ref{fig:2pt3}.  Resummation is therefore applied to
  all subprocesses listed in the middle and right columns of
  \cref{tab:LLNLLOther} plus their equivalents with $\ge 5$-jets and this gives the red dashed ``All Order Component''
  line.  The remaining subprocesses not listed are described at fixed order only
  and enter the ``Fixed Order Component'' line.
\item Relative difference: the difference between cases 1 and 2
  divided by the results of case 1. This is plotted in panel (c) of figs.~\ref{fig:2dyfb}--\ref{fig:2pt3}.
\end{itemize}
The line ``fixed-order component'' is different between Case 1 and Case 2,
simply because the fixed-order component for each case is the difference
between the full Born level at $\as^n$, $2\le n\le 6$ and the respective
resummed components expanded to the same orders in \as.

The first distributions we show (\cref{fig:2dyfb}) are for the rapidity
difference between the most-forward and most-backward jets, $\Delta y_{j_f,j_b}$,
for $pp\to (W\to \ell\nu)+\ge3j$.  For case 1 (\cref{fig:2dyfb}(a)), the all-order component
(red, dashed) of the full cross section (black, solid) lies between 30--40\%
across the range, with the rest coming from fixed-order configurations.  There
is a dramatic change once resummation is applied also to all NLL states in case 2
(\cref{fig:2dyfb}(b)), where now the all-order component begins at 65\% at
$\Delta y_{j_f,j_b}=0$ and rises to 80\% by $\Delta y_{j_f,j_b}=6.5$.  This
immediately illustrates that although the new NLL contributions are formally
suppressed, they are numerically highly significant in realistic LHC analyses.

\Cref{fig:2dyfb}(c) shows the relative difference in each of the
lines between the two cases.  The difference in the total cross section (black line) decreases linearly
from close to zero down to $-$25\% at large values of $\Delta y_{j_f,j_b}$, while
the changes in the components are much larger (between 60\% and 110\%).  The
comparatively small change in the total is a sign of the perturbative stability
of the \HEJ framework.  The linear behaviour in each case arises from the relation
$y_{j_f,j_b} \sim \log(s_{j_f,j_b}/p_{j_f,\perp}p_{j_b,\perp})$ in the MRK
limit.  Large values of $\Delta y_{j_f,j_b}$ does not guarantee MRK kinematics,
but MRK configurations only contribute at the right-hand side of the plot and
hence the logarithmic behaviour can still be seen.

In \cref{fig:2pt2}, we show the same analysis for the transverse momentum
distribution of the leading jet in $pp \to (W\to\ell\nu)+\ge 3j$.  Again one can
see a dramatic increase in the component of the cross section which is now
controlled by all-order resummation, from 30--40\% across the range in case 1 (\cref{fig:2pt2}(a)) up to
70--80\% in case 2 (\cref{fig:2pt2}(b)).  There is no systematic relation between the high-energy logarithms and
transverse momentum and we therefore see that the relative difference in each
component is quite flat in this variable.  The relative difference in the total
cross section is again comparably small at around 8\%, while the decrease in the
fixed-order component is around $-$30\% and the increase in the all-order
component is +90--100\%.

In \cref{fig:2pt3}, we again show the transverse momentum distribution of the
leading jet, but now for $pp \to (W\to\ell\nu)+\ge 4j$.  One needs at least four
hard jets before the corrections derived in \cref{sec:new-nll-components-4} are
included.  With four jets, there are many more possible final states which are neither LL
or NLL, and hence the fraction of the cross-section to which all-order
corrections are applied is less here than in \cref{fig:2pt2}.  However, one can
still see the dramatic increase in the all-order component from 20--30\% in case 1 (\cref{fig:2pt3}(a)) to
55--65\% in case 2 (\cref{fig:2pt3}(b)).  The relative
differences seen are very similar to the case of $pp \to (W\to\ell\nu)+\ge 3j$.

In \cref{sec:furth-exampl-numer}, we show and discuss more analyses of this kind for
exclusive jet rates, the invariant mass distribution of the leading jets and the
transverse momentum distribution of the leading jet in $pp \to (W\to\ell\nu)+\ge
2j$.

Having discussed the improvement in the all-order component of \HEJ predictions,
in the next section we describe a new approach to fixed-order matching such that
the fixed-order accuracy of \HEJ predictions for $pp \to (W\to\ell\nu)+\ge 2j$ is
increased to NLO.

% \begin{figure}
%   \centering
%   \includegraphics[width=0.49\linewidth]{figures/W/decomp/Plots/notuno-ht2/31_m14.pdf}
%   \includegraphics[width=0.49\linewidth]{figures/W/decomp/Plots/uno-ht2/31_m14_uno.pdf}\\
%   \includegraphics[width=0.49\linewidth]{figures/W/relative-diff/Plots/ht2/31_m14.pdf}
%   \caption{The Cross Sections for W+Jets\textbf{[To Do: Tidy up]}}
%   \label{fig:2pt4}
% \end{figure}

%%% Local Variables:
%%% mode: latex
%%% TeX-master: "main"
%%% End:

\FloatBarrier
\section{All-Order Summation and Kinematic Matching to NLO}
\label{sec:macthmerge}
%% Leading order amplitude of the previous sections are supplemented with
%% further gluon emissions in the resummation region of rapidities in-between
%% the particles exchanging a colour octet. Discussed in
%% section~\ref{sec:HEJamplitudes}.
%% Further discussion of resummation region for sub-leading processes

The perturbative accuracy obtained in the published predictions so far
obtained with
\HEJ~\cite{Andersen:2011hs,Andersen:2012gk,Andersen:2016vkp,Andersen:2017kfc,Andersen:2018tnm,Andersen:2018kjg}
is controlled by matching point-by-point to the $n$-jet Born-level
matrix element, in this study generated by
Sherpa~\cite{Bothmann:2019yzt} with the extension of
\mbox{OpenLoops}~\cite{Buccioni:2017yxi}. The $m$-parton, $n$-jet phase space
is explored using the efficient procedure described in
ref.~\cite{Andersen:2018tnm}, and the $n$-jet samples are merged with
the logarithmic accuracy obtained with the resummation. For the study
of $W+$jets at hand, fixed-order processes of up to $W+6$-jets are
considered, and thus the predictions using the method outlined
in~\cite{Andersen:2018tnm} would obtain an accuracy of $\alpha_s^n$
plus all logarithmic corrections to this of the form
$\alpha_s^n(\alpha_s \log(s))^m$ and the detailed corrections of the
form $\alpha_s^{n+1}(\alpha_s \log(s))^m$, $m>0$, for all observables
with contributions from $n=2,\ldots, 6$ jets\footnote{We note that the
suppressed interference corrections which were neglected in
\cref{sec:currents} are included at fixed order through this
matching.}. In this section we will describe a method for extending
the fixed-order input to NLO calculations and ensuring that all
distributions of $W+n$-jets obtain both full $\alpha_s^{n+1}$ (NLO)
and all-order logarithmic accuracy.

The matching of logarithmic and fixed next-to-leading order accuracy of
distributions is obtained by calculating all distributions of $W+2$-jets at both
fixed-order NLO (using again Sherpa~\cite{Bothmann:2019yzt} at fixed order) and for the NLO expansion of the leading and sub-leading
logarithmic contributions to the cross section. For the inclusive two-jet
rate, \HEJ at NLO is obtained by expanding the virtual corrections from the
reggeized propagator in the matrix elements for two-parton productions, and
truncating the real emissions at a single extra gluon.
% for the real emissions include both leading (with the additional parton
% arising from a Lipatov-vertex) and sub-leading (arising from currents
% producing two partons) contributions to the three-parton production.
The
organisation of the cancellation of the poles in $\varepsilon$ is performed
with the same subtraction method as used in the all-order
calculation. Predictions can then be obtained at full NLO and resummed
(leading and sub-leading) accuracy of each distribution by the following
procedure:
\begin{enumerate}
\item obtain histograms for the process at (a) full NLO, (b) \HEJ at NLO and (c)
  full \HEJ (including the fixed-order 2-jet and 3-jet subprocesses not subject to
  resummation which enter (b), but not including these for $\ge4$-jets).
\item scale each bin in the all-order resummed distribution 1(c) by the ratio of
  the same bin in the histograms of 1(a) and 1(b).
\end{enumerate}
The final weight in each bin, $w_{\rm HEJ2\, NLO}$, is then given by
\begin{align}
  \label{eq:binfactor}
  w_{\rm HEJ2\, NLO} = w_{\rm HEJ2} \frac{w_{\rm NLO}}{w_{\rm HEJ\,at\,NLO}} + w_{\rm FO\, W+\ge4j},
\end{align}
where $w_{\rm NLO}$ is the weight from full NLO (1(a) above), $w_{\rm HEJ\,at\,NLO}$
is the full HEJ prediction truncated at $\alpha_s^3$ (1(b) above) and $w_{\rm
  HEJ2}$ is the inclusive $W+2$-jet prediction from HEJ at all orders plus the
2-jet and 3-jet processes we must add at fixed order (1(c) above). $w_{\rm FO\, W+\ge4j}$
represents the fixed-order processes at 4-jets and above which do not enter the
NLO matching.

In practice, the matching is performed for $W^+$ and $W^-$ separately, but we find
that the
matching corrections are similar. A simple expansion of the perturbative
series will show that each bin value in the resulting distribution will have
full NLO and full logarithmic accuracy. % The procedure can straightforwardly
% be extended to include also the perturbative accuracy of $W+3$j at NLO by
% then performing the NLO matching on the 2-jet exclusive (calculated at
% $\mathcal{O}(\alpha_s^3)$) and the 3-jet inclusive (calcualted
% at $\mathcal{O}(\alpha_s^4)$) contributions separately.\todo{Not sure we want to
% mention this here because we don't do it yet.}
This matching of kinematic distributions to full NLO accuracy corrects
distributions in the sub-asymptotic kinematic regions where the leading and sub-leading
logarithmic approximation is a poor approximation to the full perturbative
result, e.g.~high-momentum regions.

\begin{figure}
  \centering
  \includegraphics[width=0.49\linewidth]{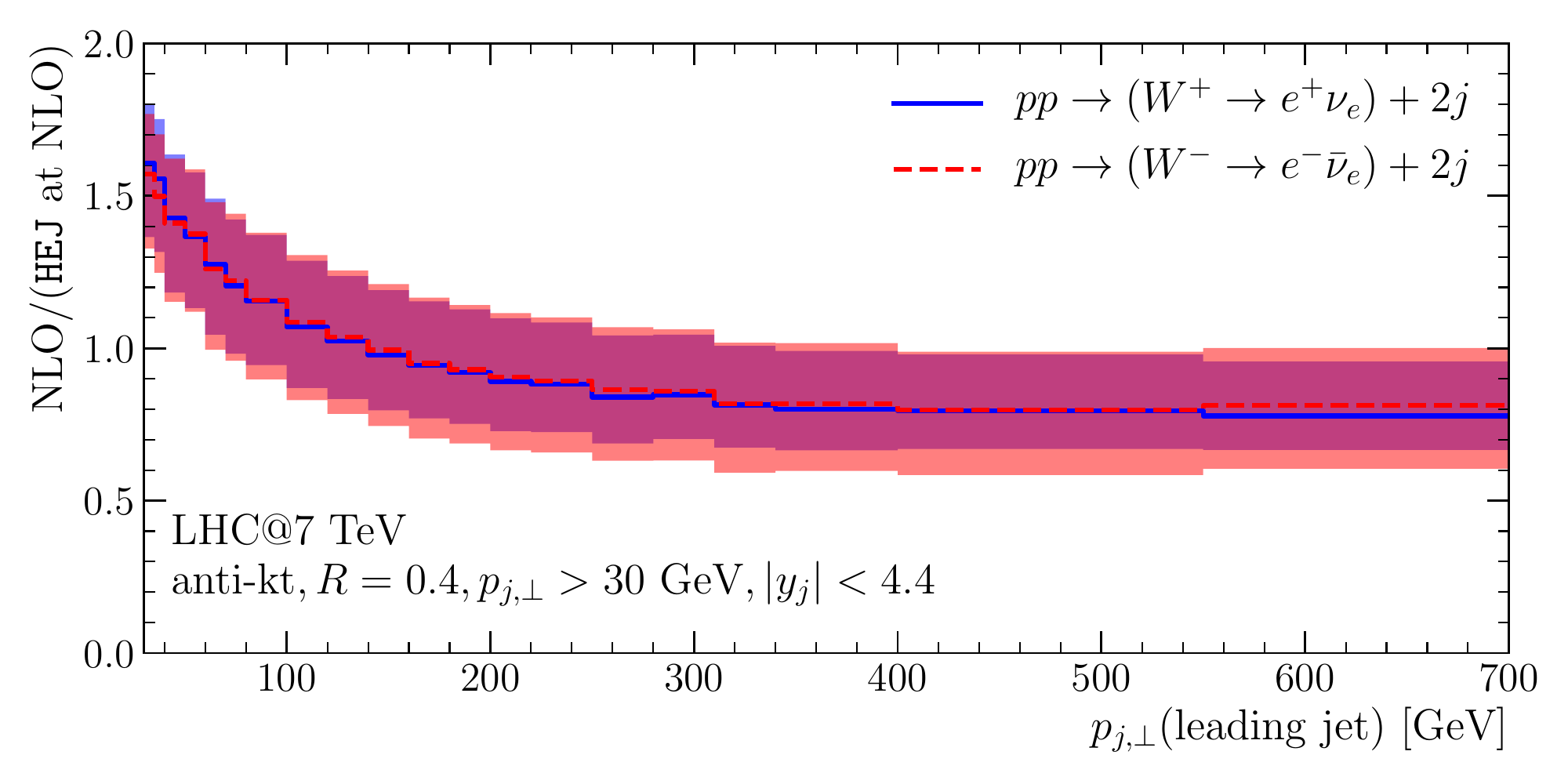}\hfill
  \includegraphics[width=0.49\linewidth]{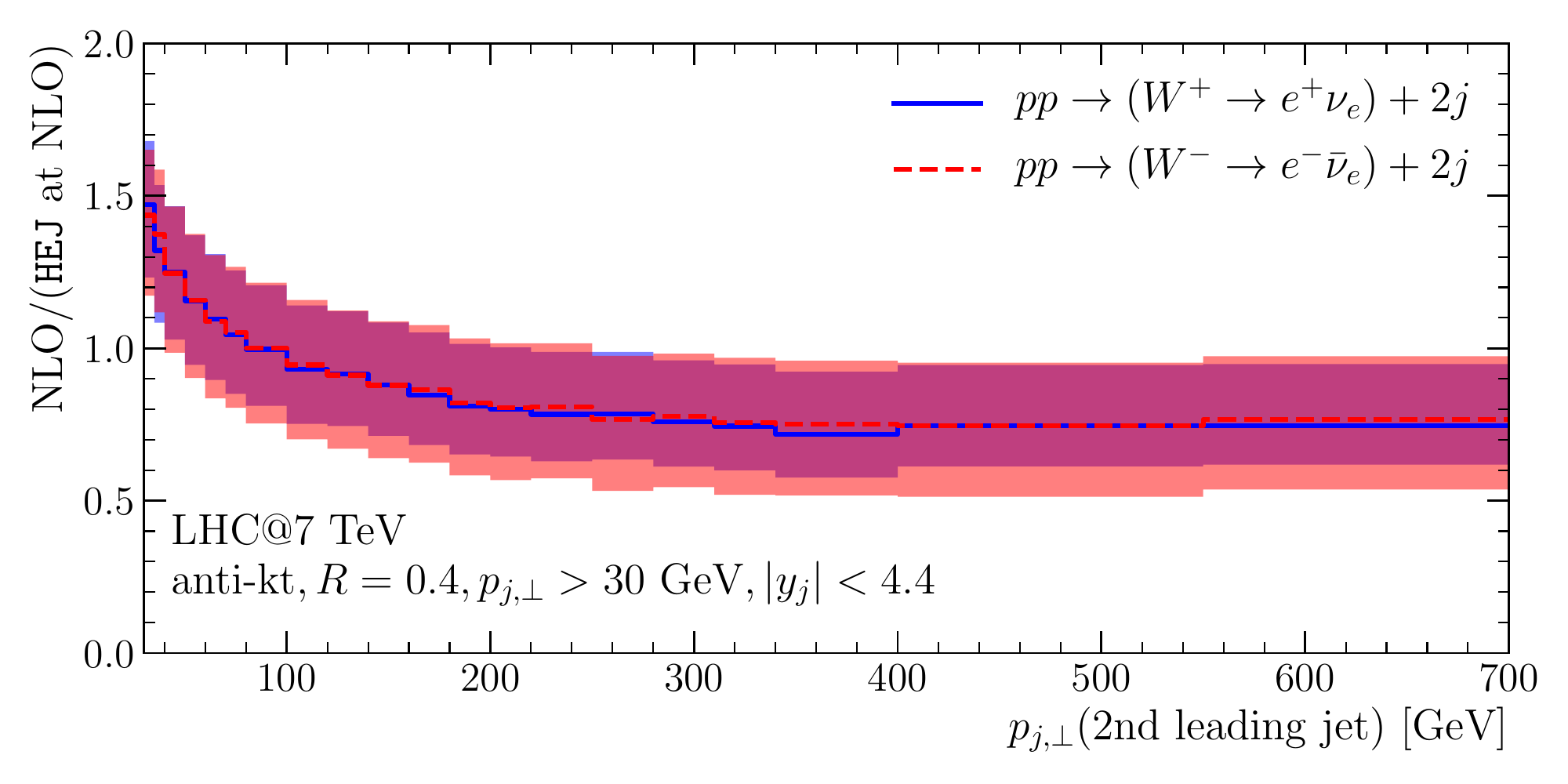}\\
  (a) \hspace{7cm} (b) \\
  \includegraphics[width=0.49\linewidth]{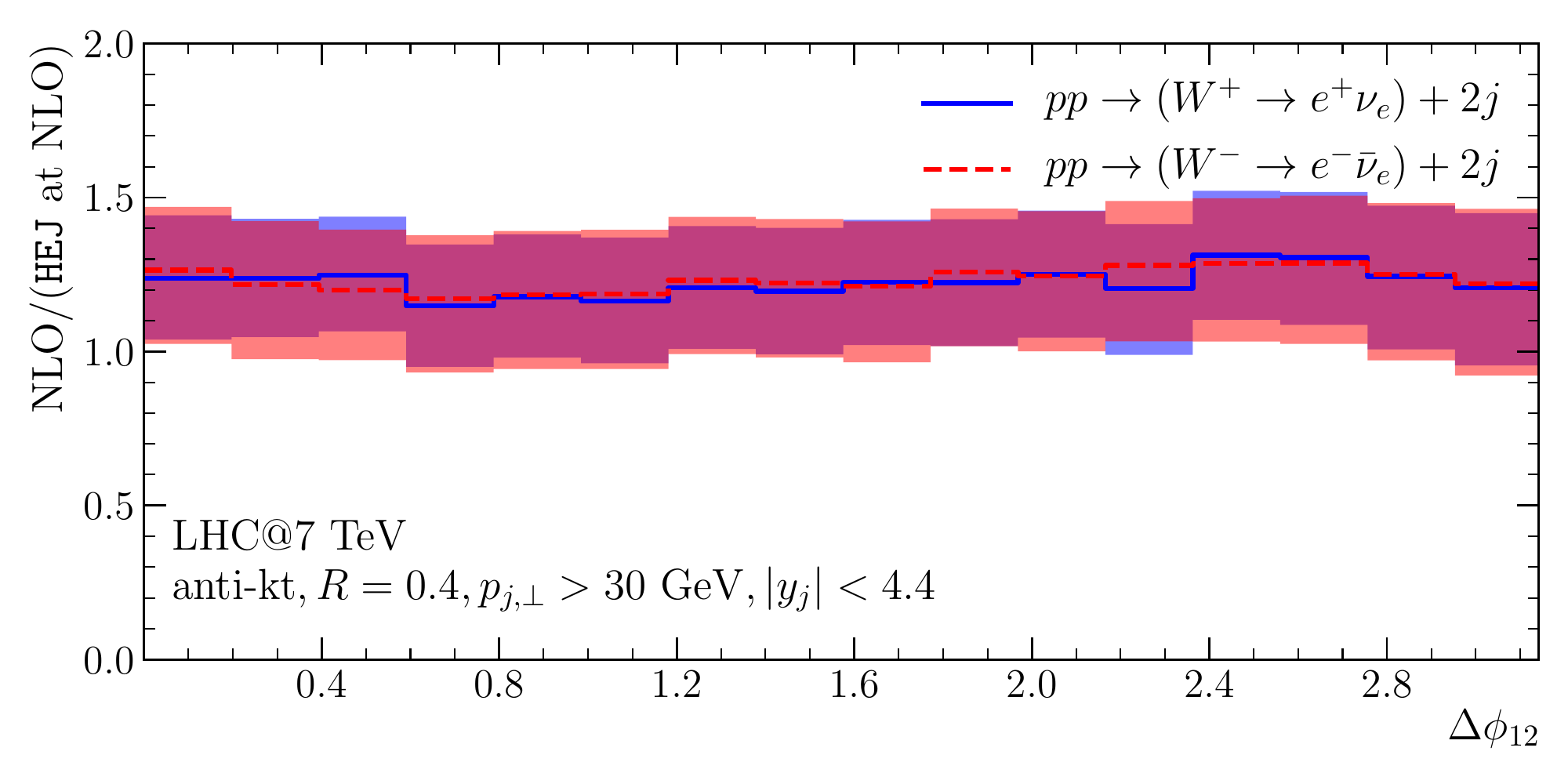}\\
  (c)
  \caption{The NLO matching corrections obtained for the following distributions
    presented in
    ref.~\cite{Aad:2014qxa} for inclusive $W^++2j$ production (blue, solid) and $W^-+2j$ production
    (red, dashed): (a) transverse momentum of
    the leading jet, (b) transverse momentum of the second jet and (c) the
    difference in azimuthal angle between the leading two jets.}
  \label{fig:NLOmatchingcorrections7}
\end{figure}

\begin{figure}
  \centering
  \includegraphics[width=0.49\linewidth]{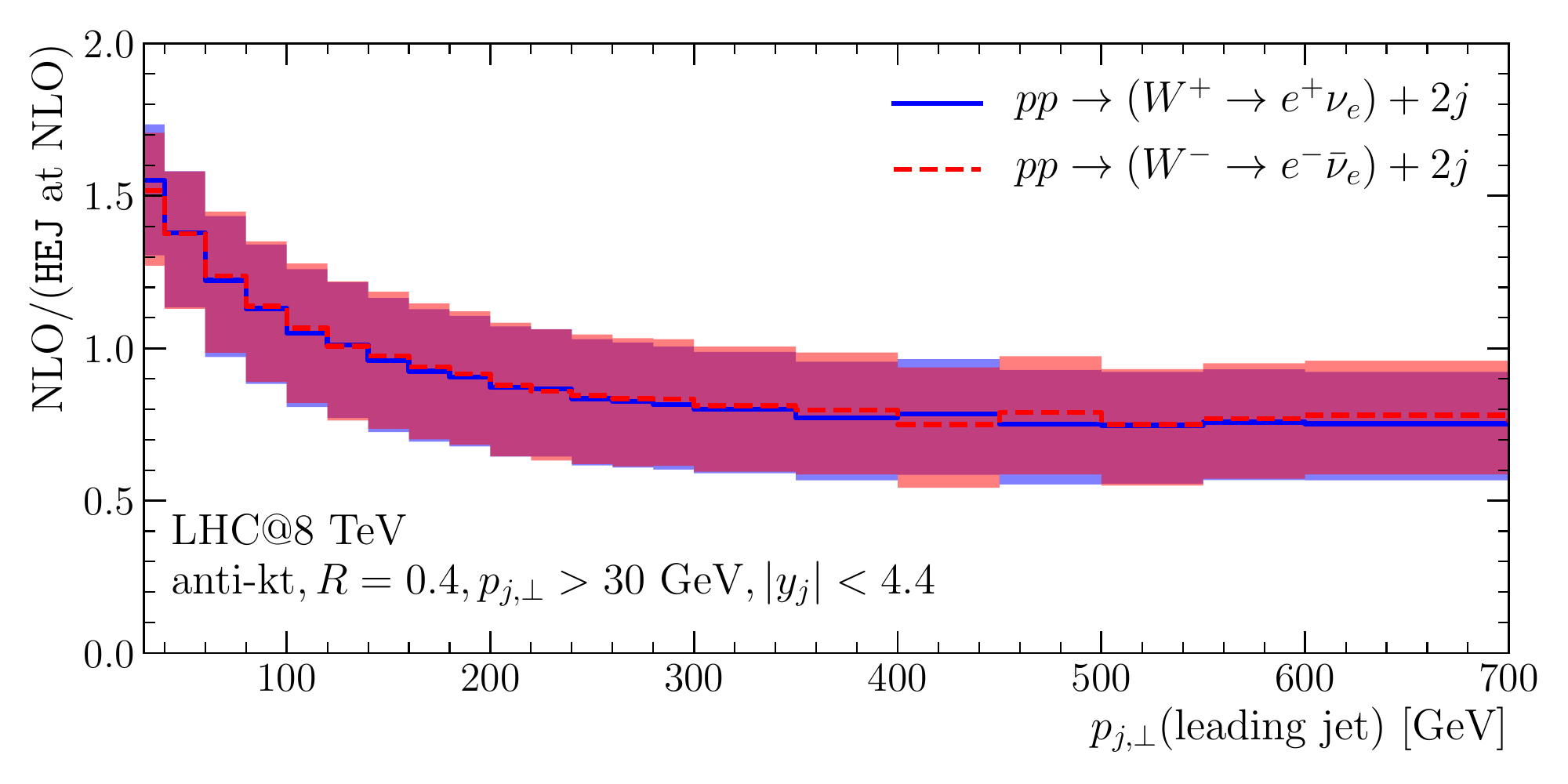}\hfill
  \includegraphics[width=0.49\linewidth]{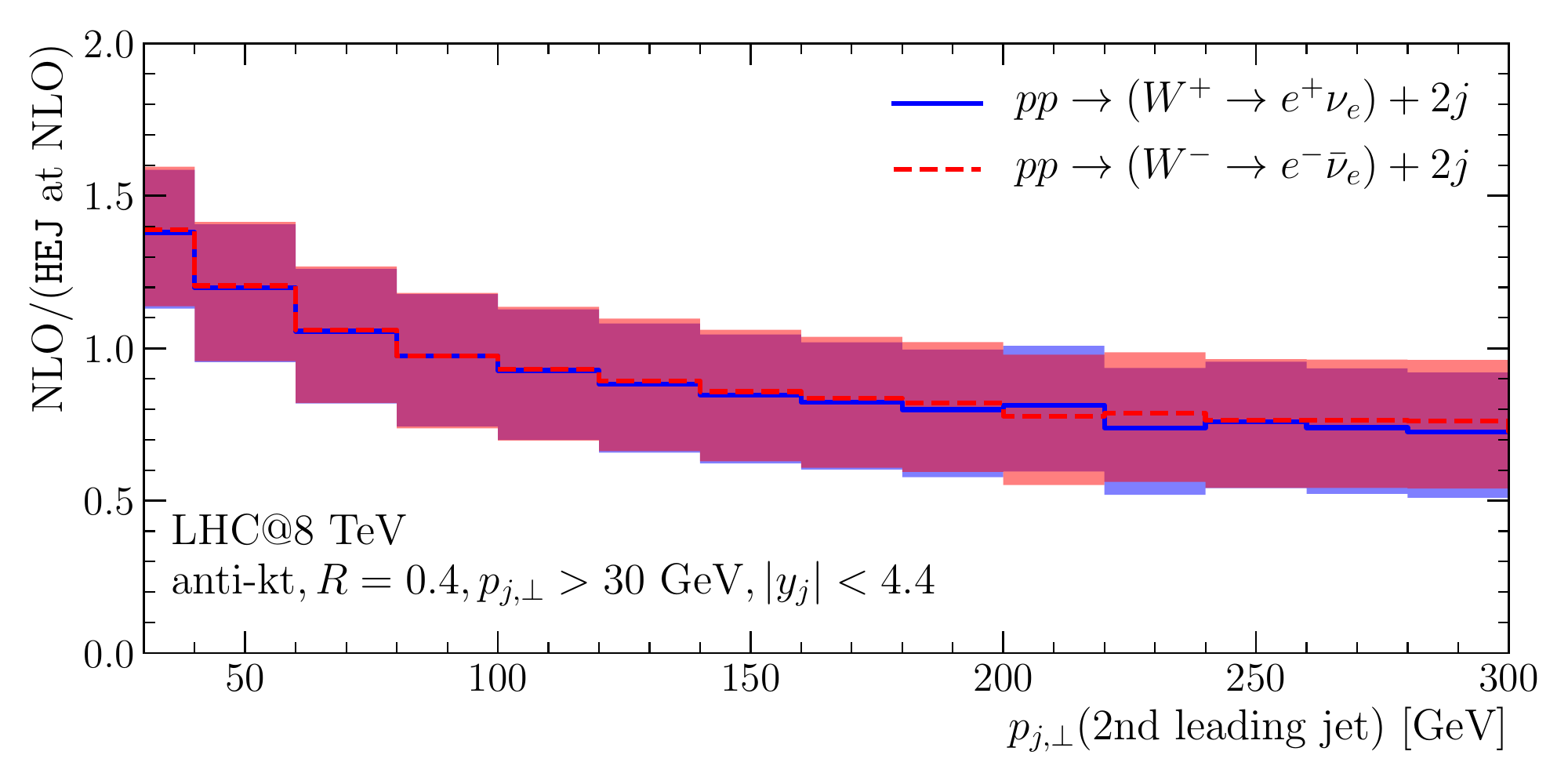}\\
  (a) \hspace{7cm} (b) \\
  \includegraphics[width=0.49\linewidth]{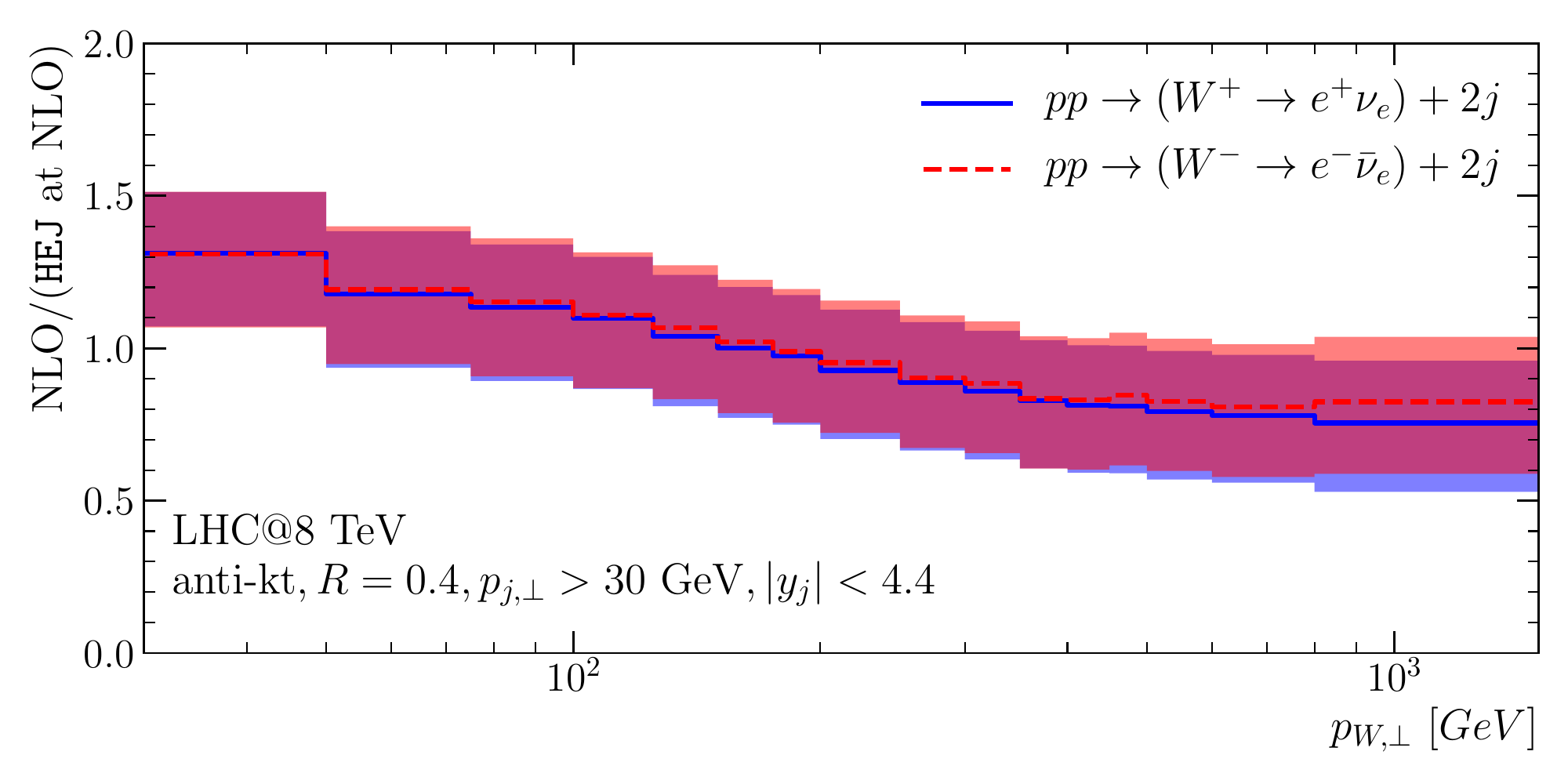}\\
  (c)
  \caption{The NLO matching corrections for the following distributions
    presented in
    ref.~\cite{Aaboud:2017soa} for inclusive $W^++2j$ production (blue, solid) and $W^-+2j$ production
    (red, dashed): (a) transverse momentum of
    the leading jet, (b) transverse momentum of the second jet and (c) the
    transverse momentum of the $W$ boson.}
  \label{fig:NLOmatchingcorrections8}
\end{figure}

We show the matching corrections used in the following section in
figs.~\ref{fig:NLOmatchingcorrections7} and
\ref{fig:NLOmatchingcorrections8}. Fig.~\ref{fig:NLOmatchingcorrections7}
shows the corrections for $W+2$-jet production at 7 TeV for distributions which
appear in ref.~\cite{Aad:2014qxa} and \cref{fig:NLOmatchingcorrections8}
shows examples for $W+2$-jet production at 8 TeV for distributions which appear in
ref.~\cite{Aaboud:2017soa}.  We chose a central scale of $\mu_r=\mu_f=H_T/2$. We varied $\mu_r$ and $\mu_f$ independently by a factor of two around this central scale choice, keeping their ratio between 0.5 and 2. The scale variation bands are obtained from the envelope of these variations. The jet parameters and cuts are given in the
figures.  We see that the $p_{j,\perp}$ distributions
(figs.~\ref{fig:NLOmatchingcorrections7}(a),
\ref{fig:NLOmatchingcorrections7}(b), \ref{fig:NLOmatchingcorrections8}(a) and
\ref{fig:NLOmatchingcorrections8}(b)) all have a similar shape; they start
around 1.5 for low $p_{j,\perp}$ values and then fall until reaching a stable value
around 0.7-0.8 for $p_{j,\perp}$ values above about $300$~GeV.  A similar drop is also
seen in the matching corrections as a function of $p_{W,\perp}$ in
fig.~\ref{fig:NLOmatchingcorrections8}(c).  On the other hand, the impact of
matching is independent of the azimuthal separation of the jets, and hence the
matching factor in fig.~\ref{fig:NLOmatchingcorrections7}(c) is relatively
flat.  None of the figures show any significant difference between $W^++2j$
production and $W^-+2j$ production.  The improvement obtained in the description of data from applying these
NLO matching corrections will be illustrated in \cref{sec:results}.

%%% Local Variables:
%%% mode: latex
%%% TeX-master: "main"
%%% End:

\FloatBarrier
\section{The NLO-matched All-Order Predictions for Measured Quantities}
\label{sec:results}
This study has introduced two elements targeted at systematically improving
the precision in regions away from asymptotic large energies compared to
transverse scales: sub-leading corrections and matching of distributions to
next-to-leading order. This section compares predictions to analyses by ATLAS
of $W$+dijets at LHC energies of 7~TeV~\cite{Aad:2014qxa} and
8~TeV~\cite{Aaboud:2017soa}, where we focus on distributions
where the new components lead to important improvements. The data was
collected using slightly different cuts for the electron and muon channel,
and then extrapolated to a ``combined'' selection contrasted with predictions
generated with the following cuts listed in table~\ref{tab:cuts7TeV}. The
rapidity selection criteria for jets, accepting jets of rapidities in the
full detector, not just the central part, is particularly important for the
study of high-energy logarithms: it the same selection criteria as that used
for the study of Higgs boson production in association with dijets,
\emph{and} it avoids the focus on jets of very large transverse momenta,
which is a result of studying large $m_{jj}$ for solely central jets.
\begin{table}
  \centering
  \begin{tabularx}{\linewidth}{ll}
    Lepton $p_\textrm{T}$& $p_\textrm{T}$$>25$~GeV\\
    Lepton rapidity & $|\eta|<2.5$\\
    Missing transverse momentum & $E^{\textrm{miss}}_{\textrm{T}}>25$~GeV\\
    Reconstructed transverse mass of boson & $m_\textrm{T}>40$~GeV\\
    Jet $p_\textrm{T}$& $p_\textrm{T}>30$~GeV\\
    Jet rapidity & $|y|<4.4$\\
    Jet isolation &$\Delta R(l,\textrm{jet})>0.5$ (7~TeV \cite{Aad:2014qxa}),
                    0.4 (8~TeV \cite{Aaboud:2017soa})\\
    &jet is removed
  \end{tabularx}
  \caption{The selection cuts used for the comparisons to data.}
  \label{tab:cuts7TeV}
\end{table}

We will compare the measured data with the fixed-order predictions for
$W$+dijets at NLO obtained using Sherpa~\cite{Bothmann:2019yzt} with
the extension of OpenLoops~\cite{Buccioni:2017yxi} and those of HEJ
obtained using the improvements presented in the present paper. We use
the NNPDF3.1nlo pdf set~\cite{Ball:2017nwa} as provided by
LHAPDF6~\cite{Buckley:2014ana}. In particular we will compare the
predictions for the inclusive and exclusive jet counts, the transverse
momenta of the leading and sub-leading jet and of the $W$-boson, and
the azimuthal angle between the two leading jets. A central scale of
$\mu_f=\mu_r=H_T/2$ is chosen for all predictions, and the scale
variation band are the envelope of an independent variation of each
scale by up to a factor up 2, except configurations with a ratio
between them larger than 2 (or smaller than $1/2$). This scale choice
is guided purely by the convention for fixed-order predictions --- it
has not been tuned in order to obtain a good fit of data to the
predictions of \HEJ.

The impact on observables of including resummation of the next-to-leading
processes and the NLO matching have been shown separately in
\cref{sec:subleading} and \cref{sec:macthmerge}
respectively. In this section, we show the new predictions combining both improvements (labelled
``\HEJtwo NLO'', shown in green).  We also show the previously available
predictions based on just the leading logarithmic
description~\cite{Andersen:2012gk} combined with matching to high-multiplicity
Born-level matrix elements (labelled ``\HEJone'', shown in blue).  The \HEJone
predictions failed systematically in the regions of large
($>100$GeV) transverse momenta of the jets --- this specifically
breaks the multi-Regge kinematic conditions that the leading logarithmic
approximation is based upon. This feature showcases one of the shortcomings
of the matching of the non-resummable part of the cross section by a na\"ive
addition of Born-level events of increasing multiplicity: these Born-level
predictions should be considered inclusive in the jet multiplicity, so simply
adding the contribution from the samples leads to overestimated cross
sections -- just as found. While this question, related to the higher-order
perturbative corrections to each sample, is not solved by including the
sub-leading corrections, the contribution from the non-resummable part of the
cross section at large transverse momenta is reduced from 60\% when
resumming only the processes contributing at leading-logarithmic accuracy to
less than 35\% when the next-to-leading subprocesses are included in the
all-order treatment (figure~\ref{fig:2pt1}). The reduction in the
contribution from fixed-order events obviously reduces the impact of their
slight perturbative mis-use.

\begin{figure}
  \centering
  \begin{subfigure}[t]{.49\linewidth}
    \centering
    \includegraphics[width=\linewidth]{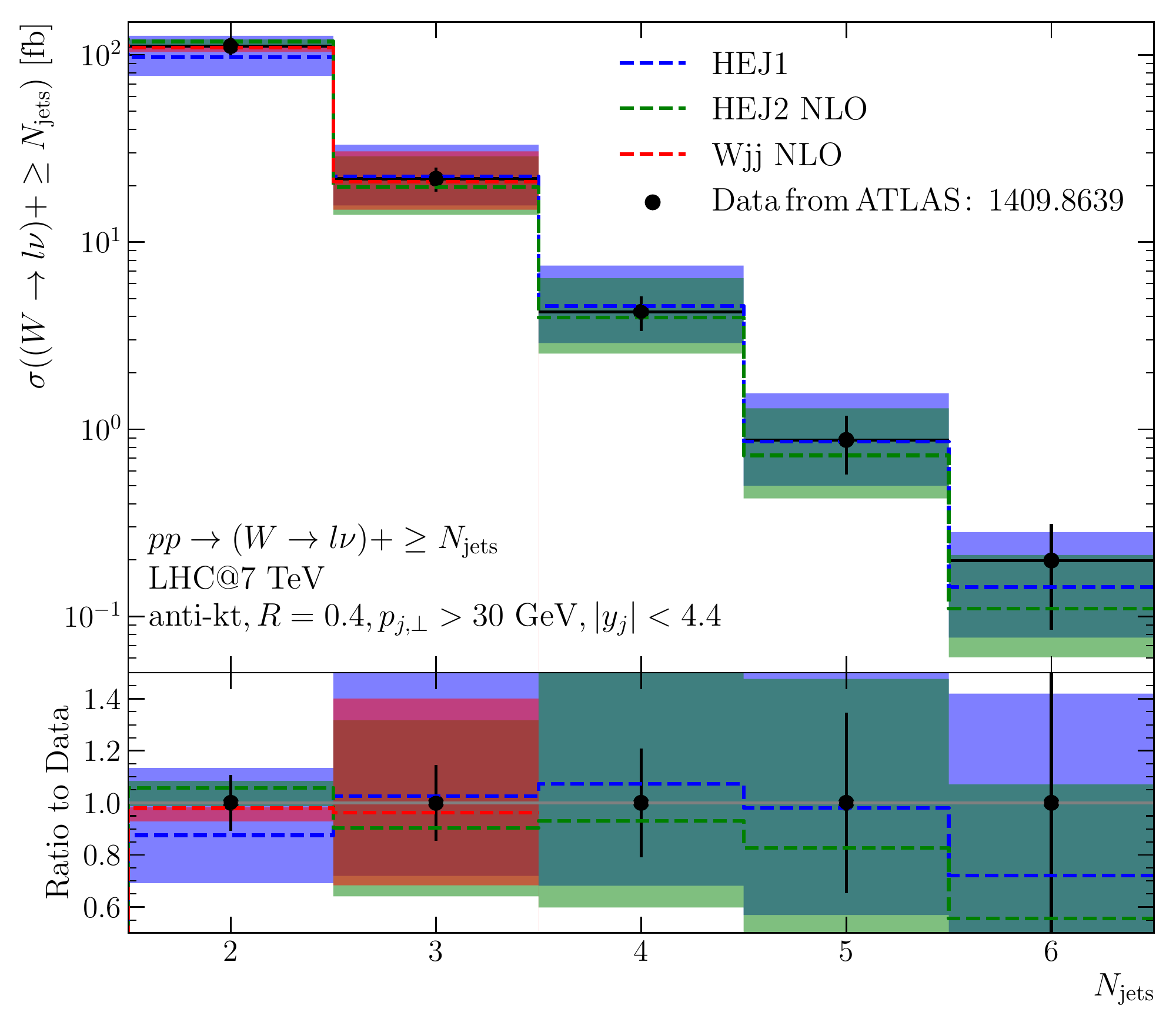}
    \caption{Inclusive jet rates.}
  \label{fig:WjetsNLOmatched7TeVa}
  \end{subfigure}\hfill
  \begin{subfigure}[t]{.49\linewidth}
    \includegraphics[width=\linewidth]{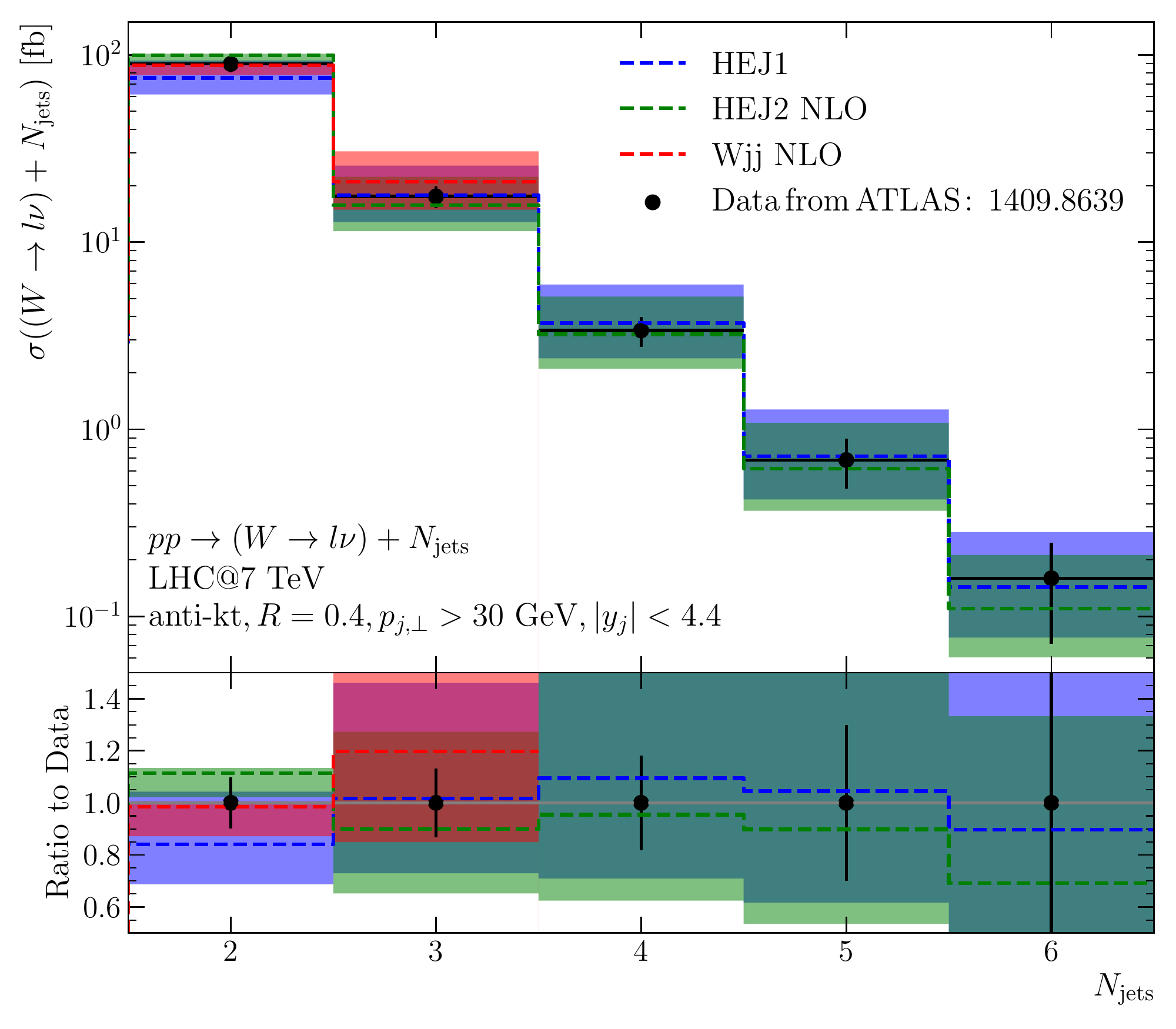}
    \caption{Exclusive jet rates.}
  \label{fig:WjetsNLOmatched7TeVb}
  \end{subfigure}\vfill
  \begin{subfigure}[t]{.49\linewidth}
    \includegraphics[width=\linewidth]{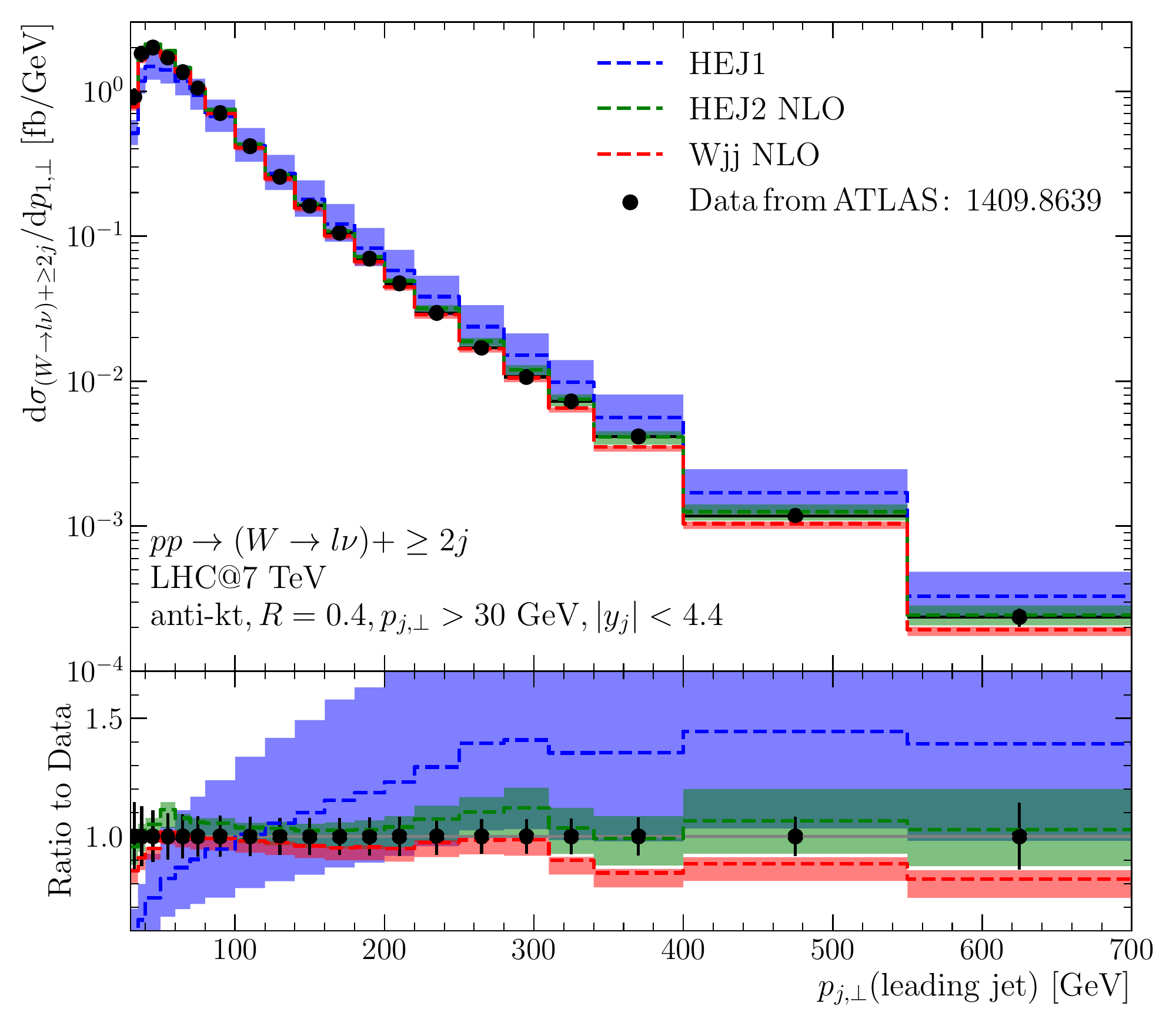}
    \caption{Differential cross section for the transverse momentum of the
      leading jet.}
  \label{fig:WjetsNLOmatched7TeVc}
  \end{subfigure}\hfill
  \begin{subfigure}[t]{.49\linewidth}
    \includegraphics[width=\linewidth]{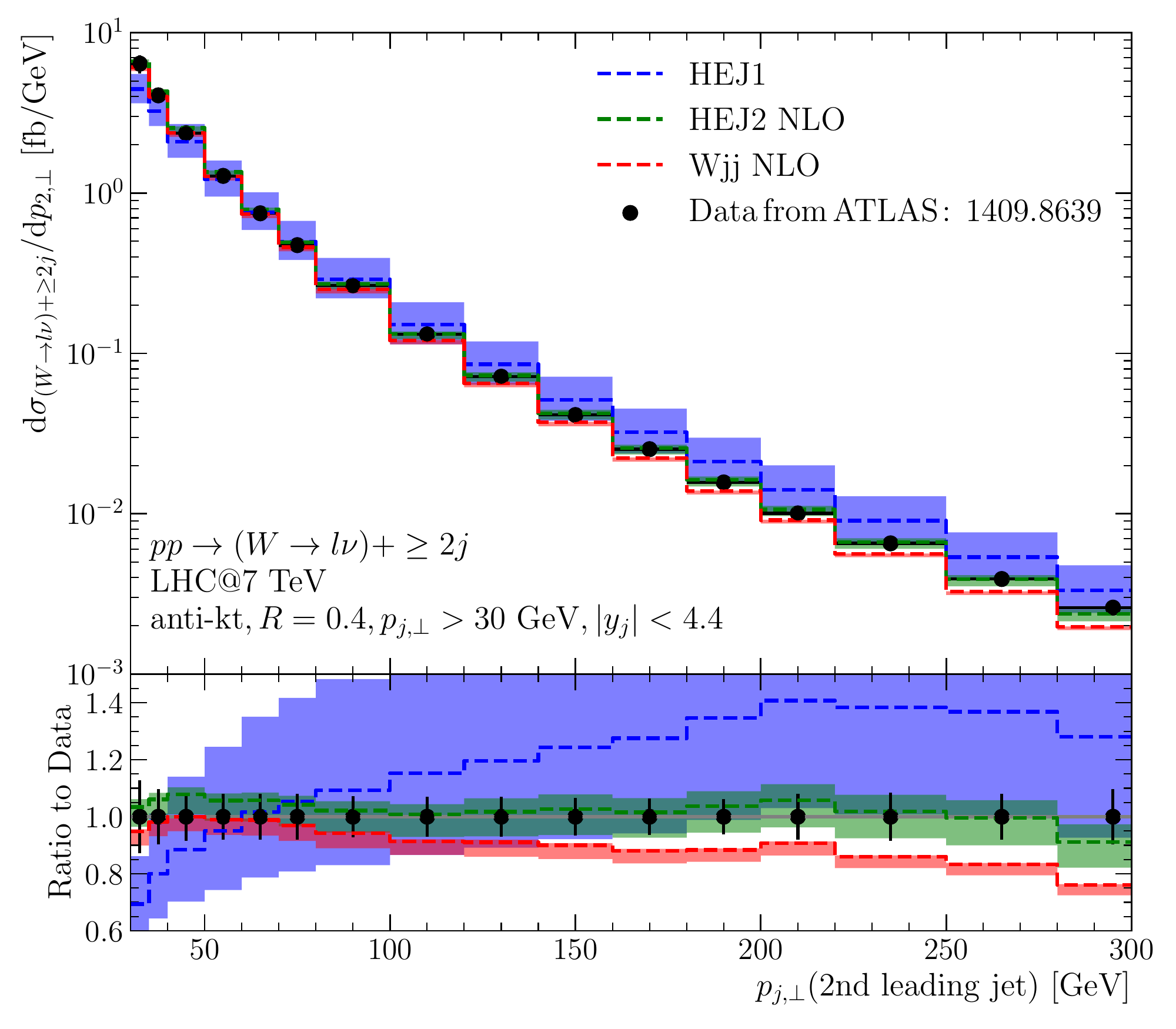}
    \caption{Differential cross section for the transverse momentum of the
      second leading jet.}
  \label{fig:WjetsNLOmatched7TeVd}
  \end{subfigure}
  \vfill
  \begin{subfigure}[t]{.49\linewidth}
    \includegraphics[width=\linewidth]{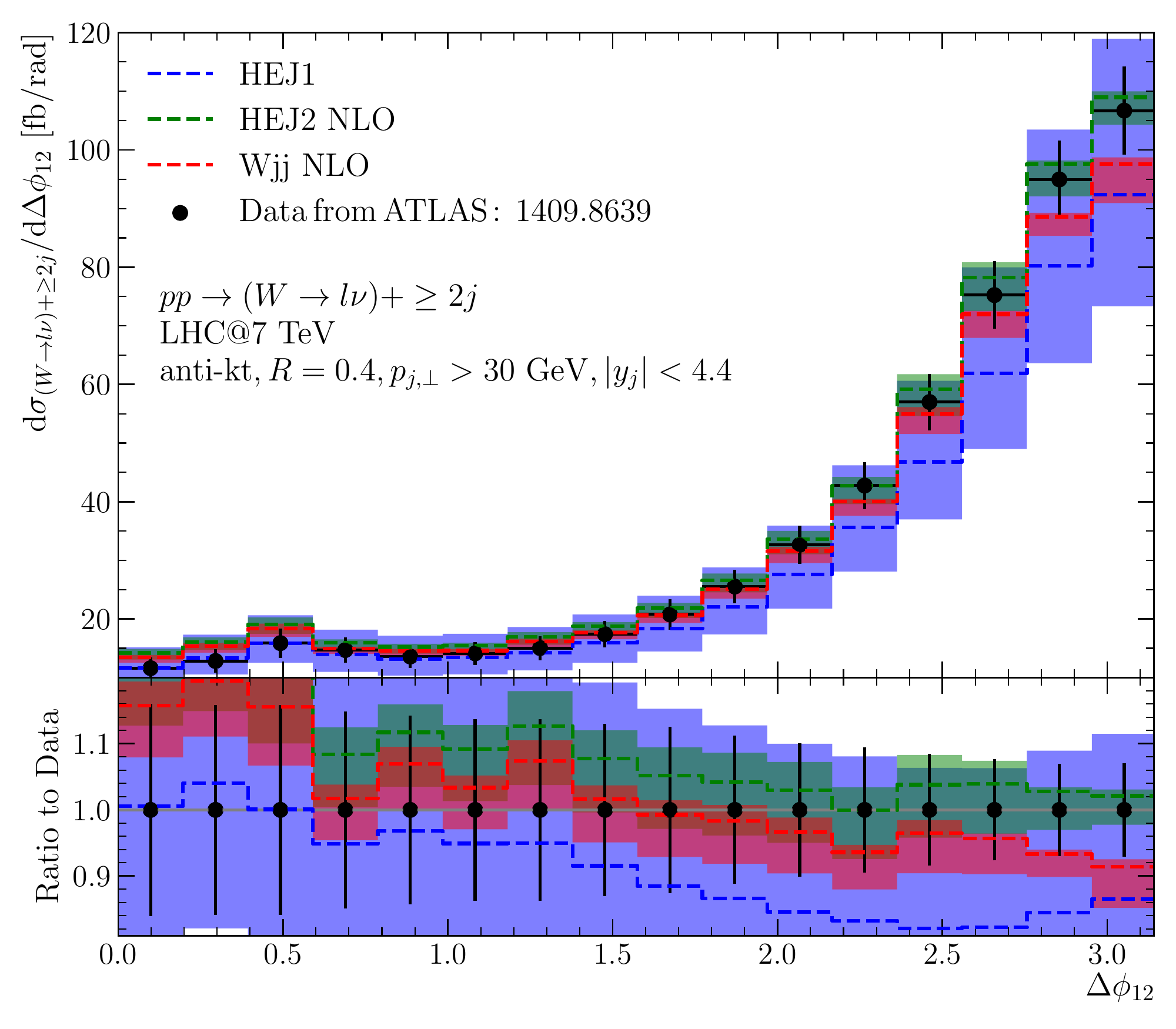}
    \caption{Differential cross section for the azimuthal angle between the
      two leading jets.}
  \label{fig:WjetsNLOmatched7TeVe}
  \end{subfigure}
  \caption{Predictions for $pp\to(W\to l\nu)+\ge2j$ for LHC@7TeV, compared to ATLAS data
    from~\cite{Aad:2014qxa}.  The original \HEJone predictions are shown in blue,
    the pure NLO predictions are shown in red and the new \HEJtwo predictions
    incorporating the methods of sections~\ref{sec:subleading} and
    \ref{sec:macthmerge} are shown in green.}
  \label{fig:WjetsNLOmatched7TeV}
\end{figure}
Figs.~\ref{fig:WjetsNLOmatched7TeVa}-\ref{fig:WjetsNLOmatched7TeVb}
compare the
predictions for the inclusive and exclusive jet rates respectively to data. The inclusive 2-jet rate is matched to NLO
accuracy, which as expected decreases the scale dependence compared to the
predictions obtained with \HEJone reported in~\cite{Aad:2014qxa}. The central
predictions for both the inclusive and exclusive rates now agree more closely
with the data. The scale variation of the 2-jet bin is
significantly reduced due to the NLO matching in that bin. It is worth noting
that while the scale variation in the inclusive 2-jet bin from pure NLO is
smaller than that of the NLO-matched \HEJ2, the scale variation in the
\emph{exclusive} 2-jet bin of fig.~\ref{fig:WjetsNLOmatched7TeVb} are similar
for the NLO-calculation and the NLO-matched \HEJtwo.

Figs.~\ref{fig:WjetsNLOmatched7TeVc}-\ref{fig:WjetsNLOmatched7TeVd}
demonstrates, as expected, that the impact of including the sub-leading
processes in the resummation is significant for the description of
scatterings at large transverse momenta. Indeed, all the problems identified
with the description in \HEJone are solved in \HEJtwo. The description of the
spectrum of the leading and sub-leading jet is better still than that of pure
$W+$dijets at NLO.  The scale variation is slightly larger which is due to the
corrections at $\alpha_s^4$ and above included in the \HEJtwo predictions.

Finally, \cref{fig:WjetsNLOmatched7TeVe} shows the differential
distribution for the azimuthal angle between the two hardest jets,
$\phi_{12}$. Again, the already decent description of \HEJone is slightly
improved in \HEJtwo, and the scale variation is reduced. Overall, the agreement
with data for this observable is similar to that reached at pure NLO.

\begin{figure}
  \centering
  \begin{subfigure}[t]{.49\linewidth}
    \centering
    \includegraphics[width=\linewidth]{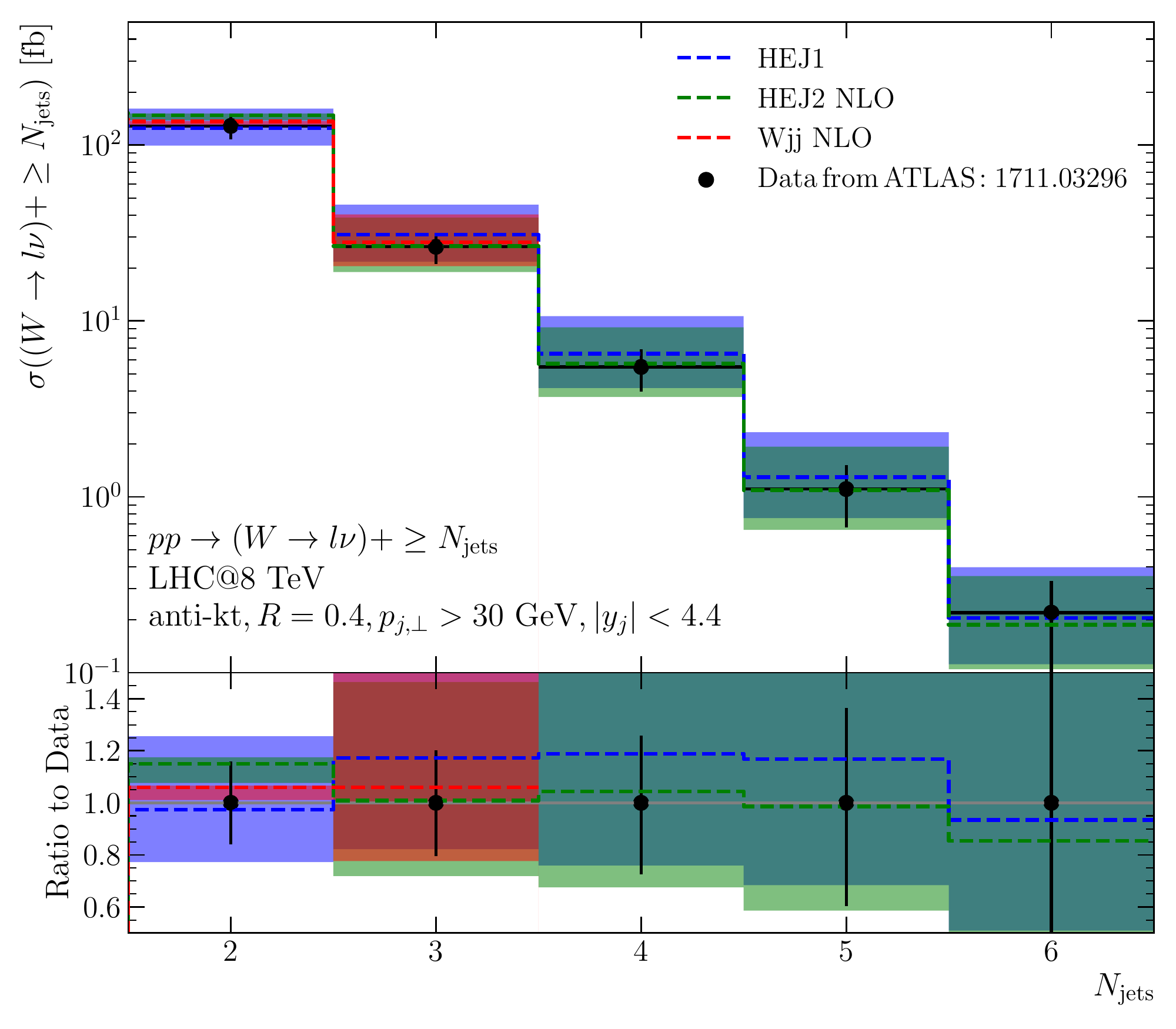}
    \caption{Inclusive jet rates.}
    \label{fig:WjetsNLOmatched8TeVa}
  \end{subfigure}\hfill
  \begin{subfigure}[t]{.49\linewidth}
    \includegraphics[width=\linewidth]{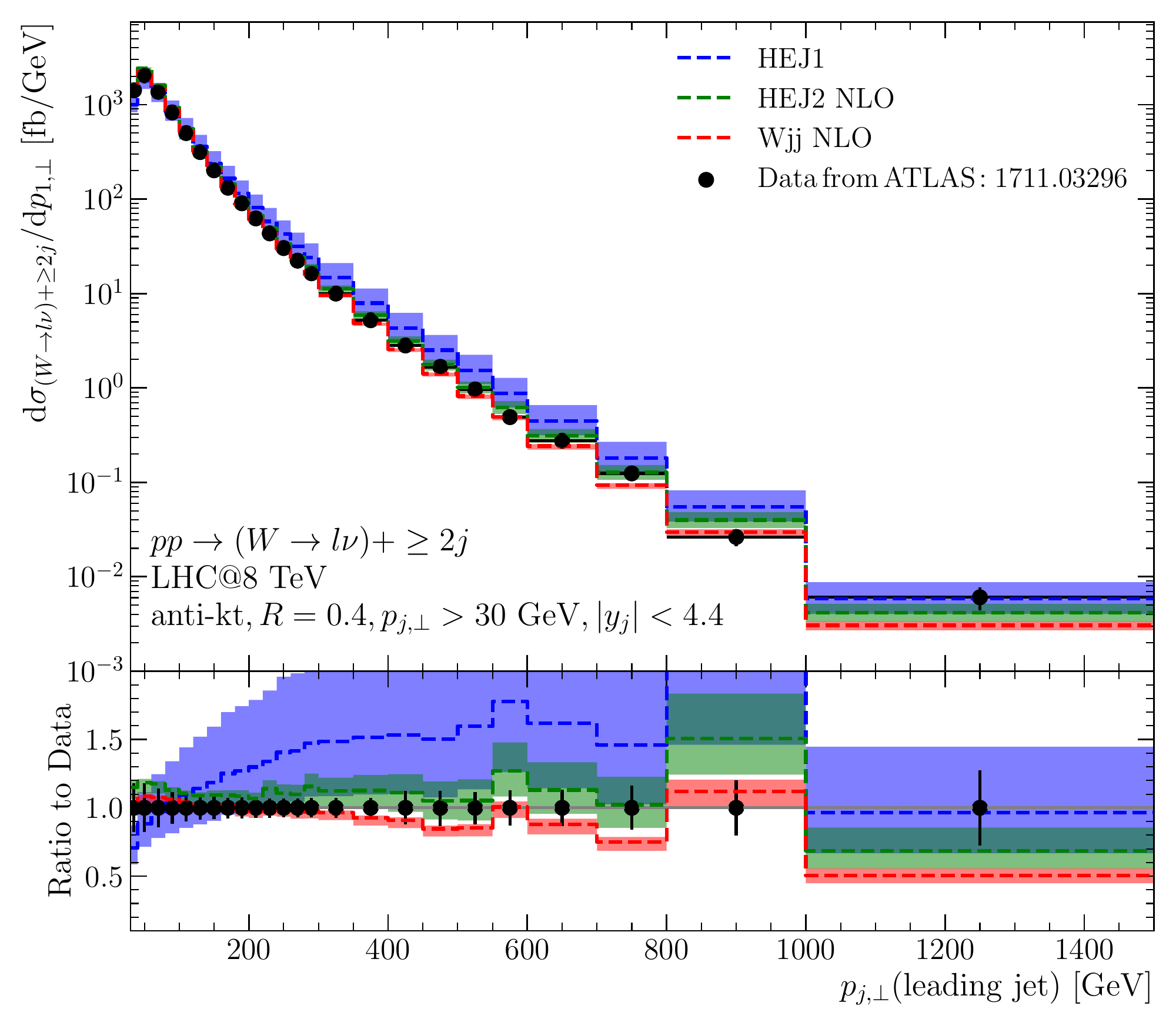}
    \caption{Differential cross section for the transverse momentum of the
      leading jet.}
  \label{fig:WjetsNLOmatched8TeVb}
  \end{subfigure}\vfill
  \begin{subfigure}[t]{.49\linewidth}
    \includegraphics[width=\linewidth]{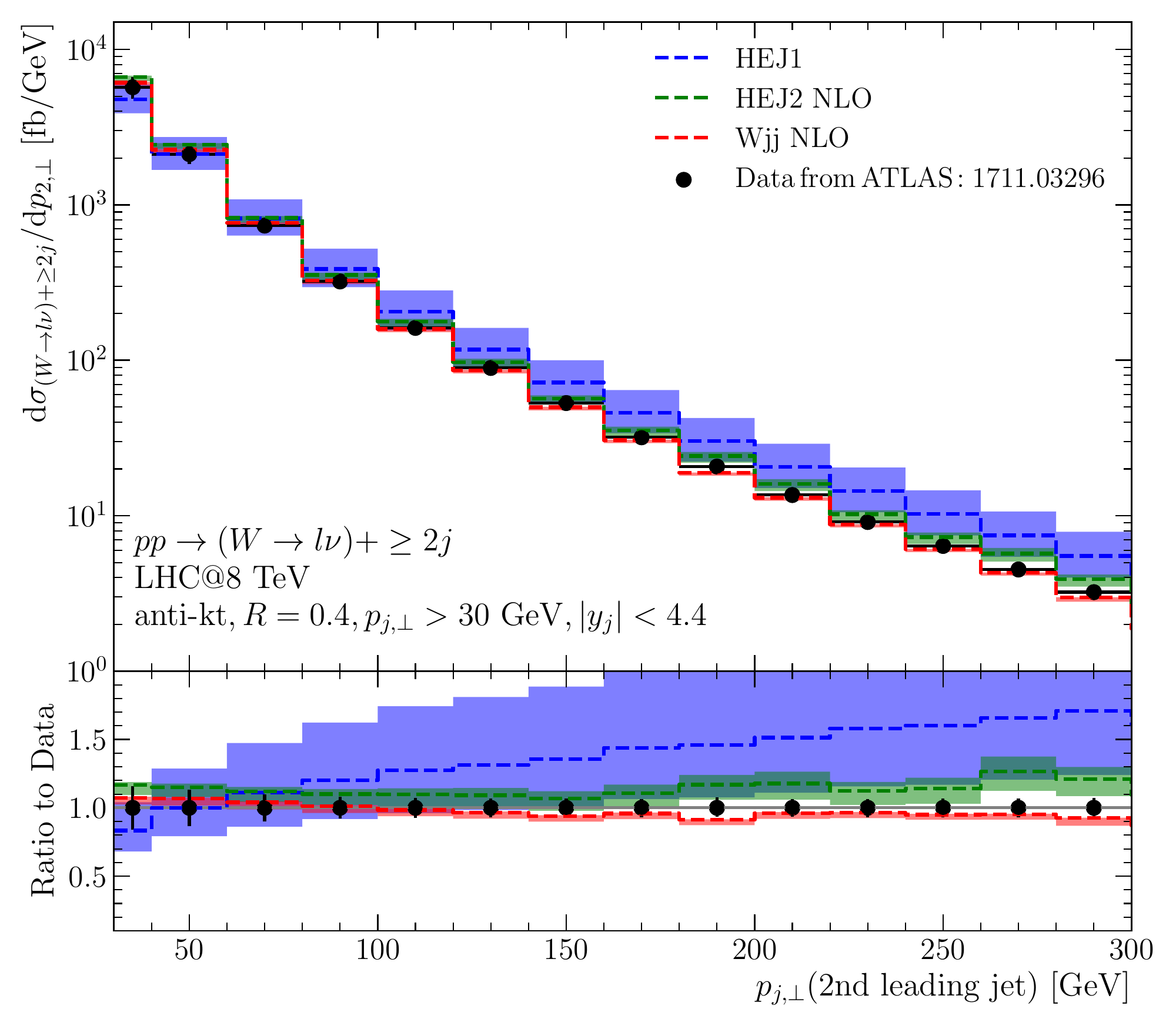}
    \caption{Differential cross section for the transverse momentum of the
      second leading jet.}
  \label{fig:WjetsNLOmatched8TeVc}
  \end{subfigure}\hfill
  \begin{subfigure}[t]{.49\linewidth}
    \includegraphics[width=\linewidth]{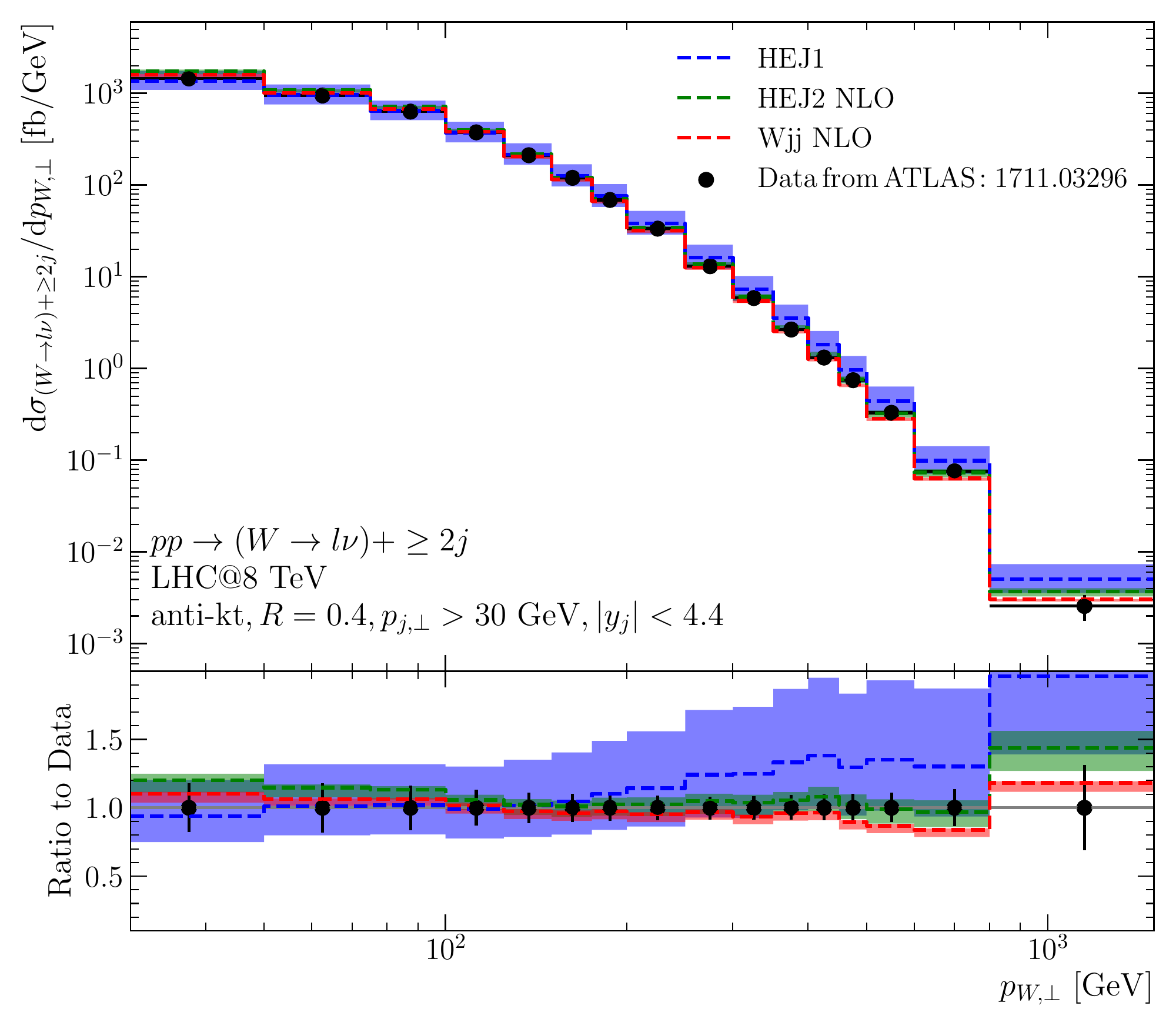}
    \caption{Differential cross section for the transverse momentum of the
      boson.}
  \label{fig:WjetsNLOmatched8TeVd}
  \end{subfigure}
  \caption{Predictions for $pp\to(W\to l\nu)+\ge2j$ for LHC@8TeV, compared to
    ATLAS data from~\cite{Aaboud:2017soa}. The original \HEJone predictions are shown in blue,
    the pure NLO predictions are shown in red and the new \HEJtwo predictions
    incorporating the methods of sections~\ref{sec:subleading} and
    \ref{sec:macthmerge} are shown in green.}
  \label{fig:WjetsNLOmatched8TeV}
\end{figure}
The comparison to the relevant results from the updated analysis for 8~TeV
collisions and similar cuts is presented in
\cref{fig:WjetsNLOmatched8TeV}. Of course very little changes in the
calculations, and the conclusions are similar to those arrived at for the
7~TeV analysis. The inclusive rates arrived at with \HEJtwo show an extremely
good agreement with data across all the matched multiplicities 2--6
inclusive. The transverse momentum is described extremely well for data available
for the first and second hardest jet, and for the $W$-boson.

The comparison with data clearly demonstrates that the combined effect of
matching to $W+$dijets at NLO and the inclusion of sub-leading channels in the
all-order treatment has repaired the short-comings of the previous approach
based solely on Born level matching and leading-logarithmic resummation. The
scale variation is reduced in line with the NLO input, and the distributions
associated with sub-leading regions of phase space, such as large transverse
momenta, are now described well.

%%% Local Variables:
%%% mode: latex
%%% TeX-master: "main"
%%% End:

\FloatBarrier
\section{Conclusions}
\label{sec:conclusions}

% \begin{itemize}
% \item Recap and summary
% \item Universality of QCD radiation (i.e.~broader applicability of results)
% \end{itemize}

% The experimental studies of dijets and W+dijets therefore also indicate that
% \emph{High Energy Jets} should be relevant for a successful description of
% the gluon-fusion production of a Higgs boson in association with dijets, in
% particular in the region of interest for the study of $CP$-properties, and
% for understanding how to use the radiation pattern to successfully suppress
% the gluon-fusion contribution to HJJ when studying Weak-Boson Fusion.
% \hrule

The High Energy Jets (\HEJ) framework provides a description of collider events
with at least two jets, which combines fixed-order accuracy with
leading-logarithmic accuracy in $\hat s/p_t^2$.  These logs dominate in the MRK
limit and have been shown to be important in describing data at large invariant
mass or large rapidity separation of jets.  However, the description
consistently struggled to give a good description of data in other regions of
phase space which are not directly related to the high energy limit, regions
with large transverse momentum for example.  In this
paper, we have directly addressed this by focussing on $pp\to (W\to \ell\nu)+\ge
2j$ and increasing the
perturbative accuracy of the \HEJtwo predictions in a two-pronged approach:
\begin{enumerate}
\item In section~\ref{sec:subleading}, we calculated the components needed to
  include logarithmic corrections to all of the partonic channels contributing
  to the first sub-leading corrections at high energies for the process
  $pp\to W+\ge2j$.  While this forms a well-defined contribution at NLL to the inclusive $W+2j$ cross
  section, it is only a part of the full NLL correction.  We showed in
  \cref{sec:impact-including-nll} that the included processes led to a significant increase in
  the fraction of the cross section which is supplemented with all-order
  corrections, for example up to at least 75\% across all values of
  $p_{j,\perp}$ for the leading jet in $pp\to (W\to\ell\nu)+\ge3j$ compared to
  35\% without the new corrections (fig.~\ref{fig:2pt2}). Most impressively,
  the change in the perturbative treatment of such a large part of the cross
  section leads to only very modest changes of a few percent. This behaviour is a
  testament to the stability introduced to the perturbative series by including
  the logarithmic corrections.
\item In section~\ref{sec:macthmerge}, we presented a new method of fixed-order
  matching which leads to next-to-leading order accuracy for the
  measured distributions, while keeping the same logarithmic accuracy.  The matching
  factors shown in
  \cref{fig:NLOmatchingcorrections7,fig:NLOmatchingcorrections8} can be seen,
  for example, to slightly suppress the
  all-order predictions for values of transverse momentum above $\sim100$~GeV.
\end{enumerate}
The new predictions from \HEJtwo, combining the two new components above, are
compared to LHC data in \cref{sec:results}. We have shown in
\cref{fig:WjetsNLOmatched7TeV,fig:WjetsNLOmatched8TeV} that the new predictions deliver an excellent
description of all the measured quantities in sub-leading regions of phase
space.  They dramatically improve upon the original predictions from \HEJone in
all distributions.  Furthermore, in many cases they also provide a better
prediction than that obtained from pure NLO, showing that \HEJtwo combines the
best of both fixed-order and leading-logarithmic descriptions.

The QCD radiation pattern arising from colour-octet exchange is known to be very
similar across QCD inclusive dijet production, $W+\ge2j$, $Z+\ge2j$ and
$H+\ge2j$.  Hence the perturbative approach developed in this paper offers an
exciting new tool to describe a large range of key processes at the LHC.

\section*{Acknowledgements}

We are grateful to our collaborators within \HEJ for useful discussions
throughout this project.  We are pleased to acknowledge funding from the UK
Science and Technology Facilities Council, the Royal Society, the ERC
Starting Grant 715049 ``QCDforfuture'' and the Marie
Sk{\l}odowska-Curie Innovative Training Network MCnetITN3 (grant
agreement no.~722104).

%%% Local Variables:
%%% mode: latex
%%% TeX-master: "main"
%%% End:

\newpage
\appendix

\section{Phase Space Slices}
\label{sec:phase-space-slices}

Here we give the parameters of the phase space slices used in the explorer plots
in \cref{sec:subleading}.  We begin with \cref{fig:3jexplore_uno}, where the
outgoing momenta are given by:
\begin{align}
  \label{eq:3jmomWZ}
  \begin{split}
        p_i&=(k_{\perp i} \cosh(y_i), k_{\perp i} \cos(\phi_i), k_{\perp i} \sin(\phi_i) ,
    k_{\perp i} \sinh(y_i) ),\\
    k_{\perp 1}&=k_{\perp 2}=k_{\perp \bar \ell}=40~\mathrm{GeV},\ k_{\perp
      \ell}=\frac{m_V^2}{2k_{\perp \bar \ell}(\cosh(y_{\bar \ell}-y_\ell)-\cos(\phi_{\bar
        \ell}-\phi_\ell))},\\
    \phi_1&=2\pi/3,\ \phi_2=0,\ \phi_{\bar \ell}=\pi/2,\ \phi_\ell=-\pi/2,\\
    y_1&=\Delta,\ y_2=0,\ y_3=-\Delta,\ y_{\bar \ell}=\Delta,\ y_\ell=\Delta,\\
    p_{3\perp}&=-p_{1\perp}-p_{2\perp}-p_{\bar \ell \perp}-p_{\ell \perp}.
  \end{split}
\end{align}
In \cref{fig:4jexplore_qqbar}, the phase space slice used is
\begin{align}
  \label{eq:4jmomWZ}
  \begin{split}
        p_i&=(k_{\perp i} \cosh(y_i), k_{\perp i} \cos(\phi_i), k_{\perp i} \sin(\phi_i) ,
    k_{\perp i} \sinh(y_i) ),\\
    k_{\perp 1}&=k_{\perp 2}=k_{\perp 3}=k_{\perp \bar \ell}=40~\mathrm{GeV},\ k_{\perp
      \ell}=\frac{m_V^2}{2k_{\perp \bar \ell}(\cosh(y_{\bar \ell}-y_\ell)-\cos(\phi_{\bar \ell}-\phi_\ell))},\\
    \phi_1&=\pi,\ \phi_2=\pi/2,\ \phi_3=-\pi/3,\ \phi_{\bar \ell}=\pi/4,\ \phi_\ell=-\pi/4,\\
    y_1&=\Delta,\ y_2= \Delta/3,\ y_3= -\Delta/3,\ y_4=-\Delta,\ y_{\bar \ell}=-\Delta/3,\
    y_\ell=-\Delta/3,\\ p_{4\perp}&=-p_{1\perp}-p_{2\perp}-p_{3\perp}-p_{\bar
      \ell\perp}-p_{\ell \perp}.
  \end{split}
\end{align}
These are provided in order to allow the results to be reproduced, but we stress
that the level of agreement does not depend on the specific choices of these points.

%%% Local Variables:
%%% mode: latex
%%% TeX-master: ../main
%%% End:

\section{Derivation of New NLL Components in $W+3$-jet Processes}
\subsection{Emission of an Unordered Gluon and a $W$ Boson from the Same Quark Line}
\label{sec:WunoDeriv}

Here we derive the current, $j_{W{\rm
    uno}\,\mu}^d(p_a,p_1,p_2,p_\ell,p_{\bar\ell})$, given in
\cref{eq:wunocurrent} in the main text.  We do this by considering the process
given in \cref{eq:Wunodef}, defined such that
the $W$ boson may only be emitted from the $p_a$--$p_2$ quark line.
We assume that the rapidities of the final-state partons obey $y_1 \sim y_2 \ll y_3$; we make no assumption for the $W$ boson or
its decay products.  We are aiming to write the amplitude in the form given in \cref{eq:SabsWuno}.

\begin{figure}[btp]
  \centering
  \begin{subfigure}{0.45\textwidth}
  \centering
  \includegraphics{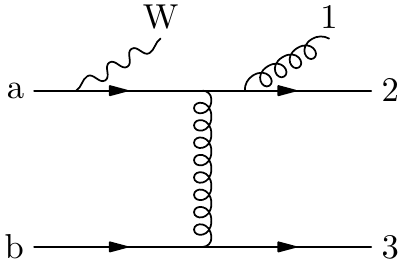}
  \caption{}
  \label{fig:U1diags}
\end{subfigure}
\begin{subfigure}{0.45\textwidth}
  \centering
\includegraphics{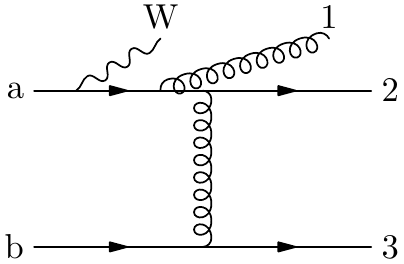}
\caption{}
\label{fig:U2diags}
\end{subfigure}
\begin{subfigure}{0.45\textwidth}
  \centering
  \includegraphics{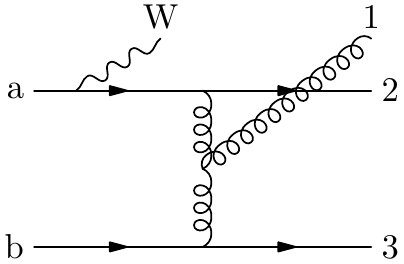}
\caption{}
\label{fig:Cdiags}
\end{subfigure}
\begin{subfigure}{0.45\textwidth}
\centering
\includegraphics{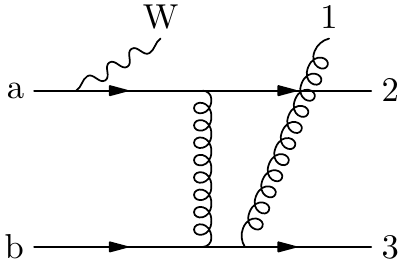}
\caption{}
 \label{fig:Ddiags}
 \end{subfigure}
  \vspace{0.4cm}
  \caption{Examples of each of the four categories of Feynman diagram which
    contribute at leading-order:  (a) where the gluon and $W$
    boson are emitted from the same quark line and the gluon comes after the
    $t$-channel propagator, (b) where the gluon and $W$ boson are emitted from
    the same quark line and the gluon comes before the $t$-channel proagator,
    (c) the gluon is emitted from the $t$-channel gluon and in (d) the gluon
    is emitted from the $b$--$3$ quark line.}
  \label{fig:Wunodiags}
\end{figure}

There are 12 Feynman diagrams in total at leading-order.  We split these by
colour structure into
four categories and give one example of each in \cref{fig:Wunodiags}.
We begin with diagrams where the $W$ boson
and unordered gluon are emitted from the same quark line, with the gluon emitted
after the $t$-channel as in the example in \cref{fig:U1diags}. The sum
of the 3 diagrams of this kind gives (using $u_i$ as shorthand for
$u^{h_i}(p_i)$, $s_{ij}=(p_i+p_j)^2$, $s_{ijk}=(p_i+p_j+p_k)^2$ and $t_{ij}=(p_i-p_j)^2$)
\begin{align}
  \label{eq:Asum}
  A= -g_s^3 T^d_{3b}\frac{ \left[
  \bar{u}_3\gamma_\mu u_b \right]}{t_{b3}} \cdot \varepsilon_\nu(p_1)\ \left[
  \bar{u}_\ell \gamma_\rho v_{\bar{\ell}} \right]
  \cdot  \left(iT^1_{2i} T^d_{ia} \tilde U_1^{\nu\mu\rho}\right),
\end{align}
where
\begin{align}
  \label{eq:U1tensor}
  \begin{split}
    \tilde U_1^{\nu\mu\rho} =& K_W \bigg( \frac{[\bar{u}_2 \gamma^\nu (\slashed{p}_2+
      \slashed{p}_1)\gamma^\mu (\slashed{p}_a - \slashed{p}_W)\gamma^\rho u_a]
    }{s_{12}t_{aW}}  + \frac{[\bar{u}_2\gamma^\nu (\slashed{p}_2+
      \slashed{p}_1)\gamma^\rho (\slashed{p}_2+\slashed{p}_1 +
      \slashed{p}_W)\gamma^\mu u_a ]}{s_{12}s_{12W}}   \\ 
    &+ \frac{[\bar{u}_2 \gamma^\rho (\slashed{p}_2+ \slashed{p}_W) \gamma^\nu
      (\slashed{p}_1 + \slashed{p}_2+\slashed{p}_W) \gamma^\mu u_a]}{s_{2W}s_{12W}} \bigg),
  \end{split}
\end{align}
and we have introduced the factor
\begin{align}
	\label{eq:KW}
	K_W= \left(\frac{ig_W}{\sqrt{2}}\right)^2\frac{1}{\left(p_W^2-m_W^2+im_W \Gamma_W\right)}.
\end{align}

These diagrams already have the colour and kinematic structure of
\cref{eq:SabsWuno} and therefore no further approximation is applied.

The next category of Feynman diagrams is where the $W$ and gluon are emitted
from the same quark line, but now the gluon is emitted before the $t$-channel
gluon.  An example is shown in \cref{fig:U2diags}.  This is very similar to the
first kind and the sum of the
three diagrams in this category is given by
\begin{align}
  \label{eq:Bsum}
  B= -g_s^3 T^d_{3b} \frac{ [ \bar{u}_3 \gamma_\mu
  u_b] }{t_{b3}} \cdot \varepsilon_\nu(p_1)\ [\bar{u}_\ell \gamma_\rho v_{\bar{\ell}}]
  \cdot \left(iT^d_{2i} T^1_{ia} \tilde U_2^{\nu\mu\rho}\right),
\end{align}
where
\begin{align}
  \label{eq:U2tensor}
  \begin{split}
    \tilde U_2^{\nu\mu\rho} =& K_W \bigg( \frac{[\bar{u}_2 \gamma^\mu
      (\slashed{p}_a-\slashed{p}_W-\slashed{p}_1)\gamma^\nu (\slashed{p}_a -
      \slashed{p}_W) \gamma^\rho u_a] }{t_{aW1}t_{aW}}  + \frac{[\bar{u}_2
      \gamma^\mu (\slashed{p}_a-\slashed{p}_W- \slashed{p}_1) \gamma^\rho
      (\slashed{p}_a-\slashed{p}_1) \gamma^\nu u_a] }{t_{a1W}t_{a1}}  \\
    &+ \frac{[\bar{u}_2 \gamma^\rho (\slashed{p}_2+ \slashed{p}_W) \gamma^\mu
      (\slashed{p}_a-\slashed{p}_1) \gamma^\nu u_a]}{s_{2W}t_{a1}} \bigg).
  \end{split}
\end{align}
Here we have used the further shorthand notation $t_{ijk}=(t_i-t_j-t_k)^2$.
These diagrams also already have the structure shown in \cref{eq:SabsWuno}
and hence no further approximation is made.

The next category of Feynman diagrams we will consider are the two where the
outgoing gluon is emitted from the $t$-channel gluon propagator,
e.g. \cref{fig:Cdiags}.  The sum of the two gives
\begin{align}\label{eq:sum3}
  \begin{split}
 C =&\  g_s^3 K_W T^d_{3b}\frac{ [\bar{u}_3
   \gamma_\mu u_b]}{t_{b3}} \cdot \varepsilon_\nu(p_1)\ [\bar{u}_\ell
   \gamma_\rho v_{\bar\ell} ] \cdot f^{e d 1  }T^e_{2a} \\
 &\times \frac{1}{t_{aW2}}\left(\frac{[\bar{u}_2 \gamma^\sigma (\slashed{p}_a - \slashed{p}_W) \gamma^\rho
     u_a]}{t_{aW}} + \frac{[\bar{u}_2 \gamma^\rho (\slashed{p}_2 + \slashed{p}_W)
     \gamma^\sigma u_a] }{s_{2W}} \right) \\
 &\times \left( g^{\sigma \mu} (q_1 +q_2)^\nu + g^{\mu \nu}(-q_2 +p_1)^\sigma-
   g^{\nu \sigma}(p_1 +q_1)^\mu \right),
 \end{split}
\end{align}
where $q_1$, $q_2$ are the $t$-channel momenta flowing into and out of the
3-gluon vertex: $q_1 = p_a - p_2 -p_W$ and $q_2 = q_1-p_1$.  These diagrams also
immediately satisfy the structure of \cref{eq:SabsWuno}.  There is some freedom
in how one writes the expression in the last term of \cref{eq:sum3} in $W+3j$,
because any term proportional to $q_2^\mu$ gives zero.  We will take
$-(p_1+q_1)^\mu=-2p_1^\mu-q_2^\mu$ and remove the zero-contribution from
$q_2^\mu$.  This is the unique choice which remains gauge invariant when we
generalise our current to processes with an arbitrary number of extra emissions.

Finally we consider the diagrams in which the gluon and $W$ are emitted from
different quark lines, as in the example in \cref{fig:Ddiags}.  The
full expression for that diagram is given by
%We shall neglect terms of form $\langle 2 | \mu \slashed{p}_g \nu | b \rangle $ which are suppressed in the MRK limit, and obtain:
\begin{align}
	\begin{split}
%  C_3 &= i (ig_s)^3\left(\frac{i g_W}{\sqrt{2}}\right)T^d_{1a}T^d_{2i}T^g_{ib}\varepsilon_{g\nu}\varepsilon_{W\rho}\frac{\langle 1 |\mu (\slashed{p}_a - \slashed{p}_W) \rho P_L | a\rangle \langle 2 |\mu | b\rangle 2p_b^{\nu} }{t_{aW}t_{bg}t_{b2g}} \\
% C_4 &= i (ig_s)^3\left(\frac{i g_W}{\sqrt{2}}\right)T^d_{1a}T^d_{2i}T^g_{ib}\varepsilon_{g\nu}\varepsilon_{W\rho}\frac{\langle 1 |\rho P_L (\slashed{p}_1 + \slashed{p}_W) \mu | a\rangle \langle 2 |\mu | b\rangle 2p_b^{\nu} }{s_{1W}t_{bg}t_{b2g}}\\
 D_{\rm (d)} &=\   -ig_s^3 K_W T^d_{2a}T^1_{3i}T^d_{ib}\
       \ \varepsilon_\nu(p_1)\ [\bar{u}_\ell \gamma_\rho v_{\bar \ell}] \\ &\ \times \left( \frac{[\bar{u}_2 \gamma^\mu (\slashed{p}_a
       - \slashed{p}_W) \gamma^\rho u_a] \ [\bar{u}_3 \gamma^\nu
  (\slashed{p}_1+\slashed{p}_3)\gamma_\mu u_b]}{t_{aW}s_{13}t_{b13}} \right).
% C_6 &= i (ig_s)^3\left(\frac{i g_W}{\sqrt{2}}\right)T^d_{1a}T^g_{2i}T^d_{ib}\varepsilon_{g\nu}\varepsilon_{W\rho}\frac{\langle 1 |\rho P_L (\slashed{p}_1 + \slashed{p}_W) \mu | a\rangle \langle 2 |\mu | b\rangle 2p_2^{\nu} }{s_{1W}s_{2g}t_{b2g}}
	\end{split}
 \end{align}
 This is our first example which does not satisfy the structure of
 \cref{eq:SabsWuno} and so we need to make some approximation. In the high
 energy limit, $p_b$ and $p_3$ are dominated by their forward lightcone
 components, while $p_a$, $p_1$ and $p_2$ are dominated by their backward
 lightcone components.  This automatically means that e.g.~$s_{a1} \ll s_{a3}$.
 By considering the contractions of the $\mu$-index, and of the $\nu$-index for
 a generic reference vector in $\varepsilon_\nu(p_1)$, in this limit the
 $\slashed{p}_1$-term is suppressed compared to the $\slashed{p}_3$-term in the
 second spinor term by ratios of invariants like $s_{a1}/s_{a3}$.  We therefore
 drop this term % (analogously to the approximation made in
 % \cref{eq:labelAs})
 and find
 \begin{align}\label{eq:Ddapprox}
 \begin{split}
   D_{\rm (d)} \approx &\   -ig_s^3 K_W T^d_{2a}T^1_{3i}T^d_{ib}\
       \varepsilon_\nu(p_1)\ [\bar{u}_\ell \gamma_\rho v_{\bar \ell}] \\ &\ \times \left( \frac{[\bar{u}_2 \gamma^\mu (\slashed{p}_a
       - \slashed{p}_W) \gamma^\rho u_a]\ [\bar{u}_3 \gamma_\mu u_b]\ 2p_3^\nu}{t_{aW}s_{13}t_{b13}} \right).
   \end{split}
 \end{align}
 We perform the same approximation for the other 3 diagrams in this category, to
 get the sum of the four to be
\begin{align} \label{eq:sum1}
  \begin{split}
    D \approx &\  -ig_s^3 K_W \varepsilon_\nu(p_1)\ [\bar{u}_\ell \gamma_\rho v_{\bar \ell}] \ \frac{
      [\bar{u}_3 \gamma_\mu u_b]}{t_{b13}} \left( T^d_{2a}
      T^d_{3i}T^1_{ib} \frac{2p_b^{\nu}}{t_{b1}}+ T^d_{2a}T^1_{3i}T^d_{ib}
      \frac{2p_3^{\nu}}{s_{13}} \right)   \\
    &\ \times \left(\frac{[\bar{u}_2 \gamma^\mu (\slashed{p}_a - \slashed{p}_W) \gamma^\rho
        u_a]}{t_{aW}} + \frac{[\bar{u}_2 \gamma^\rho (\slashed{p}_2 + \slashed{p}_W)
        \gamma^\mu u_a] }{s_{2W}} \right).
  \end{split}
\end{align}
In the QMRK, $p_3$ is dominated by its positive lightcone component, taken to be
roughly equal to $p_b$: $p_3 \sim p_b = p_+$.  We use this approximation to combine the colour factor:
\begin{align}\label{eq:sum2}
  \begin{split}
    D \approx &\  -ig_s^3 K_W \varepsilon_\nu(p_1)\ [\bar{u}_\ell \gamma_\rho v_{\bar \ell}] \ \frac{
      [\bar{u}_3 \gamma_\mu u_b] }{t_{b13}} \left( i T^d_{2a} f^{1de} T^e_{3b}\ \frac{2p_+^\nu}{s_{1+}}
       \right)   \\
    &\ \times \left(\frac{[\bar{u}_2 \gamma^\mu (\slashed{p}_a - \slashed{p}_W) \gamma^\rho
        u_a]}{t_{aW}} + \frac{[\bar{u}_2 \gamma^\rho (\slashed{p}_2 + \slashed{p}_W)
        \gamma^\mu u_a] }{s_{2W}} \right).
  \end{split}
\end{align}
We can now choose to symmetrise the $p_+$-term to reinstate the original
symmetry of the process.  Our final expression for the sum of these diagrams is therefore
\begin{align}
  \label{eq:symmD}
    \begin{split}
    D \approx &\  g_s^3 K_W \varepsilon_\nu(p_1)\ [\bar{u}_\ell \gamma_\rho v_{\bar \ell}] \ \frac{
      [\bar{u}_3 \gamma_\mu u_b] }{t_{b13}}\cdot   T^e_{2a} f^{1ed}
    T^d_{3b}\cdot \frac12\left( \frac{p_b^\nu}{p_b.p_1} +
      \frac{p_3^\nu}{p_1.p_3} \right)   \\
    &\ \times \left(\frac{[\bar{u}_2 \gamma^\mu (\slashed{p}_a - \slashed{p}_W) \gamma^\rho
        u_a]}{t_{aW}} + \frac{[\bar{u}_2 \gamma^\rho (\slashed{p}_2 + \slashed{p}_W)
        \gamma^\mu u_a] }{s_{2W}} \right).
  \end{split}
\end{align}
We notice that this has an identical factor to the third category of Feynman
diagram and we therefore write their sum as
\begin{align}
\label{eq:CplusD}
 C+D   \approx \ g_s^3 T^d_{3b}  \frac{ [\bar{u}_3 \gamma_\mu u_b] }{t_{b3}} \varepsilon_\nu(p_1)\ [\bar{u}_\ell \gamma_\rho v_{\bar \ell}] \left( f^{1de}T^e_{2a}\tilde L^{\nu\mu\rho} \right)
\end{align}
where
\begin{align}
\label{eq:Ltensor}
 \tilde L^{\nu\mu\rho} &=
 \frac{K_W}{t_{aW2}}\left(\frac{[\bar{u}_2 \gamma_\sigma (\slashed{p}_a - \slashed{p}_W)
  \gamma^\rho u_a]}{t_{aW}} + \frac{[\bar{u}_2 \gamma^\rho (\slashed{p}_2 +
  \slashed{p}_W) \gamma_\sigma u_a] }{s_{2W}} \right) \\ \nonumber
 &\vphantom{+\frac{1}{t_{aW2}}}\quad\times \left( \left( \frac{p_b^{\nu}}{s_{1b}}
   + \frac{p_3^{\nu}}{s_{13}} \right) (q_1-p_1)^2g^{\sigma\mu} +
g^{\sigma \mu} (2q_1 -p_1)^\nu +
   g^{\mu \nu}(2p_1-q_1)^\sigma -2 g^{\nu \sigma}p_1^\mu \right).
\end{align}
The presence of $p_b$ and $p_3$ in this equation at first looks like it spoils
the factorised structure we were looking for.  However, dependence on outer
partons mirrors our Lipatov vertex description of extra emissions: the
traditional form would contain projection operators for lightcone components,
where we choose to restore some of the underlying physics by using the average
of a contraction with $p_b$ and $p_3$ instead~\cite{Andersen:2009nu}.  The
choice will only affect sub-leading terms.  The critical point is that we do not
have dependence on any other emission, so the complexity does not increase with
the number of final state particles.

We can now deduce the current we need, $j^\mu_{\rm
  Wuno}(p_a,p_1,p_2,p_\ell,p_{\bar{\ell}})$, by comparing the sum of \cref{eq:Asum,eq:Bsum,eq:CplusD} to \cref{eq:SabsWuno}.  We find
\begin{align}\label{eq:total}
  \begin{split}
  j^\mu_{\rm Wuno}(p_a,p_1,p_2,p_\ell,p_{\bar{\ell}}) =&\ -i\varepsilon_\nu(p_1)\
  [\bar{u}_\ell \gamma_\rho v_{\bar \ell}] \\  &\ \times
\left(T^1_{2i} T^d_{ia} (\tilde U_1^{\nu\mu\rho}-\tilde L^{\nu\mu\rho}) + T^d_{2i} T^1_{ia} (\tilde U_2^{\nu\mu\rho}+\tilde L^{\nu\mu\rho}) \right),
  \end{split}
\end{align}
where the tensors $\tilde U_1^{\nu\mu\rho}$, $\tilde U_2^{\nu\mu\rho}$ and
$\tilde L^{\nu\mu\rho}$ are given in \cref{eq:U1tensor,eq:U2tensor,eq:Ltensor}.
Later in \cref{sec:calculation-new-nll}, in \cref{eq:BWuno},  these tensors are used in processes with
$n-3$ additional gluon emissions.  In this case,
$p_3$ should be mapped to $p_n$.

\subsection{Emission of an Unordered Gluon from a Quark Line}
\label{sec:pure-uno}
Here we give $j^{d,h_b,h_2,h_3\, \mu}_{\rm uno}(p_b, p_2, p_3)$, originally derived
in~\cite{Andersen:2017kfc}, and used
in this paper in \cref{eq:SWoppuno} to describe an unordered gluon emission from an incoming-outgoing quark line in
processes where the $W$ boson is emitted from a different quark line:
\begin{align}
  \label{eq:unoderivapp}
  j^{d,h_b,h_2,h_3\, \mu}_{\rm uno}(p_b, p_2, p_3) = -i \varepsilon_\nu(p_3)
  \left(T^3_{2i}T^d_{ib}\left(U_1^{\mu \nu} - L^{\mu \nu}\right) +
  T^d_{2i}T^3_{ib}\left(U_2^{\mu \nu} + L^{\mu \nu}\right) \right),
\end{align}
with
\begin{align}
  \label{eq:unocomps}
  \begin{split}
    U_1^{\mu \nu} &= \frac{1}{s_{23}} \left([\bar{u}_2^{h_2} \gamma^\nu
      u^{h_2}_3] [\bar{u}_3^{h_b} \gamma^\mu u^{h_b}_{b}] + 2 p_2^\nu
      [\bar{u}_2^{h_b} \gamma^\mu u_b^{h_b}] \right), \\
    U_2^{\mu \nu} &= \frac{1}{t_{b3}} \left( 2p_b^\nu [\bar{u}^{h_2}_2
      \gamma^\mu u^{h_2}_{b}] - [\bar{u}_2^{h_2} \gamma^\mu u^{h_2}_{3}]
      [\bar{u}^{h_b}_3 \gamma^\nu u^{h_b}_{b}] \right), \\
    L^{\mu \nu} &= \frac{1}{t_{b2}} \left(-2p^\mu_3 [\bar{u}_2^{h_b} \gamma^\nu
    u^{h_b}_b] + 2 p_3^\rho [\bar{u}_2^{h_b} \gamma_\rho
    u^{h_b}_b]g^{\mu \nu} \vphantom{\frac{p_a^\nu}{t_{a3}}} \right. \\ %\vphantom matches bracket sizes
    & \qquad\quad \left. + (q_2 + q_3)^\nu [\bar{u}_2^{h_b} \gamma^\mu
    u^{h_b}_b] + q_2^2 [\bar{u}_2^{h_b} \gamma^\mu
    u^{h_b}_b] \left(\frac{p_1^\nu}{s_{13}} - \frac{p_a^\nu}{t_{a3}} \right) \right),
  \end{split}
\end{align}
where the derived $t$-channel momenta here are $q_2 = p_b - p_2 - p_3$ and $q_3 = p_b - p_2$.
The separate spinor strings in
$U_1$ and $U_2$ arise from splitting a longer spinor string using completeness
relations which means that the helicities do not always match the helicity of
the corresponding particle.  Note these expressions are zero unless $h_b=h_2$.

In order to generalise this expression
to events with $n-3$ additional gluon emissions, one should make the replacements
$p_2 \to p_{n-1}$ and $p_3\to p_n$, where the numbering runs over all outgoing
coloured particles in increasing order of rapidity.

\section{Derivation of New NLL Components in $W+4$-jet Processes}
\subsection{Emission of a Central $Q\bar{Q}$ Pair}
\label{sec:cenqqbar}
\begin{figure}[btp]
\centering
\begin{subfigure}{0.45\textwidth}
\centering
\includegraphics{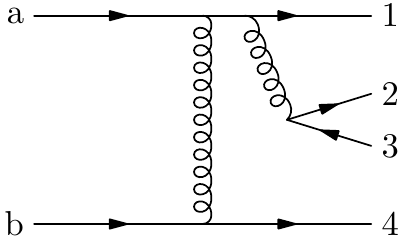}
\end{subfigure}
\begin{subfigure}{0.45\textwidth}
\centering
\includegraphics{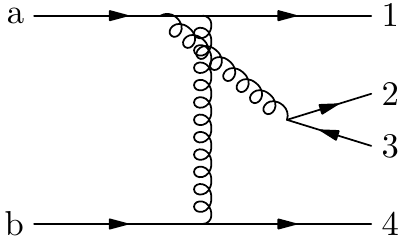}
\end{subfigure}
\begin{subfigure}{0.45\textwidth}
\centering
\includegraphics{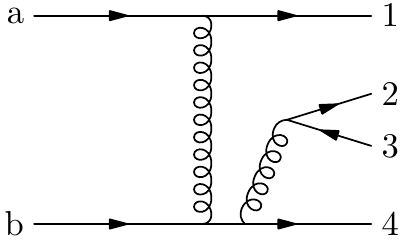}
\end{subfigure}
\begin{subfigure}{0.45\textwidth}
\centering
\includegraphics{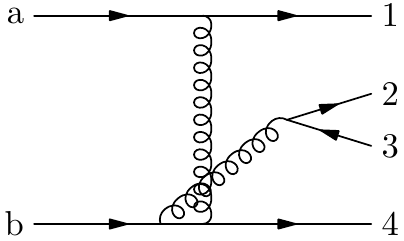}
\end{subfigure}
\begin{subfigure}{0.45\textwidth}
\centering
\includegraphics{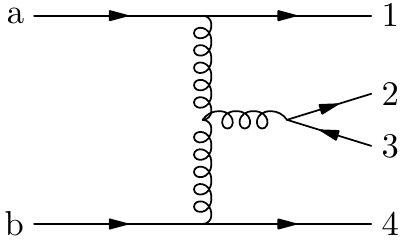}
\end{subfigure}
\begin{subfigure}{0.45\textwidth}
\centering
\includegraphics{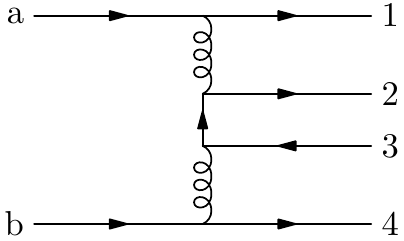}
\end{subfigure}
\begin{subfigure}{0.45\textwidth}
\centering
\includegraphics{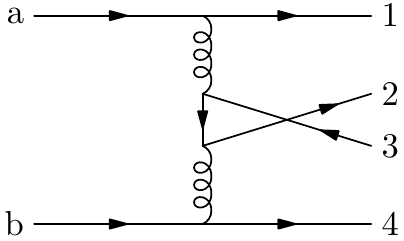}
\end{subfigure}
\caption{All Feynman diagrams which contribute to $q\tilde{q} \to qQ\bar{Q} \tilde{q}$ at
  leading order.}
\label{fig:qq_qQQq_graphs}
\end{figure}

%
%
% We keep the usual conventions for
% $t$-channel momenta and therefore define $q_1 = p_a - p_1 $ and
% $q_3 = p_a - p_1 - p_2 - p_3= p_4 - p_b $.
In this section, we outline the derivation of the tensor $X_{\rm cen}^{de\,
  \mu\nu}$, which is introduced in \cref{eq:Mcentral} in the main text to describe processes with a $Q\bar{Q}$
emission between the most forward and backward partons,
e.g.~\cref{4jetqqbar}(a).  We derive the necessary component by considering
$q\tilde{q}\to qQ\bar{Q} \tilde{q}$, \cref{eq:cenproc}.
There are seven
Feynman diagrams at leading-order, shown in \cref{fig:qq_qQQq_graphs}.  We
define $t_i=q_i^2$ where $q_i$ are defined as in \cref{sec:subleading}, which in
this context is $q_1=p_a-p_1$ and $q_3=p_4-p_b$.  We choose not to use $q_2$
here as it would depend on the rapidity ordering of $Q$ and $\bar{Q}$ which we do
not specify.

The first diagram is given by
\begin{align}
iA_1 = \frac{- i g_s^4 T^e_{1i}T^g_{ia}T^e_{23}T^g_{4b}}{s_{23} t_3} \left[
  \bar{u}_1 \gamma^\mu \frac{(\slashed{p}_1+\slashed{p}_2 + \slashed{p}_3)}{(p_1
  + p_2 + p_3)^2} \gamma^\rho u_a \right] \left[\bar{u}_2 \gamma_\mu v_3 \right]
  \left[ \bar{u}_4 \gamma_\rho u_b \right].
\label{eq:A1cenqqbar}
\end{align}
As in \cref{sec:WunoDeriv}, we will expand the square bracket using
completeness relations:
\begin{align}
\frac{1}{s_{12} + s_{13} + s_{23}} \left(2p_1^\mu \left[ \bar{u}_1 \gamma^\rho
  u_a\right] + \left[ \bar{u}_{1}\gamma^\mu u_2\right] \left[
  \bar{u}_2\gamma^{\rho}u_{a}\right] + \left[ \bar{u}_{1}\gamma^{\mu} u_{3}
  \right] \left[\bar{u}_{3} \gamma^{\rho} u_{a}\right] \right).
\label{eq:complet}
\end{align}
In this equation the spinors relating to $p_2$ and $p_3$ come from $\slashed{p}_i=\sum_h
{u}^h(p_i)\bar{u}^h(p_i)$ and hence the helicity of these spinors is determined
by the helicity of quarks $a$ and $1$, and not by the helicity of $2$ or $3$.
This relation also gives e.g.~$u_3$ instead of $v_3$.  Considering the contraction of \cref{eq:complet} with the other two spinor strings in
\cref{eq:A1cenqqbar} for all possible helicity choices of external particles shows the first term is
always dominant because both $s_{13}/s_{23}$ and $s_{14}/s_{24}$ tend to
infinity in the QMRK limit, \cref{eq:cenqqbarraporder}. We therefore approximate
this diagram as:
\begin{align}
  iA_1 \approx  \frac{- i g_s^4 T^e_{1q}T^g_{qa}T^e_{23}T^g_{4b}}{s_{23} t_3
  (s_{12} + s_{13} + s_{23})} \left[\bar{u}_1 \gamma^\rho u_a\right] \left[ \bar{u}_4 \gamma_\rho
      u_b\right] \times 2p_1^\mu \left[\bar{u}_2 \gamma_\mu v_3\right].
\end{align}
A similar analysis of the remaining three diagrams in the first two lines of
\cref{fig:qq_qQQq_graphs} gives
\begin{align}
  \begin{split}
    iA_2 &\approx \frac{ i g_s^4 T^g_{1q}T^e_{qa}T^e_{23}T^g_{4b}}{s_{23} t_3
      (s_{a2}+s_{a3} - s_{23})} \left[\bar{u}_1 \gamma^\rho u_a\right] \left[ \bar{u}_4 \gamma_\rho
      u_b\right]\times 2p_a^\mu \left[\bar{u}_2 \gamma_\mu v_3\right],\\
    iA_3 &\approx \frac{-i g_s^4 T^g_{1a} T^e_{4q}T^g_{qb}T^e_{23}}{t_1
      s_{23}(s_{24}+s_{34} +s_{23})}\left[\bar{u}_1 \gamma^\rho u_a\right] \left[ \bar{u}_4 \gamma_\rho
      u_b\right] \times 2     p_4^\mu \left[\bar{u}_2 \gamma_\mu v_3\right], \\
    iA_4 &\approx \frac{i g_s^4 T^g_{1a} T^g_{4q} T^e_{qb} T^e_{23}}{t_1
      s_{23}(s_{2b}+s_{3b} - s_{23})} \left[\bar{u}_1 \gamma^\rho u_a\right] \left[ \bar{u}_4 \gamma_\rho
      u_b\right] \times 2  p_b^\mu \left[\bar{u}_2 \gamma_\mu v_3\right].
  \end{split}
\end{align}
These diagrams have the desired kinematic form, but not yet the correct colour
factors.  In the QMRK limit, we can neglect the $s_{23}$ factor in each
of the denominators as this remains finite while each of
the other $s_{2i}$ and $s_{3i}$ factors in the brackets are large.  We may also
take $p_1$ ($p_4$) to be dominated by its negative
(positive) lightcone component which is taken roughly equal to that of the
incoming parton.  We may therefore take $p_1 \simeq p_a=p_-$ and $p_4 \simeq p_b=p_+$
which leads to
\begin{align}
  \label{eq:final1to4}
  \begin{split}
    i(A_1 + A_2) &\approx \frac{g_s^4\ f^{egc}T^c_{1a}T^e_{23}T^g_{4b}}{s_{23} t_3 (s_{-2} + s_{-3})}
    \left[\bar{u}_1 \gamma^\rho u_a\right] \left[ \bar{u}_4 \gamma_\rho
      u_b\right] \times 2 p_-^\sigma \left[ \bar{u}_2 \gamma_\sigma u_3 \right], \\
    i(A_3 + A_4) &\approx \frac{-g_s^4\ f^{egc}T^c_{1a}T^e_{23}T^g_{4b}}{s_{23}t_1 (s_{+2}+s_{+3})}
    \left[ \bar{u}_1 \gamma^\rho u_a \right] \left[ \bar{u}_4 \gamma_\rho u_b
    \right] \times 2 p_+^\sigma \left[ \bar{u}_2 \gamma_\sigma u_3 \right].
\end{split}
\end{align}
These four diagrams now all yield something proportional to the colour factor of
the fifth diagram in \cref{fig:qq_qQQq_graphs}.  We now choose to reinstate
the dependence on the external partons, analogously to
\cref{eq:Ltensor}. As in \cref{sec:WunoDeriv}, this dependence on external partons does not
damage the factorisation property as it only appears as an alternative to a
projection operator, and does not grow in complexity with additional emissions.
We therefore define
\begin{align}
  X_{sym}^\sigma = p_a^\sigma \left( \frac{t_1}{s_{a2}+s_{a3}} \right) + p_1^\sigma \left( \frac{t_1}{s_{12}+s_{13}} \right)
- p_b^\sigma \left( \frac{t_3}{s_{b2}+s_{b3}} \right) - p_4^\sigma \left( \frac{t_3}{s_{42}+s_{43}} \right),
\end{align}
such that
\begin{align}
i \sum_{i=1}^4 A_i \approx \frac{g_s^4\ f^{egc}T^c_{1a}T^e_{23}T^g_{4b}}{s_{23}
  t_1 t_3}
    \left[\bar{u}_1 \gamma^\rho u_a\right] \left[ \bar{u}_4 \gamma_\rho
      u_b\right] \left[ \bar{u}_2 \gamma_\sigma u_3 \right] X_{sym}^\sigma.
\label{eq:A1to4cenqqbar}
\end{align}

By inspection, one can see that the three remaining graphs will immediately have
the colour and kinematic structure given in \cref{eq:Mcentral} as they have
simple spinor strings between $p_1$ and $p_a$ and between $p_4$ and $p_b$.  We
therefore include them exactly:
\begin{align}
  \label{eq:final567}
  \begin{split}
    iA_5 &= \frac{g_s^4 f^{geg'}T^g_{1a} T^{g'}_{4b}T^e_{23}}{t_1 s_{23}
      t_3}\ \left[\bar{u}_1 \gamma_\nu u_a \right]  \left[\bar{u}_4 \gamma_\lambda u_b
    \right] \left[\bar{u}_2 \gamma_\sigma v_3 \right]\ X_{5}^{\nu\lambda\sigma}, \\
    iA_6 &= \frac{i g_s^4 T^g_{1a} T^{g}_{2q} T^{g'}_{q3}  T^{g'}_{4b}}{t_1
       t_3}\left[\bar{u}_1 \gamma^\mu u_a \right]
    \left[\bar{u}_4 \gamma^\sigma u_b \right] \ X_{6\;\mu \sigma}, \\
    iA_7 &= \frac{-i g_s^4 T^g_{1a} T^{g'}_{2q} T^{g}_{q3}  T^{g'}_{4b}}{t_1
       t_3}\left[\bar{u}_1 \gamma^\mu u_a \right]
    \left[\bar{u}_4 \gamma^\sigma u_b \right] \ X_{7\;\mu\sigma},
  \end{split}
\end{align}
where
\begin{align}
  \begin{split}
\label{eq:cencomponents}
    X_{5}^{\nu \lambda \sigma} &= (q_1 + p_2 + p_3)^\lambda \eta^{\nu \sigma} + (q_3 -
    p_2 -p_3)^\nu \eta^{\lambda \sigma} - (q_1 + q_3)^\sigma \eta^{\nu \lambda},
    \\
    X_{6\;\mu \sigma} &= \frac{\bar{u}_2 \gamma_\mu (\slashed{q}_1 -
      \slashed{p}_2)\gamma_\sigma v_3}{(q_1-p_2)^2}, \quad
    X_{7\;\mu\sigma } =\frac{\bar{u}_2 \gamma_\sigma (\slashed{q}_1 - \slashed{p}_3)\gamma_\mu v_3}{(q_1-p_3)^2}.
    \end{split}
\end{align}

The effective vertex $X^{de\, \mu \nu}_{\rm cen}$ is obtained by combining
\cref{eq:A1to4cenqqbar,eq:final567} to get:
\begin{align}
X^{de\,\mu \nu}_{\rm cen} =  i C_2 \left( X_s^{\mu\nu} + X^{\mu \nu}_{6} \right) -
  i C_3  \left( X_s^{\mu\nu} + X^{\mu \nu}_{7} \right),
\end{align}
where we have defined the following colour factors:
\begin{equation}
\label{eq:centralcolourfactors}
C_2 = T^d_{2q} T^{e}_{q3}
\quad {\rm and} \quad C_3 = T^{e}_{2q}T^d_{q3}.
\end{equation}
and introduced
\begin{align}
\begin{split}
\label{eq:Xsdefinition}
X_s^{\mu\nu} &= \frac{1}{s_{23}}\left(\eta^{\mu \nu} X_{sym}^\sigma + X^{\mu \nu \sigma}_{5}
\right) \left[ \bar{u}_{2} \gamma_{\sigma} v_{3}\right].
\end{split}
\end{align}
These are used later in \cref{sec:new-nll-components-4} in \cref{eq:Bcen} to construct the skeleton
function for processes of the form
\begin{align}
  \label{eq:qqbaratnapp}
  q(p_a) f_b(p_b) \to (W\to)\ell(p_\ell) \bar{\ell}(p_{\bar\ell})\  q'(p_1) g(p_2) \ldots g(p_{i-1}) Q(p_i) \bar{Q}(p_{i+1})
  g(p_{i+2}) \ldots g(p_{n-1}) f_b(p_n).
\end{align}
In this case, the following substitutions should be made within all the tensors:
$p_2\to p_i$, $p_3\to p_{i+1}$, $p_4\to p_n$, $q_1\to q_{i-1}$,
$q_3\to q_{i+1}$.

\FloatBarrier
%%% Local Variables:
%%% mode: latex
%%% TeX-master: "../main"
%%% End:

\subsection{Emission of a Central $Q\bar{Q}'$ Plus $W$ }
\label{sec:cenWqqbar}
In this section, we outline the derivation of the tensor $X_{\rm cenW}^{de\,
  \mu\nu}$, which is necessary for describing processes with a $Q\bar{Q}'W$
emission between the most forward and backward partons,
e.g.~\cref{4jetqqbar}(b) and \cref{eq:McentralW}.  We derive the necessary
component by considering $q \tilde{q} \to q Q\bar{Q}^\prime
(W\to)\ell \bar{\ell} \tilde{q}$, \cref{eq:processWcen}.
There are 16 diagrams which contribute at LO. These are the diagrams
of \cref{fig:qq_qQQq_graphs} with a $W$ boson at each possible point on the
2--3 quark line.  There is no added complexity in the colour
factors here compared to the pure QCD case.

%%%%%% J_V
We begin with the calculation of a building block
which will be present in most diagrams; namely, a gluon splitting into
a $Q\bar{Q}$ pair which subsequently emits a $W$ boson. This
building block takes into account both the emission from the quark and
the anti-quark line:
\begin{align}
  \label{eq:JV}
  \begin{split}
    J_V^\mu(p_2,p_\ell,p_{\bar{\ell}},p_3)=&K_W \
[\bar{u}_\ell \gamma_\nu u_{\bar{\ell}}] \\
& \quad \times \left( \frac{ [\bar{u}_2 \gamma^\nu (\slashed{p}_2 +
        \slashed{p}_l +
        \slashed{p}_{\bar{l}}) \gamma^\mu u_3]}{s_{2\ell\bar{\ell}}} -
       \frac{[\bar u_2
        \gamma^\mu(\slashed{p}_3 + \slashed{p}_l + \slashed{p}_{\bar{l}}) \gamma^\nu
        u_3]}{s_{3\ell\bar{\ell}}} \right) .
  \end{split}
\end{align}
We use $s_{ij...k}=(p_i+p_j+...+p_k)^2$ throughout this section and
$t_i=q_i^2$ where $q_i$ are defined as in \cref{sec:subleading}, which in
this context is $q_1=p_a-p_1$ and $q_3=p_4-p_b$.  We choose not to use $q_2$
here as it would depend on the rapidity ordering of $Q$ and $\bar{Q}'$ which we do
not specify.

%%%%%% A_1
We take the first subset of diagrams, corresponding to
the first in \cref{fig:qq_qQQq_graphs}. These give
\begin{align}
  \label{eq:leg1WEmit}
  \begin{split}
    iA_{1}\approx &\frac{-ig_s^4\, T_{1q}^eT^{g}_{qa}
      T_{23}^{e}T_{4b}^g}{s_{23\ell\bar{\ell}}\,t_3}
    \bigg[\bar{u}_1\gamma^\mu\frac{\big(\slashed{p}_a+\slashed{p}_b
      -\slashed{p}_4\big)}{(p_a+p_b-p_4)^2}\gamma^\rho u_a\bigg]
    J_{V\mu}(p_2,p_\ell,p_{\bar{\ell}},p_3)\, [\bar{u}_4\gamma_\rho u_b].
  \end{split}
\end{align}
This is not yet in the required form, \cref{eq:Mcentral}. We expand the
spinor-string $[\bar{u}_1 ~...~ u_a]$ with the use of completeness relations to find:
\begin{align}
  \begin{split}
     \left[\bar{u}_1\gamma^\mu (\slashed{p}_a+
     \slashed{p}_b-\slashed{p}_4)
     \gamma^\rho u_a\right] =&
     2p_1^\mu [\bar{u}_1\gamma^\rho u_a] +
       [\bar{u}_1\gamma^\mu u_2][\bar{u}_2\gamma^\rho u_a] +
       [\bar{u}_1\gamma^\mu u_\ell][\bar{u}_l \gamma^\rho u_a]  \\ & \quad +
       [\bar{u}_1\gamma^\mu u_{\bar{\ell}} ][u_{\bar{\ell}} \gamma^\rho u_a] +
       [\bar{u}_1\gamma^\mu u_3][\bar{u}_3 \gamma^\rho u_a].
  \end{split}
   \label{eq:u1uacurrent}
\end{align}
Using a similar argument to that after \cref{eq:complet}, we find after
contracting with the current $[\bar{u}_4 \gamma^\rho u_b]$, that the
first term is dominant in the QMRK.  We therefore approximate this contribution
as
\begin{align}
  \label{eq:leg1WApprox}
  \begin{split}
    iA_{1}\approx &\frac{-ig_s^4\, T_{1q}^eT^{g}_{qa}
      T_{23}^{e}T_{4b}^g}{s_{23l\bar{l}}
      \,t_3(s_{123l\bar{l}})}\ [\bar{u}_1\gamma^\rho u_a][\bar{u}_4 \gamma_\rho u_b]
    \times 2p^\mu_1 J_{V\mu}.
  \end{split}
\end{align}
One can follow a very similar process for
the next three subsets of diagrams, corresponding to adding a $W$ boson where
allowed to diagrams 2--4 in \cref{fig:qq_qQQq_graphs}:
\begin{align}
  \label{eq:legaWApprox}
  \begin{split}
    iA_{2}\approx&\frac{-ig_s^4\, T_{1q}^gT^{e}_{qa}
      T_{23}^{e}T_{4b}^g}{s_{23l\bar{l}}
      \,t_3(s_{a23l\bar{l}})}\ [\bar{u}_1\gamma^\rho u_a][\bar{u}_4 \gamma_\rho u_b]
    \times 2p^\mu_a J_{V\mu}, \\
    iA_{3}\approx&\frac{-ig_s^4\, T_{1a}^gT^{e}_{4q}
      T_{qb}^{g}T_{23}^e}{s_{23l\bar{l}}
      \,t_1(s_{234l\bar{l}})}\ [\bar{u}_1\gamma^\rho u_a][\bar{u}_4 \gamma_\rho u_b]
    \times 2p^\mu_4 J_{V\mu}, \\
    iA_{4}\approx &\frac{-ig_s^4\, T_{1a}^gT^{g}_{4q}
      T_{qb}^{e}T_{23}^e}{s_{23l\bar{l}}
      \,t_1(s_{b23l\bar{l}})}\ [\bar{u}_1\gamma^\rho u_a][\bar{u}_4 \gamma_\rho u_b]
    \times 2p^\mu_b J_{V\mu}.
  \end{split}
\end{align}
As in the previous case without a $W$, these do not yet have the correct colour
structure so further approximations are required.  As in \cref{eq:final1to4},
working in the QMRK limit allows us to approximate
\begin{align}
  \begin{split}
    i(A_1+A_2) & \approx \frac{g_s^4\, f^{egc}T^c_{1a}
      T_{23}^{e}T_{4b}^g}{s_{23l\bar{l}}
      \,t_3}\ [\bar{u}_1\gamma^\rho u_a][\bar{u}_4 \gamma_\rho u_b]
    J_{V\mu}\ \frac{2p^\mu_-}{s_{-2}+s_{-3}+s_{-\ell}+s_{-\bar\ell}}, \\
    i(A_3+A_4) & \approx \frac{g_s^4\, f^{egc} T_{1a}^gT^{c}_{4b}T_{23}^e}{s_{23l\bar{l}}
      \,t_1}\ [\bar{u}_1\gamma^\rho u_a][\bar{u}_4 \gamma_\rho u_b]
    J_{V\mu} \ \frac{2p^\mu_+}{s_{+2}+s_{+3}+s_{+\ell}+s_{+\bar\ell}}.
  \end{split}
\end{align}
In line with our treatment of the earlier processes, we reinstate the
dependence on $p_a$, $p_b$, $p_1$ and $p_4$ to write the sum of these four contributions as
\begin{align}
  \begin{split}
\label{eq:Xsymdefniition}
    i\sum_{i=1}^4 A_i &\approx \frac{g_s^4\, f^{ceg}T^c_{1a}
      T_{23}^{e}T_{4b}^g}{s_{23l\bar{l}}
      \,t_1t_3}\ [\bar{u}_1\gamma^\rho u_a][\bar{u}_4 \gamma_\rho u_b]
    J_{V\mu}\ \tilde{X}_{\rm sym}^{\mu}, \\
    \tilde{X}_{\rm sym}^{\mu} &= \left(
      \frac{p^\mu_a}{s_{a2}+s_{a3}+s_{a\ell}+s_{a\bar\ell}}t_1 +
      \frac{p^\mu_1}{s_{12}+s_{13}+s_{1\ell}+s_{1\bar\ell}}t_1 \right. \\
    & \qquad \quad \left. -
      \frac{p^\mu_b}{s_{b2}+s_{b3}+s_{b\ell}+s_{b\bar\ell}}t_3 -
      \frac{p^\mu_4}{s_{24}+s_{34}+s_{4\ell}+s_{4\bar\ell}}t_3\right).
  \end{split}
\end{align}

%%%%%%% A_5
The $t$-channel pole is already apparent in the next subset of diagrams in
\cref{fig:qq_qQQq_graphs}, so we include these contributions exactly:
\begin{align}
  \label{eq:3GluonWEmit}
  \begin{split}
  iA_{5}=&\ \frac{g_s^4\, f^{ceg}T_{1a}^cT_{23}^eT_{4b}^{g}}{
    t_1s_{23l\bar{l}}\,t_3} \left[\bar{u}_1\gamma_\rho
    u_a\right]\left[\bar{u}_4\gamma_\lambda
    u_b\right]J_{V\mu}(p_2,p_l,p_{\bar{l}},p_3) \\
  &\quad \times \bigg(\left(q_1+p_2+p_3+p_l+p_{\bar{l}}\right)^\lambda\eta^{\rho\mu}
    +\left(q_3-p_2-p_3-p_l-p_{\bar{l}}\right)^\rho\eta^{\mu\lambda}
      \\ & \qquad \quad -\left(q_1+q_3\right)^\mu\eta^{\rho\lambda}\bigg),
  \end{split}
\end{align}
where $q_1 = p_a-p_1$ and $q_3=p_4-p_b$.  From this
we define
\begin{align}
  \label{eq:X5W}
  \tilde X_5^{\mu\rho\lambda} = \left(q_1+p_2+p_3+p_l+p_{\bar{l}}\right)^\lambda\eta^{\rho\mu}
  +\left(q_3-p_2-p_3-p_l-p_{\bar{l}}\right)^\rho\eta^{\mu\lambda}
      -\left(q_1+q_3\right)^\mu\eta^{\rho\lambda}.
\end{align}

We have two final subsets of
diagrams corresponding to the last two diagrams in \cref{fig:qq_qQQq_graphs}.
Neither of these require any approximation and so are included exactly. The
kinematic parts of the contributions to $X_{\rm cenW}^{de\, \mu\nu}$ from these are
\begin{align}
  \begin{split}
    \label{eq:uncrossContr}
      \tilde X^{\mu\nu}_{6}=-K_W[\bar{u}_\ell \gamma_\sigma v_{\bar{\ell}}] \times \bar{u}_2&\left[-\frac{
  \gamma^\sigma(\slashed{p}_2+\slashed{p}_l+\slashed{p}_{\bar{l}})\gamma^\mu
  (\slashed{q}_3+ \slashed{p}_3)\gamma^\nu}{(p_2 + p_l +
  p_{\bar{l}})^2(q_3 +p_3)^2}\right.
  \\
    &\qquad\left. +\frac{\gamma^\mu(\slashed{q}_1-\slashed{p}_2)\gamma^\sigma(\slashed{q}_3+\slashed{p}_3)\gamma^\nu}{(q_1 - p_2)^2(q_3+p_3)^2}\right.
  \\
    &\qquad\qquad\left. +\frac{\gamma^\mu(\slashed{q}_1-\slashed{p}_2)\gamma^\nu(\slashed{p}_3+\slashed{p}_l+\slashed{p}_{\bar{l}})\gamma^\sigma
  }{(q_1 - p_2)^2(p_3 + p_l + p_{\bar{l}})^2}\right]v_3
  \end{split}
\end{align}
and
\begin{align}
  \begin{split}
    \label{eq:crossContr} \tilde X^{\mu\nu}_{7}=K_W[\bar{u}_\ell
      \gamma_\sigma v_{\bar{\ell}}] \times \bar{u}_2&\left[-\frac{
\gamma^\nu(\slashed{q}_3+\slashed{p}_2)\gamma^\mu
(\slashed{p}_3+\slashed{p}_l+\slashed{p}_{\bar{l}})\gamma^\sigma}{(q_3
+p_2)^2(p_3+p_l+p_{\bar{l}})^2}\right.
\\
&\qquad\left. +\frac{\gamma^\nu (\slashed{q}_3+\slashed{p}_2) \gamma^\sigma
    (\slashed{q}_1-\slashed{p}_3) \gamma^\mu}{(q_3 + p_2)^2(q_1 - p_3)^2}\right.
\\ &\qquad\qquad\left
. +\frac{\gamma^\sigma(\slashed{p}_2+\slashed{p}_l+\slashed{p}_{\bar{l}})\gamma^\nu(\slashed{q}_1-\slashed{p}_3)\gamma^\mu
}{(p_2+p_l+p_{\bar{l}})^2(q_1 - p_3)^2}\right]v_3.
  \end{split}
\end{align}

Putting the resuls of \cref{eq:Xsymdefniition,eq:X5W,eq:uncrossContr,eq:crossContr} together means that the
full result for the effective tensor representing central production of  $Q\bar{Q}'W$
within \HEJ, is given by:
\begin{align}
  \label{eq:EffectiveVertexWqqbar}
  \begin{split}
  X^{de\, \mu\nu}_{\rm cenW}&= iC_2 \left( \tilde X_s^{\mu\nu}  + \tilde
    X^{\mu\nu}_{6} \right) -iC_3\left(\tilde X_s^{\mu\nu}  + \tilde X^{\mu\nu}_{7}\right), \\
\tilde X_s^{\mu\nu} &= \frac{1}{s_{23l\bar{l}}}\left(\tilde{X}_{\rm
    sym}^{\sigma} \eta^{\mu\nu} + \tilde X_5^{\sigma\mu\nu}\right)J_{\text{V}
  \sigma}(p_2,p_l,p_{\bar{l}},p_3),
  \end{split}
\end{align}
where the colour factors $C_2$ and $C_3$ are defined as in \cref{eq:centralcolourfactors}. We have chosen to include the electroweak coupling factors in  $X^{de\, \mu\nu}_{\rm cenW}$. This lets us adapt results derived for a central $Q\bar{Q}$ emission to describe a central $Q\bar{Q}'W$ emission by the simple replacement of $X^{de\, \mu\nu}_{\rm cen}$ with $X^{de\, \mu\nu}_{\rm cenW}$. 

This tensor may also be used to describe a process with additional gluons
before and after the central $Q\bar{Q}'$ in rapidity, but within the extremal
partons.  In this case, if the $Q$ and $\bar{Q}'$ appear in positions $i$ and
$i+1$ respectively in the rapidity-ordered list of coloured particles, the following
substitutions should be made within all the tensors: $p_2\to p_i$,
$p_3\to p_{i+1}$, $p_4\to p_n$, $q_1\to q_{i-1}$, $q_3\to q_{i+1}$.

%%% Local Variables:
%%% mode: latex
%%% TeX-master: "../main"
%%% End:

\section{Further Examples of the Numerical Impact of NLL Components}
\label{sec:furth-exampl-numer}

\begin{figure}
  \centering
  \includegraphics[width=0.49\linewidth]{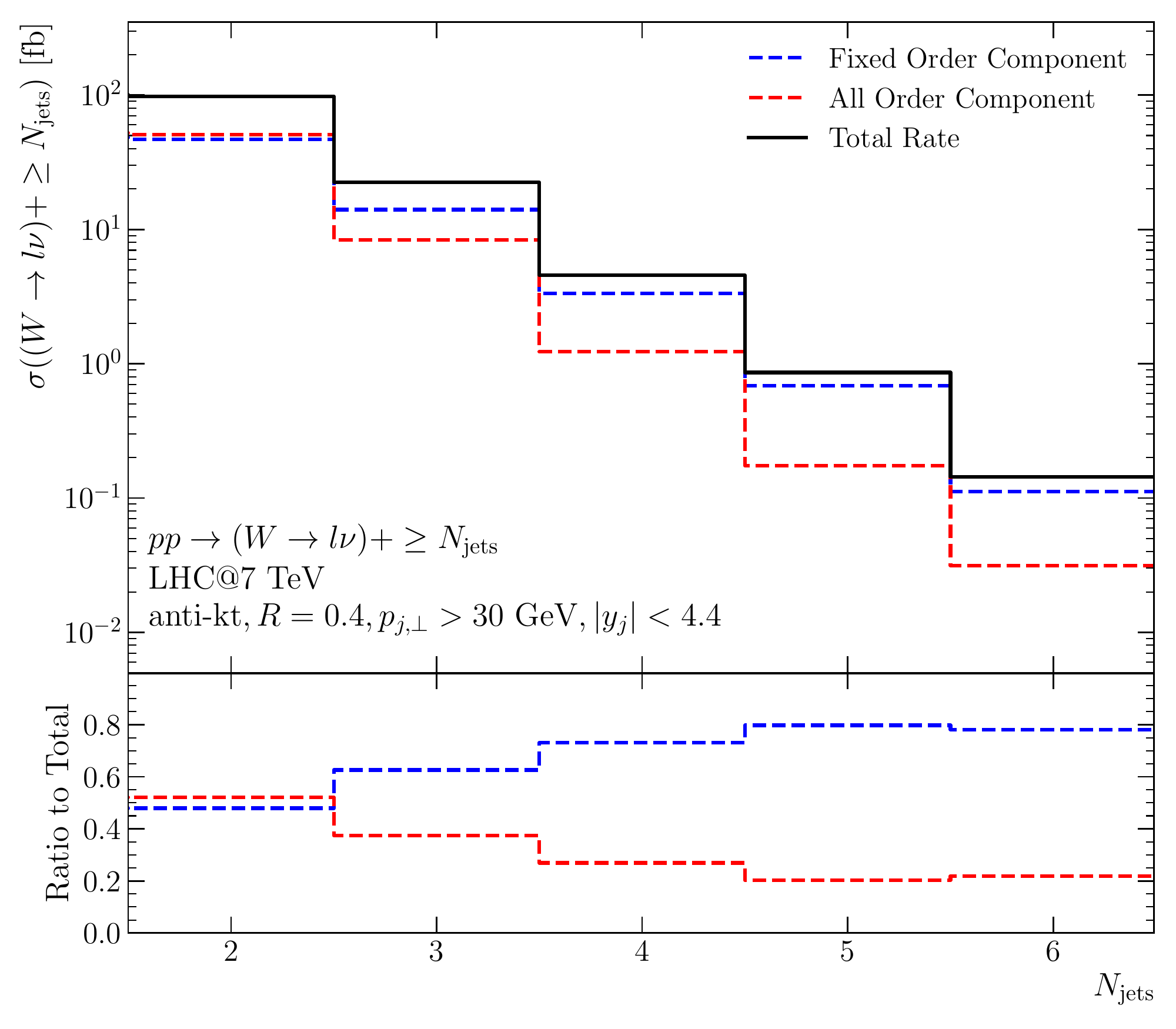}\hfill
  \includegraphics[width=0.49\linewidth]{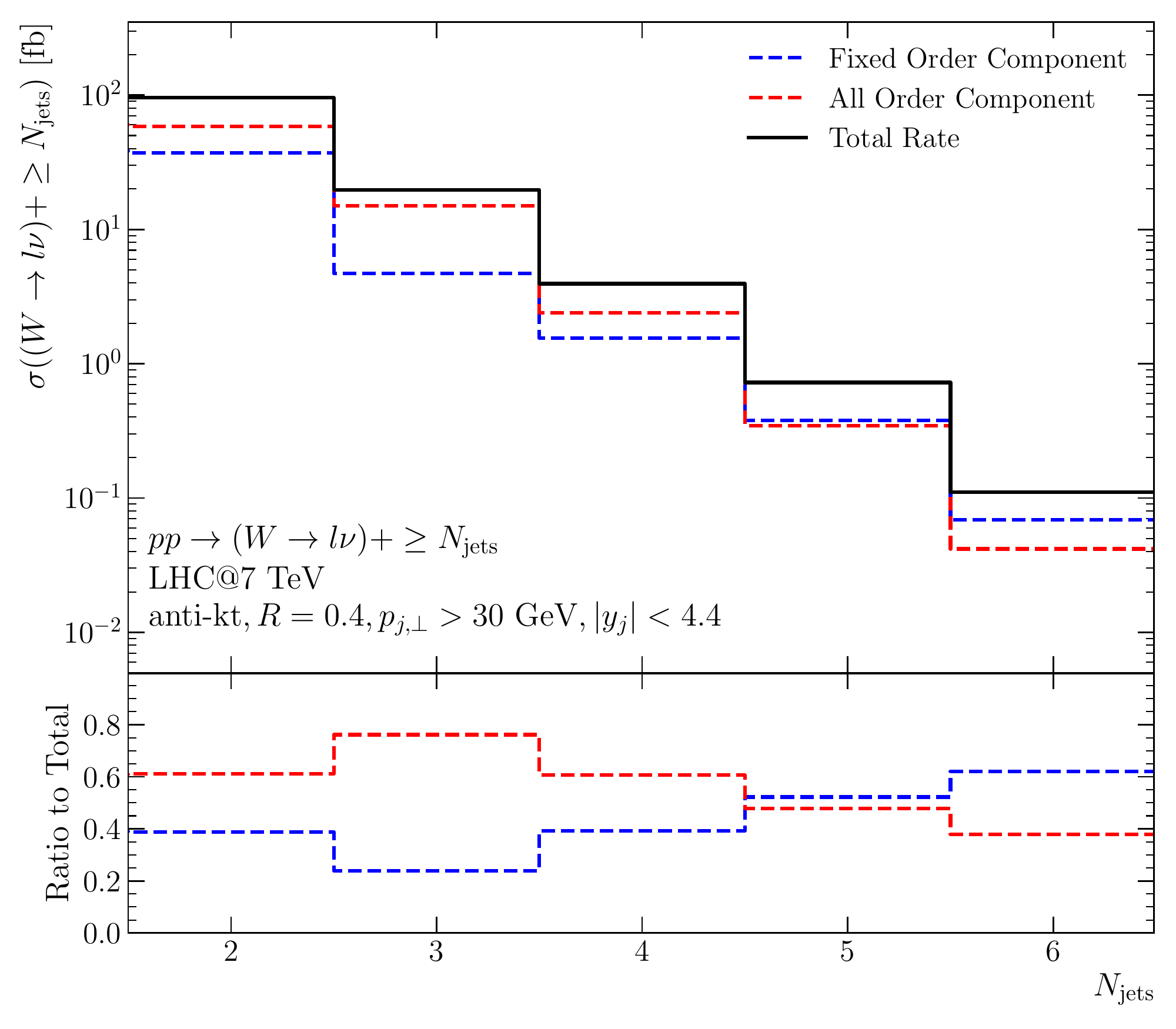}\\
  (a) \hspace{7cm} (b) \\
  \includegraphics[width=0.49\linewidth]{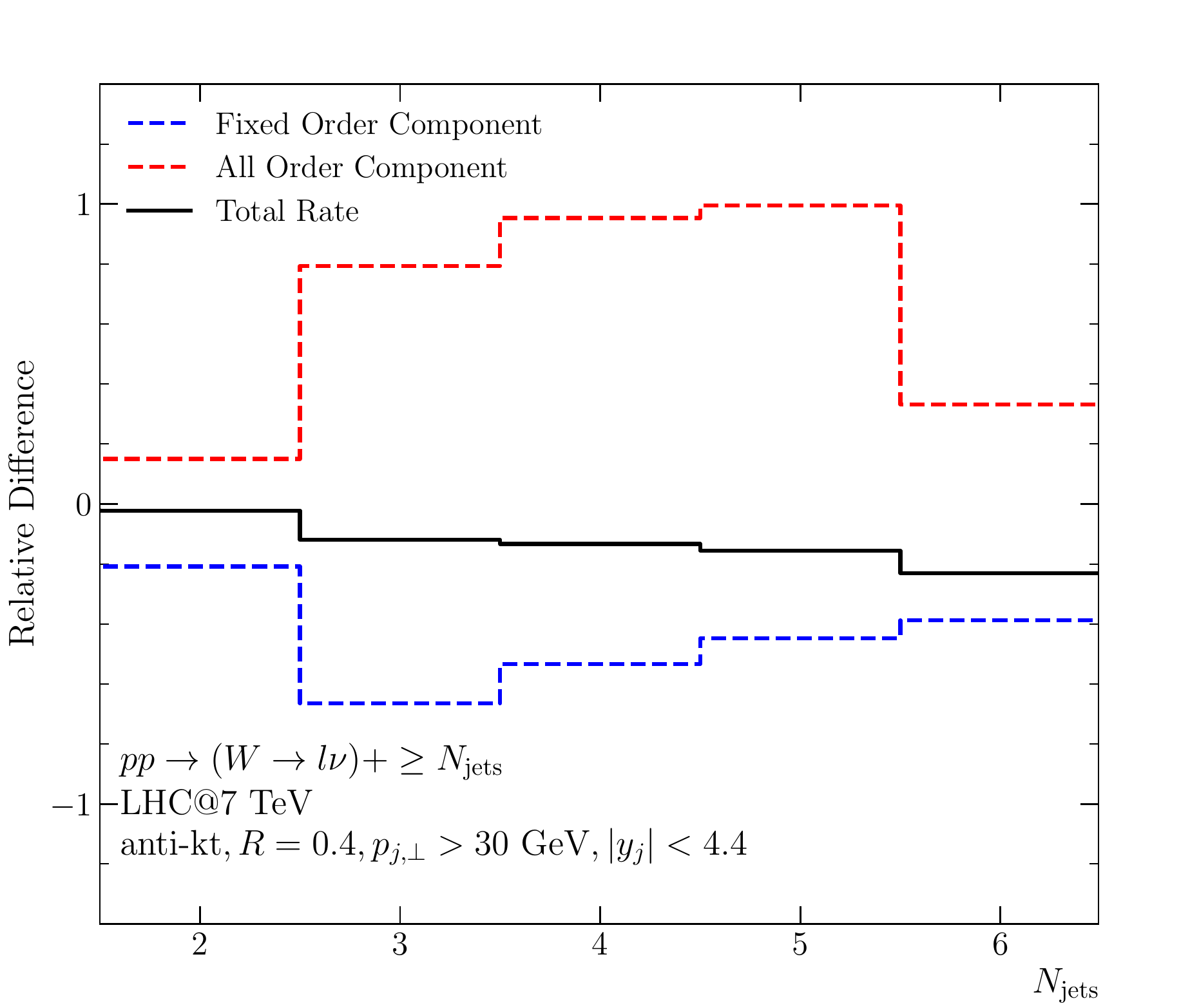}\\
  (c)
  \caption{Inclusive $N_\mathrm{jets}$ cross sections for $pp\to
    (W\to\ell\nu)+\ge N_\mathrm{jets}$ where resummation is applied only to
    LL states (a) and where resummation is applied to all LL and
    NLL states (b).  Also shown in each case is the breakdown into the
    component where all-order resummation is applied (red, dashed) and the
    component which remains described at fixed-order only (blue, dashed).  Panel (c) shows the relative change in each line. Further details are given in \cref{sec:impact-including-nll}.}
  \label{fig:Winclusivejet}
\end{figure}

\begin{figure}
  \centering
  \includegraphics[width=0.49\linewidth]{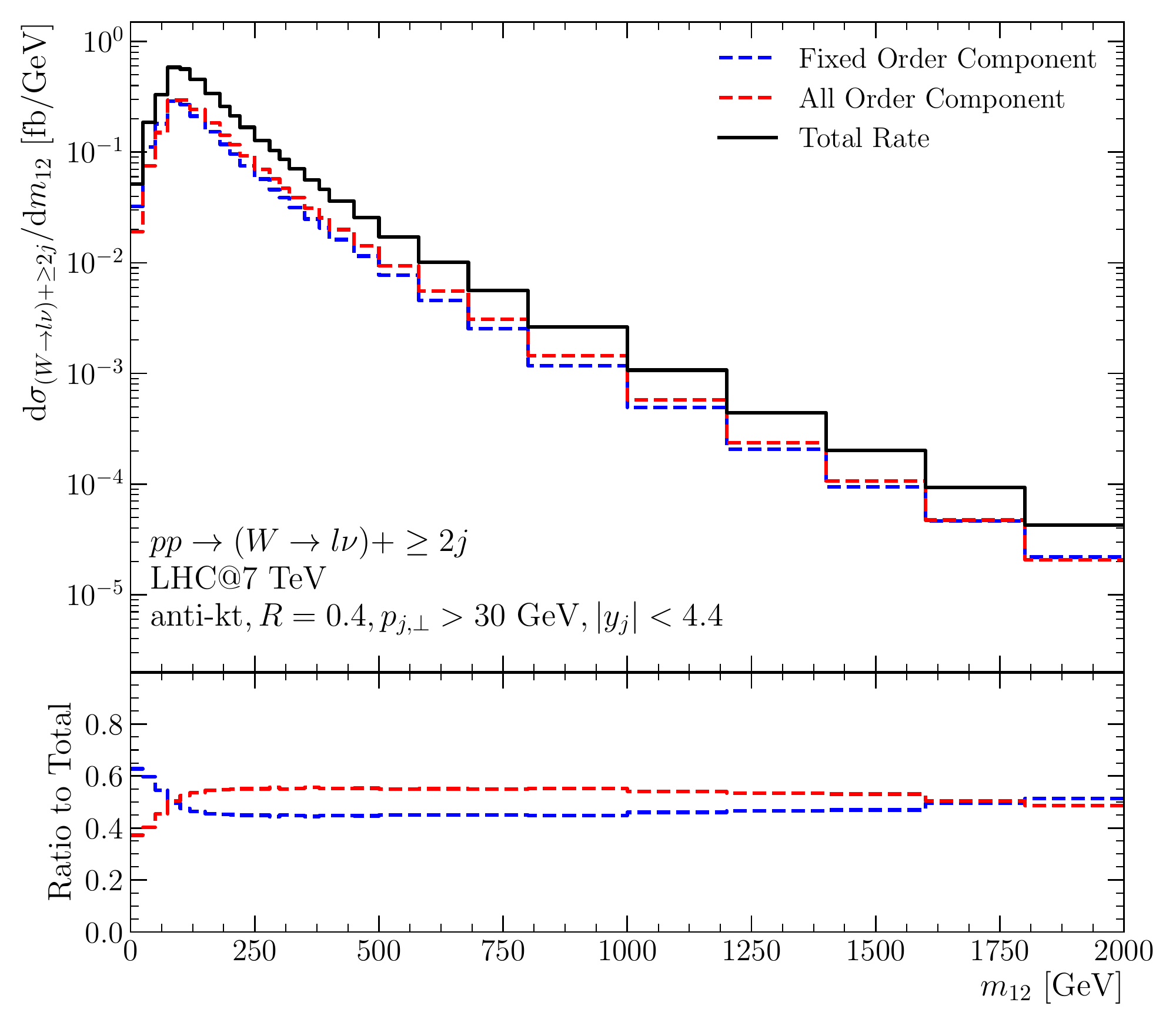}\hfill
  \includegraphics[width=0.49\linewidth]{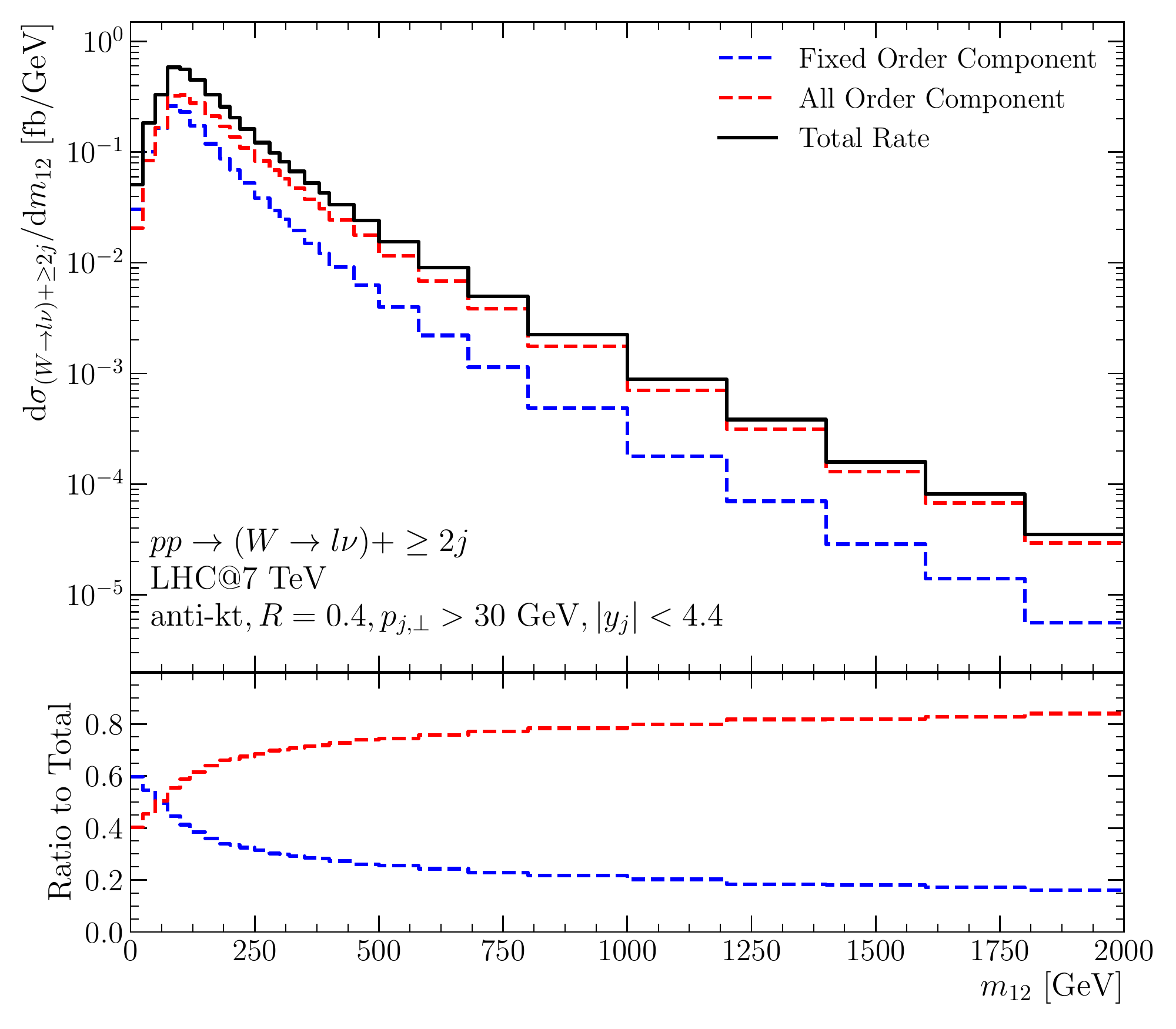}\\
  (a) \hspace{7cm} (b) \\
  \includegraphics[width=0.49\linewidth]{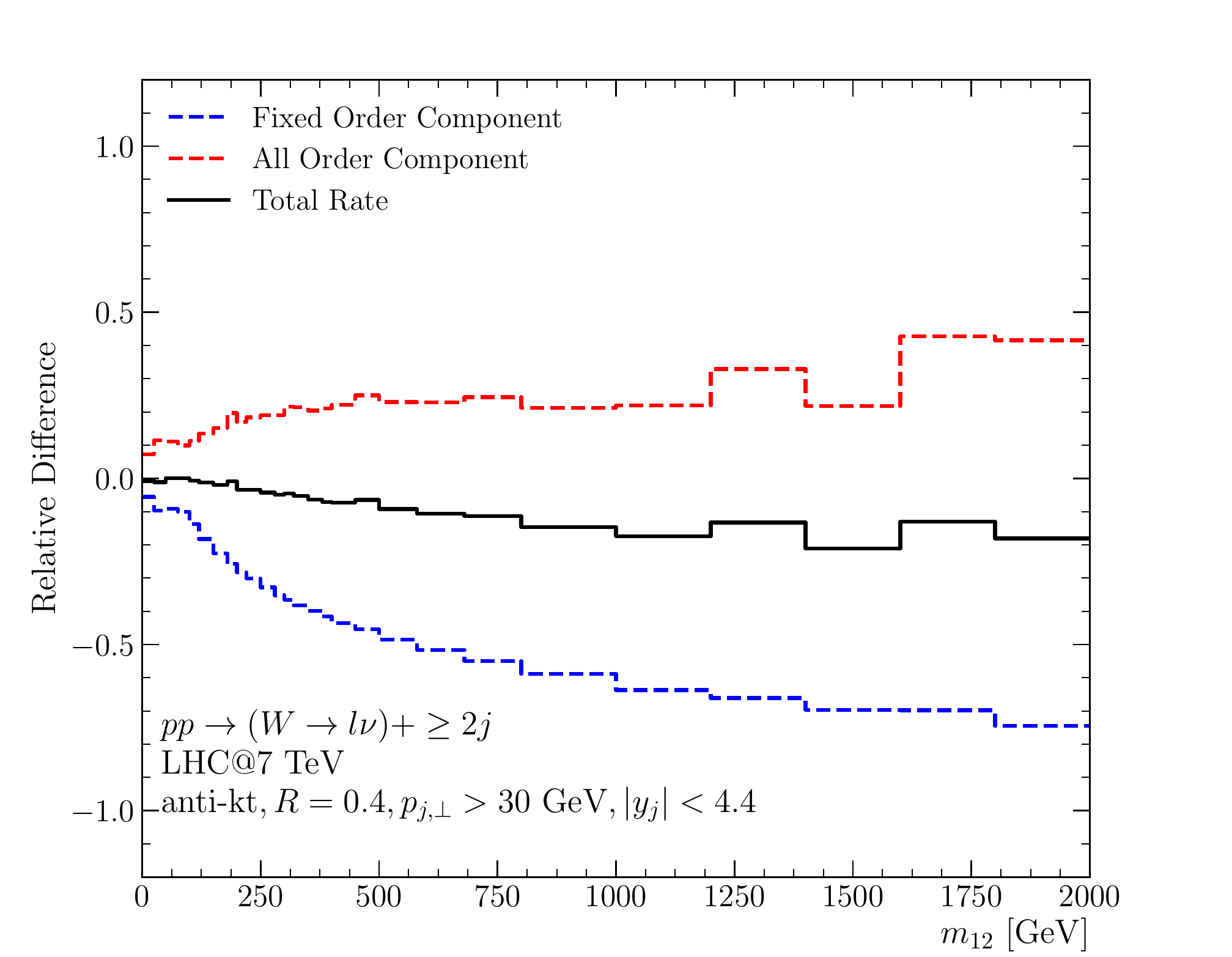}\\
  (c)
  \caption{The differential distribution (black, solid) in the invariant mass of
    the two leading jets in $pp \to (W\to\ell\nu)\ge 2j$, without and with resummation applied to
    NLL states.  The panels and lines are as in \cref{fig:Winclusivejet}.}
  \label{fig:2m12}
\end{figure}

\begin{figure}
  \centering
  \includegraphics[width=0.49\linewidth]{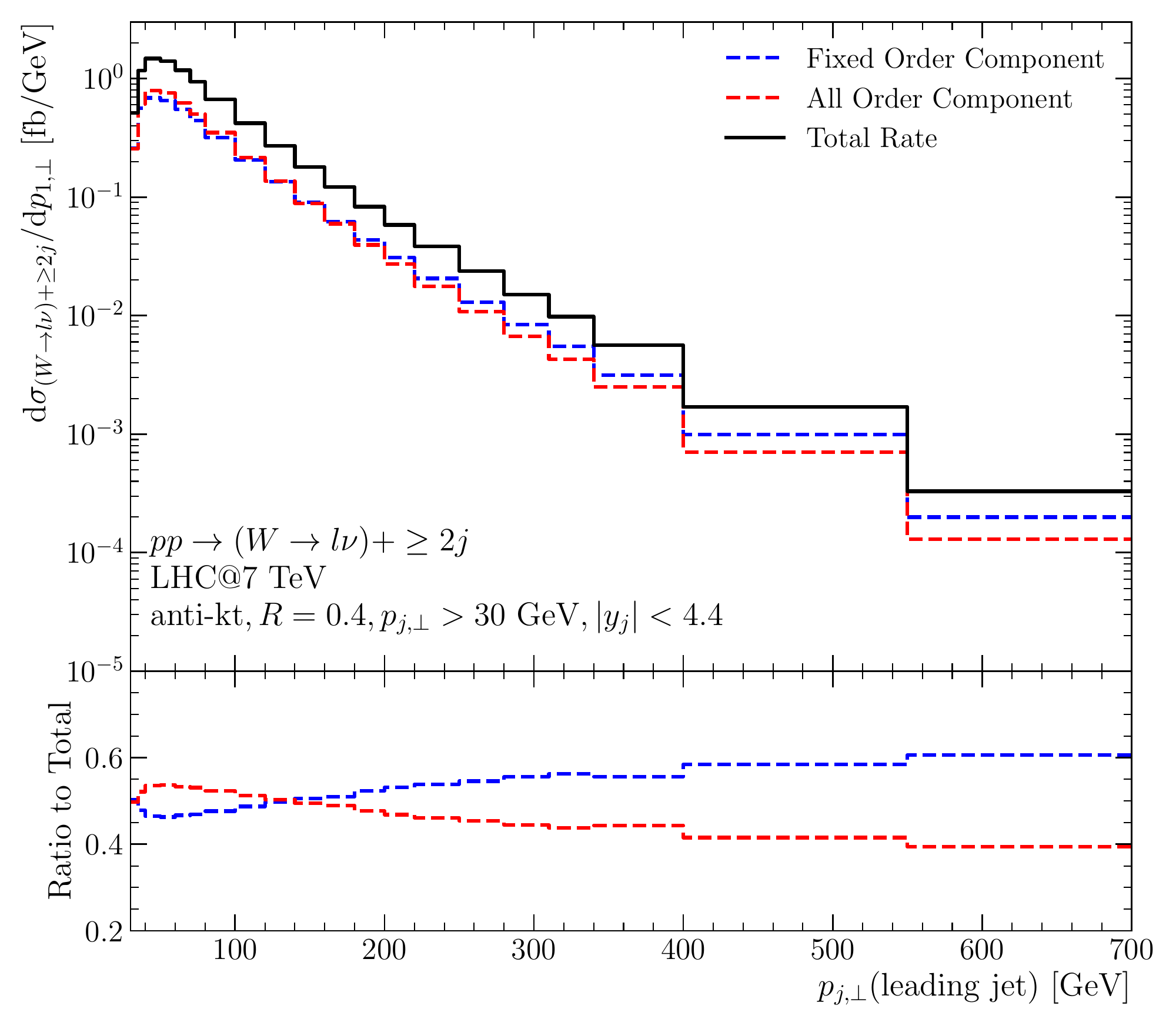}\hfill
  \includegraphics[width=0.49\linewidth]{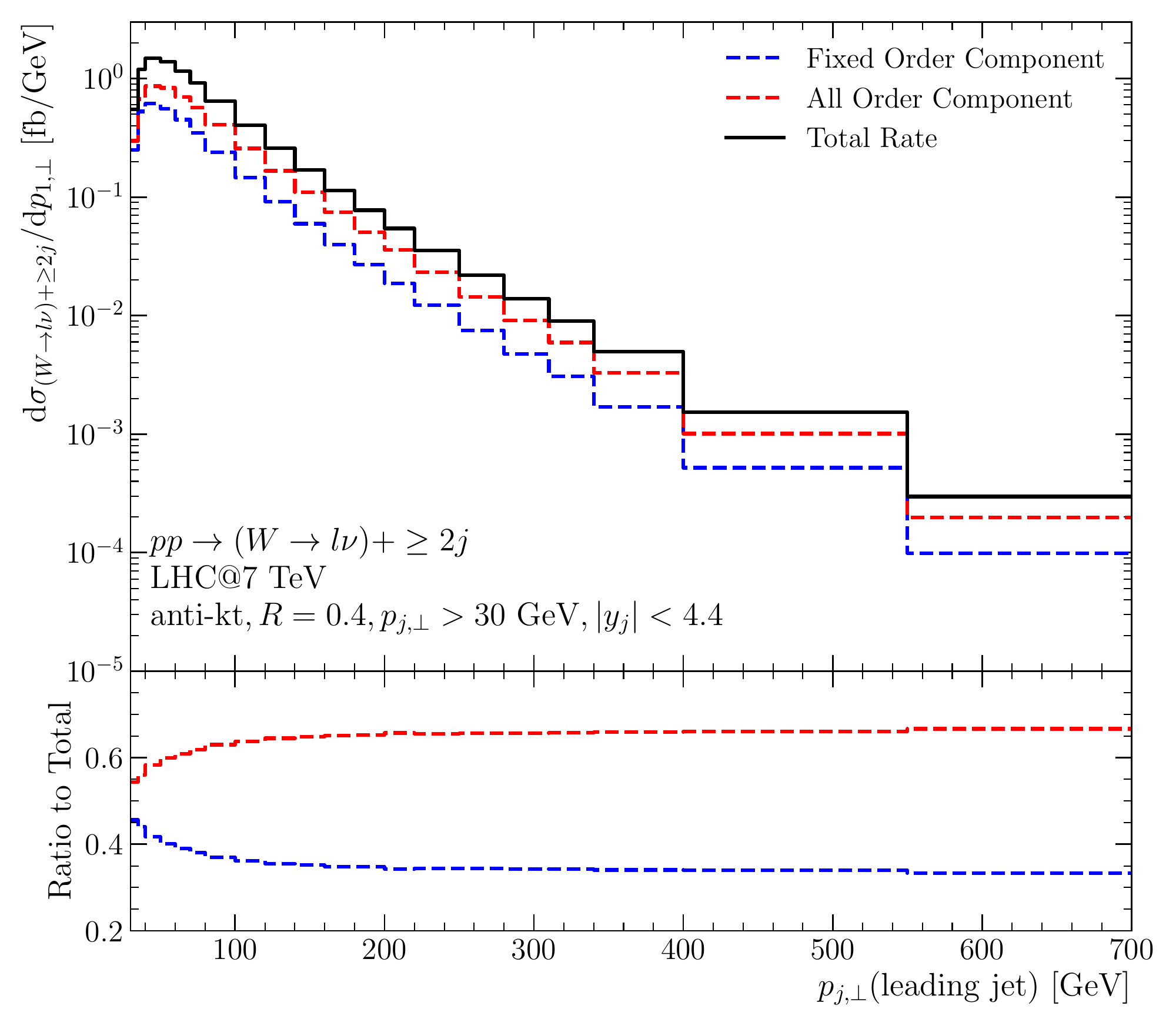}\\
  (a) \hspace{7cm} (b) \\
  \includegraphics[width=0.49\linewidth]{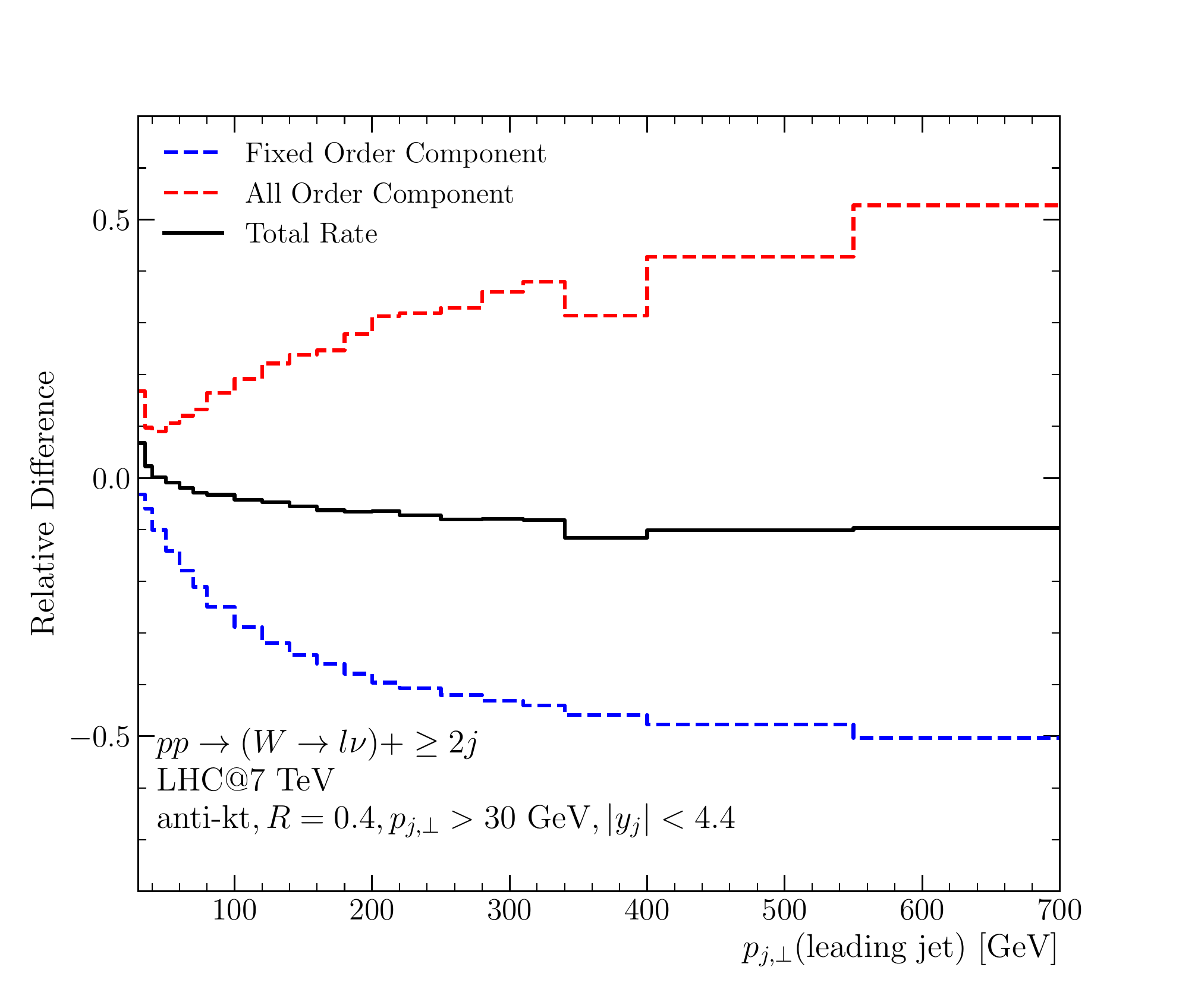}\\
  (c)
  \caption{The differential distribution (black, solid) in the transverse momentum of the
    leading jet in $pp \to (W\to\ell\nu)\ge 2j$, without and with resummation applied to
    NLL states.  The panels and lines are as in \cref{fig:Winclusivejet}.}
  \label{fig:2pt1}
\end{figure}

In this appendix, we further illustrate the numerical impact of the new NLL
components developed in \cref{sec:subleading}, adding to the discussion in
\cref{sec:impact-including-nll}.  They are presented in the same way as in that
section: the total rate is shown by the black, solid line.  It comprises a
component where all-order corrections are applied (red, dashed) and a component
which is taken from fixed-order matching (blue, dashed).  Panel (a) of figs.~\ref{fig:Winclusivejet}--\ref{fig:2pt1} shows the results where all-order resummation is applied only to LL
states.  Panel (b) of figs.~\ref{fig:Winclusivejet}--\ref{fig:2pt1} shows the results where all-order resummation is applied to LL and NLL states.  The lower plot in each case shows the relative difference in each of
the lines.

Fig.~\ref{fig:Winclusivejet} shows the inclusive jet rates for $pp\to
(W\to\ell\nu)+\ge N_\mathrm{jets}$. When resummation is only applied to LL states (\cref{fig:Winclusivejet}(a)), one can see that the all-order component successively decreases with
each multiplicity which can be understood from a combinatoric argument: for a
given set of incoming particles, the
number of non-LL states increases with the number of jets, where there remains
only one LL state.  The new NLL components included in
this paper only apply to events with three or more jets, and once included
lead to an increase in the all-order component from $W+\ge2j$ to $W+\ge3j$ (\cref{fig:Winclusivejet}(b)).  The rate
then decreases for the same reason as above, but remains above half for $W+\ge4j$.
There is only a modest relative effect
on the all-order and fixed-order components of the inclusive $W+\ge2j$ rate (of
about 20\%) as the change only affects events with three or more jets.  Above
this, the relative increase in the all-order component is around 80\% for
$W+\ge3j$, $W+\ge4j$ and $W+\ge5j$ clearly illustrating how much more of the
cross section is now controlled by the resummation.

In \cref{fig:2m12}, we show the comparison now for the invariant mass
distribution of the two leading jets in $pp\to (W\to\ell\nu)+\ge 2j$.  When
all-order corrections are applied to just LL states (\cref{fig:2m12}(a)), we see that the all-order component begins at about 37\%, then quickly rises to around 55\% by $m_{12}=125$~GeV and remains
flat right up to $m_{12}=2$~TeV.  When all-order corrections are also added to
NLL states (\cref{fig:2m12}(b)), the all-order component begins at a similar value but now quickly
rises to much larger values, finally reaching a plateau at about 80\% of the
total rate for large values of $m_{12}$.  Large $m_{12}$ is closely related to
the MRK limit, but not identical because the MRK limit requires large invariant
mass between \emph{all} jets and not just between the leading two.  Again \cref{fig:2m12}(c) shows the relative difference in each line with the change in the total
rate being much smaller than in each component.  All three lines show very little
relative difference at low $m_{12}$, but then the all-order component shows a relative increase
of up to 40\% while the fixed-order component increasingly decreases down to
about -60\%.  As the logarithmic corrections grow with invariant mass, we expect
this increasing impact with $m_{12}$.

Finally in \cref{fig:2pt1} we show the transverse momentum distribution of the
leading jet in $pp\to (W\to\ell\nu)+\ge2j$ events.  The same distribution for
$pp\to (W\to\ell\nu)+\ge3j$ and $pp\to (W\to\ell\nu)+\ge4j$ was shown in
\cref{fig:2pt2,fig:2pt3}.  Unlike the higher-multiplicity processes, for
inclusive 2-jet events one can see that the all-order component is larger than
the fixed-order component for low values of $p_{j,\perp}$, which is consistent
with the jet rates in \cref{fig:Winclusivejet}.  It does steadily fall as
$p_{j,\perp}$ increases down to 40\%.  After all-order corrections are applied
also to the NLL states though (\cref{fig:2pt1}(b)), the behaviour changes and the all-order component
rises as a fraction of the total, up to 67\% by $p_{j,\perp}=700$~GeV.  Again in
this variable, the relative difference in the lines is very small at low values
of $p_{j,\perp}$ and then the components show much larger differences, up to
50\% for the all-order component and down to -50\% for the fixed-order
component.

\clearpage

%%% Local Variables:
%%% mode: latex
%%% TeX-master: "../main"
%%% End:

\bibliographystyle{JHEP}
\bibliography{papers}

\end{document}